\documentclass[twoside, a4paper, notoc, 10pt ]{JHEP3}
\usepackage{amsfonts}
\usepackage{amssymb}
\usepackage{comment}
\usepackage{epsfig}
\usepackage{graphicx}
\usepackage{amsmath}
\usepackage{amsxtra}

\newcommand{\bea}{\begin{eqnarray}}
\newcommand{\eea}{\end{eqnarray}}
\newcommand{\vo}{{\cal V}}
\newcommand{\be}{\begin{equation}}
\newcommand{\ee}{\end{equation}}

\title{Reheating for closed string inflation}

\author{Michele Cicoli~$^{1}$, and Anupam Mazumdar~$^{2,~3}$\\
$^1$~Deutsches Elektronen-Synchrotron DESY,
Notkestrasse 85, 22603 Hamburg, Germany \\
$^{2}$~Physics Department, Lancaster University, Lancaster, LA1 4YB, United Kingdom\\
$^{3}$~Niels Bohr Institute, Copenhagen University, Blegdamsvej-17, Copenhagen-2100, Denmark}

\abstract{We point out some of the outstanding challenges for embedding inflationary cosmology within string theory
studying the process of reheating for models where the inflaton is a closed string mode parameterising the size of an internal
cycle of the compactification manifold. A realistic
model of inflation must explain the tiny perturbations in the cosmic microwave background radiation and also how to
excite the ordinary matter degrees of freedom after inflation, required for the success of Big Bang Nucleosynthesis.
We study these issues focusing on two promising inflationary models embedded in LARGE volume type IIB flux compactifications.
We show that phenomenological requirements and consistency of the effective field theory treatment imply the
presence at low energies of a \textit{hidden sector} together with a {\it visible sector}, where
the Minimal Supersymmetric Standard Model fields are residing. A detailed calculation of
the inflaton coupling to the fields of the hidden sector, visible sector, and moduli sector,
reveals that the inflaton fails to excite primarily the visible sector fields, instead hidden sector fields
are excited copiously after the end of inflation. This sets severe constraints on hidden sector model building
where the most promising scenario emerges as a pure $N=1$ SYM theory, forbidding the kinematical decay
of the inflaton to the hidden sector.
In this case it is possible to reheat the Universe with the visible degrees of freedom even though in some cases
we discover a new tension between TeV scale SUSY and reheating on top of the well-known tension between TeV scale SUSY and
inflation.}

\preprint{DESY 10-072}
\begin{document}

\tableofcontents

\bigskip

\section{Introduction}

Primordial inflation has been a very successful paradigm which can explain the observed temperature anisotropy
in the cosmic microwave background (CMB) radiation~\cite{WMAP7} (for a recent review see~\cite{RM}). Any successful model of inflation must also explain how to excite the Standard Model (SM) quarks and leptons after the end of inflation through the process of reheating (for a review see~\cite{Reheat-rev}) required for the success of Big Bang Nucleosynthesis~\cite{BBN}. In this regard only visible sector models of inflation can be considered to be safe where the inflaton carries the SM charges~\cite{MSSM-inf}. Such models are primarily based on the Minimal Supersymmetric Standard Model (MSSM), where supersymmetry (SUSY) helps
in obtaining a {\it cosmologically flat} potential via $D$-flat directions (for a review on MSSM flat directions see~\cite{MSSM-rev}). The advantage of
visible sector models  embedded within the MSSM is that it can also reproduce the required dark matter abundance necessary for the structure formation~\cite{ADM}.

On the contrary there exists a plethora of models of inflation where the inflaton belongs to the hidden sector since it is a SM gauge singlet. Typically
these models can explain the temperature anisotropy in the CMB~\cite{RM}, but fail to explain how the inflaton energy density gets transferred primarily
to visible sector degrees of freedom (\textit{dof}), and not to hidden sector ones. Note that \textit{a priori} a gauge singlet inflaton has no preference to either the visible or the hidden sector. Therefore, it is a challenge to construct a model of inflation where the inflaton belongs to the hidden sector and motivate the inflaton couplings to hidden and visible \textit{dof}.

It is typical to encounter such a scenario in string theory which provides
many closed string modes that are good inflaton candidates. In particular,
the low energy effective action of string compactifications on
Calabi-Yau three-folds typically has a large number of uncharged massless scalar fields with a flat potential, called moduli.
These moduli can be the ideal candidates to drive inflation. Since their vacuum expectation value (VEV) determines various parameters (like masses and coupling constants) of the four-dimensional effective theory, it is important to stabilise them by providing
masses to these moduli. Otherwise, the presence of these massless scalars with effective gravitational coupling would mediate unobserved long range fifth forces.
It is well known by now that nonzero background fluxes induce potentials for some of the moduli~\cite{gvw,drs,gkp}.
However these fluxes severely backreact on the Calabi-Yau geometry generically modifying the internal
space to a manifold with a more complicated structure. This is not the case for type IIB compactifications where
the fluxes induce just a warp factor. For this reason, moduli stabilisation is best understood in the context
of type IIB which is therefore well suited for the discussion of phenomenology and cosmology. However the stabilisation of
all the geometric moduli requires to take into account various perturbative and non-perturbative effects \cite{kklt, LVS, GenAnalofLVS}.

One of the nice properties of type IIB flux compactifications is that global (or bulk)
issues decouple from local (or brane) issues. The former are model independent issues
which depend on moduli stabilisation and some of them are:
(a)~Different hierarchical scales,~(b)~SUSY breaking,~(c)~Soft SUSY breaking mass terms,~(d)~Cosmological constant,
and ~(e)~Inflation. On the other hand, the local issues depend on the particular local $D$-brane construction
and some of them are: (a)~ Gauge group,~(b)~Chiral spectrum,~(c)~Yukawa couplings,~(d)~Gauge coupling unification,
~(e)~Proton stability, and ~(f)~Baryogenesis.

An excellent example of type IIB compactifications with stabilised moduli
is given by the LARGE Volume Scenario (LVS) originally proposed in \cite{LVS}.
This is very appealing for particle physics phenomenology and cosmology. In these
compactifications $\alpha'$ and $g_s$ corrections can be combined with non-perturbative effects in order
to generate a potential for all the K\"{a}hler moduli \cite{GenAnalofLVS}, whereas the background fluxes induce a potential for the dilaton and the
complex structure moduli. Unlike the KKLT set-up \cite{kklt}, the moduli stabilisation is performed without fine tuning of the
values of the internal fluxes and the Calabi-Yau volume is fixed at an exponentially large value (in string units). As a consequence, one
has a very reliable four-dimensional effective description, as well as a tool for the generation of phenomenologically desirable
hierarchies.

There are two different choices to embed the visible sector within LVS:
\begin{enumerate}
\item {\it Geometric regime}: the visible sector is built via magnetised
intersecting $D7$ branes wrapping a blow-up 4-cycle which is stabilised at a value larger than the string scale~\cite{GeomRegime};

\item {\it Quiver locus}: the visible sector is built via fractional $D$-branes at the singularity obtained by shrinking down a blow-up mode~\cite{quiver}.
\end{enumerate}
Both of them have different phenomenology, the scales and hierarchies are different and we will briefly review them in the following section.

One of the interesting features for all these setups is that at low energies they always come with a hidden and a visible sector.
Therefore these models will serve as a perfect playground to study the inflaton couplings to both the \textit{dof}.
Furthermore, the two sectors do not have any direct coupling among themselves, except indirectly via the closed string sector.
In this respect the two sectors are isolated from each other, nevertheless their total energy density will govern the dynamics of the Universe.
It is then crucial to find an answer to each of the following questions:
how does the inflaton couple to each sector? how does perturbative reheating take place? is it at all possible to excite predominantly the MSSM \textit{dof}? Thankfully, these questions can be answered within this setup
which involves an interesting interplay between local and global issues. We shall stress that reheating
can set severe constraints both on the hidden sector physics and the scale of inflation which should be compatible with CMB predictions,
successful reheating, and TeV scale SUSY\footnote{Reheating and thermalisation have been studied previously in the context of closed string excitations in brane inflation~\cite{Many-reheat}, in the case of warped geometry~\cite{Warped-reheat}, and in closed string models of inflation~\cite{Green-reheat, Preheating}. In all these cases, no clear distinction has been made between hidden and visible \text{dof}. In this respect such studies are interesting
but leave little impact on how the MSSM \textit{dof} are excited, whose answer is relevant for the success of BBN.}.

In order to study these issues, we shall focus on two very promising models of string inflation embedded in LVS, where the
inflaton is a closed string mode, more precisely a K\"{a}hler modulus parameterising the size of an internal 4-cycle:
(1)~{\it Blow-up inflation} (BI) \cite{kahlerinfl}: in this case the inflaton is the size of a blow-up mode yielding a small field inflationary model
which is in good agreement with current observational data;
and (2)~{\it Fibre inflation} (FI) \cite{fiberinfl}: in this case the inflaton is the size of a K3 fibre over
a $\mathbb{C}P^1$ base producing a large field inflationary model which can be a potential candidate if primordial gravity waves are discovered.
These models represent some of the few known
examples where inflation can be achieved without fine-tuning any parameter in the potential. In fact it is the
typical no-scale structure of the potential that allows to solve the $\eta$-problem in a natural way.

In spite of the successes of these models we shall discover two outstanding challenges for closed string inflationary scenarios:
\begin{enumerate}
\item Most of the inflaton energy gets dumped to hidden sector \textit{dof}. This will lead to
overproduction of hidden sector dark matter if there exists a stable species. This problem
seems to rule out all the inflationary models unless the hidden sector consists just
of an $N=1$ pure SYM theory. This severe constraint on hidden sector model building can be relaxed only
for the geometric regime case with the inflaton wrapped by the visible sector $D7$-stack.

\item Incompatibility between TeV scale SUSY and reheating
on top of the well-know tension between inflation and TeV scale SUSY \cite{InfvsSUSY}: assuming that
the density perturbations can be generated in a non-standard way by a second curvaton or modulating field \cite{BCGQTZ},
one could set the scale of the scalar potential in order to get inflation, the correct amount of CMB temperature
fluctuations and TeV scale SUSY at the same time. However it might still be impossible to achieve an efficient
reheating of the visible sector due to the fact that the inflaton might decay after BBN.
This problem seems to rule out all the FI models (with the possible exception of FI at the quiver locus)
and BI in the geometric regime with the
inflaton not wrapped by the visible sector. The only models which are left over are BI
in the geometric regime with the inflaton wrapped by the visible sector or BI at the quiver locus.
It is important to stress that both these models require fine-tuning to have a successful inflationary scenario.
%
\end{enumerate}

This paper is organised as follows. In section 2, we give a brief review of LVS stressing the
different features of the geometric regime and the quiver locus case. We then discuss the dynamics
of BI and FI presenting their predictions for the cosmological observables. We finally briefly describe the process of perturbative and non-perturbative
reheating and thermalisation. In section 3, we discuss the dynamics of the hidden sector and the constraints on hidden sector
\textit{dof}. In section 4, we describe the inflaton couplings to both hidden and visible \textit{dof} estimating the maximal reheating temperature
in the approximation of sudden thermalisation of the visible sector for BI and FI, respectively. In section 5, we discuss our main results and conclude.
In appendix \ref{appendice}, we present a detailed derivation of the moduli canonical normalisation, mass spectrum, couplings and decay rates to any
particle present in our models both for BI and FI. More precisely, we have derived the moduli couplings
to fermions/sfermions, Higgs/Higgsinos, gauge bosons/gauginos of the visible and hidden sector, and the moduli self couplings.


\section{Inflation in the LARGE Volume Scenario}

In this section we shall present a brief review of the LVS and then proceed to
describe the main features of the two inflationary models under consideration: blow-up and fibre inflation.

\subsection{LARGE Volume Scenario}

The LVS emerges naturally in the context of type IIB
flux compactifications on Calabi-Yau orientifolds in the presence of space-time filling $D3-D7$ branes and
$O3-O7$ planes. The overall volume of the Calabi-Yau is stabilised exponentially large by
the interplay of non-perturbative and $\alpha'$ corrections without fine-tuning the background fluxes \cite{LVS}.
Two big advantages of this scenario are that the effective field theory is under good control
and one can generate hierarchies via the exponentially large volume.
Expressing the K\"{a}hler moduli as $T_i=\tau_i+i b_i$, $i=1,...,h_{1,1}$, with $\tau_i$ the volume of the internal 4-cycle
$\Sigma_i$ and $b_i= \int_{\Sigma_i}C_4$, the only two conditions on an arbitrary Calabi-Yau to obtain the LVS are \cite{GenAnalofLVS}:
\begin{enumerate}
\item $h_{1,2}>h_{1,1}>1$, where $h_{1,2}$ gives the number of complex structure moduli which are flux stabilised at tree-level.

\item The existence of at least one blow-up mode $\tau_s$ resolving a point-like singularity that gets non-perturbative
corrections: $W=W_0+A_s e^{-a_s T_s}$. Then the overall volume can be fixed such that $\vo\simeq \sqrt{\tau_s} \,e^{a_s \tau_s}$.
\end{enumerate}
The simplest examples of LVS are:
\begin{itemize}
\item Swiss-Cheese Calabi-Yau three-folds with volume given by \cite{ExplicitSC}:
\be
\vo=\alpha\left(\tau_b^{3/2}-\sum_{i} \gamma_i \tau_i^{3/2}\right).
\ee
In this case all the $\tau_i$ are rigid cycles which are fixed small by non-perturbative corrections,
whereas the large cycle $\tau_b$ is stabilised exponentially large due to $\alpha'$ and non-perturbative
effects.

\item K3-fibered Calabi-Yau three-folds with volume of the form \cite{ExplicitK3}:
\be
\vo=\alpha\left(\sqrt{\tau_1}\tau_2-\sum_{i} \gamma_i \tau_i^{3/2}\right).
\ee
In this case still all the rigid cycles $\tau_i$ are fixed small by non-perturbative corrections,
whereas the overall volume $\vo\simeq \sqrt{\tau_1}\tau_2$ is stabilised exponentially large due to $\alpha'$ and non-perturbative
effects. The remaining flat direction can be frozen via the inclusion of subleading string loop corrections \cite{stringloops} which naturally
fix both $\tau_1$ and $\tau_2$ large.
\end{itemize}
Particular attention has to be paid on the stabilisation of the cycle supporting the {\it visible sector}
which we shall call $\tau_4$ for later convenience in the notation. In fact $\tau_4$ has to be
one of the small cycles in order not to get a visible gauge coupling $g^{-2}\sim \tau_4$ which is too small.
However this cycle cannot get any non-perturbative correction since any instanton wrapped around $\tau_4$ will
have chiral intersections with the visible sector inducing prefactors that depend on visible chiral fields \cite{blumen}:
\be
W_{np}\supset \left(\Pi_i \Phi_i \right)e^{-a_4 T_4}=0\text{ \ \ for \ \ }\langle\Phi_i\rangle=0.
\ee
Hence $\tau_4$ cannot be stabilised by non-perturbative effects but it can still be fixed by either $D$-terms \cite{quiver, ExplicitK3, blumen, IntersectingD} or
string loop corrections to the K\"{a}hler potential \cite{GenAnalofLVS}. However $\tau_4$ can be stabilised
either in the geometric regime, i.e. at values larger than the string scale,
or it can shrink down at the singularity. These two regimes are characterised
by different effective field theories with different features, which we discuss below.

\subsubsection{Geometric Regime}

In this case the cycle supporting the visible sector is stabilised in the geometric regime. There are two ways to do it:
\begin{enumerate}
\item via $D$-terms for intersecting rigid cycles \cite{ExplicitK3, blumen, IntersectingD}.

\item via string loop corrections to the K\"{a}hler potential \cite{GenAnalofLVS}.
\end{enumerate}
In this case the visible sector is built via magnetised intersecting $D7$-branes wrapping $\tau_4$.
Supersymmetry is broken due to non-vanishing background fluxes by the $F$-terms of the K\"{a}hler moduli \cite{GeomRegime}. Hence we are in the
presence of gravity, or better moduli-mediated supersymmetry breaking. In particular, the $F$-term of $\tau_4$ is non-zero:
\be
F^4\simeq e^{K/2}K^{4\bar{4}}\left(W_{\bar{4}}+W_0 K_{\bar{4}}\right)
=e^{K/2}K^{4\bar{4}}W_0 K_{\bar{4}}\sim \frac{W_0\tau_4}{\vo}\neq 0.
\ee
Then the main scales in the model are \cite{GeomRegime}:
\begin{itemize}
\item Planck-scale: $M_P\sim 10^{18}$ GeV.

\item String-scale: $M_s\sim \frac{M_P}{\sqrt{\vo}}$.

\item Kaluza-Klein scale: $M_{KK}\sim \frac{M_s}{\tau_4^{1/4}}\sim \frac{M_P}{\tau_4^{1/4}\sqrt{\vo}}$.

\item Gravitino mass: $m_{3/2}\sim \frac{M_P}{\vo}$.

\item Soft-terms: $M_{soft}\sim m_{3/2}$.

\item Small blow-up modes: $m_{\tau_i}\sim m_{3/2}$.

\item Volume mode: $m_{\vo}\sim \frac{M_P}{\vo^{3/2}}$.

\item Large mode orthogonal to the volume (only for K3-fibrations) \cite{GenAnalofLVS}: $m_{\chi}\sim \frac{M_P}{\vo^{5/3}}$.
\end{itemize}

Setting the volume $\vo \simeq 10^{14- 15}$ in string units, corresponding to an intermediate
$M_s\simeq 10^{11- 12}$ GeV, we have the following good phenomenological properties:
\begin{itemize}
\item TeV-scale SUSY: $M_{soft}\sim 1$ TeV.

\item Right QCD axion scale \cite{axion}: $f_a \sim  M_s \sim 10^{11}$ GeV.

\item Right Majorana scale for right handed neutrinos \cite{neutrino}: $M_{\nu_{R}}\sim M_P \vo^{-1/3}\sim 10^{14}$ GeV,
required for generating the observed light neutrino masses via the see-saw mechanism.
\end{itemize}
\noindent
We also point out that there are a few shortcomings:
\begin{enumerate}
\item No standard GUT theory since the scale for minimal $SU(5)$ or $SO(10)$ is much larger than the string scale.

\item No viable inflationary scenario given that BI works for $\vo\simeq 10^{6- 7}$, whereas
FI works for $\vo\simeq 10^{3- 4}$. The latter scales are fixed by the observation of temperature anisotropies in
the cosmic microwave background radiation~\cite{WMAP7}.

\item Cosmological moduli problem for the volume mode (and $\chi$ in the case of compactifications on K3-fibered
Calabi-Yau three-folds).
\end{enumerate}

All these three problems could be solved at the same time by setting the volume smaller, of the order $\vo\simeq 10^4$,
and then fine-tuning the back-ground fluxes to give rise to a highly warped region where the visible sector sits.
In this way $M_{soft}$ would redshift down to the TeV scale. However warping is usually negligible in constructions
with exponentially large overall volume, and so this possibility does not seem very attractive.

As far as GUT theories are concerned, it is fair to say that there is no string model with $M_s\sim M_{GUT}$
which is able to reproduce the standard picture of gauge coupling unification, and so we do not consider LVS with
an intermediate string scale less promising than other models in this regard. In addition there are left-right
symmetric constructions where the gauge couplings unify at $10^{11-12}$ GeV \cite{StrangeGUTs}.

The authors of \cite{VolumeInflation} proposed a model to reconcile the high scale of inflation with the lower scale of the
soft SUSY breaking terms, by invoking inflation driven by the volume mode, even though this model requires
a good amount of fine-tuning. Another way to get inflation and TeV scale SUSY at the same time, could be to generate the
density perturbations via a non-standard mechanism, like a curvaton or a modulating field \cite{BCGQTZ}. In this way, one
might be able to lower the inflationary scale and solve, or more likely, attenuate this tension between cosmology and
particle phenomenology. Finally it would be interesting to investigate if,
in the case of compactifications with several moduli, the
cosmological moduli problem could be solved due to the dilution factor induced by the out-of-equilibrium decay of
a heavy modulus.

\subsubsection{Quiver locus}

When the cycle supporting the visible sector, $\tau_4$, is a rigid cycle which does not intersect with other cycles,
it can shrink down at the singularity, $\tau_4\to 0$, due to $D$-terms, in the absence of visible sector singlets which can
get a non-vanishing VEV \cite{quiver}.

In this case the visible sector is built via $D3$-branes at the quiver locus.
Supersymmetry is still broken due to non-vanishing background fluxes by the $F$-terms of the K\"{a}hler moduli which
then mediate this breaking to the visible sector. However in this case the $F$-term of $\tau_4$ is vanishing:
\be
F^4\simeq e^{K/2}K^{4\bar{4}}\left(W_{\bar{4}}+W_0 K_{\bar{4}}\right)
=e^{K/2}K^{4\bar{4}}W_0 K_{\bar{4}}\sim K^{4\bar{4}}W_0 \xi_{FI}= 0,
\ee
given that the $\tau_4$-cycle is stabilised by requiring a vanishing Fayet-Iliopoulos term $\xi_{FI}=0$.
Therefore there is no local SUSY-breaking and the visible sector is sequestered. This implies that now
the soft terms do not scale as the gravitino mass anymore but they can even be suppressed with respect to $m_{3/2}$
by an inverse power of the volume\footnote{This might not be the case in the presence of a 1-loop
redefinition of $\tau_4$ \cite{Pedro}.}. Then the main scales in the model are \cite{quiver}:
\begin{itemize}
\item Planck-scale: $M_P\sim 10^{18}$ GeV.

\item GUT-scale: $M_{GUT}\sim \vo^{1/6} M_s \sim \frac{M_P}{\vo^{1/3}}$.

\item String-scale: $M_s\sim \frac{M_P}{\sqrt{\vo}}$.

\item Kaluza-Klein scale: $M_{KK}\sim \frac{M_P}{\vo^{2/3}}$.

\item Gravitino mass: $m_{3/2}\sim \frac{M_P}{\vo}$.

\item Small blow-up modes: $m_{\tau_i}\sim m_{3/2}$.

\item Soft-terms: $M_{soft}\sim \frac{m_{3/2}^2}{M_P}\sim\frac{M_P}{\vo^2}$.

\item Volume mode: $m_{\vo}\sim \frac{M_P}{\vo^{3/2}}$.

\item Large mode orthogonal to the volume (only for K3-fibrations): $m_{\chi}\sim \frac{M_P}{\vo^{5/3}}$.
\end{itemize}
Setting the volume $\vo \simeq 10^{6- 7}$ in string units, corresponding to
$M_s\simeq 10^{15}$ GeV, one finds several nice features \cite{quiver}:

\begin{itemize}
\item Standard GUTs: $M_{GUT}\sim 10^{16}$ GeV.

\item TeV scale SUSY: $M_{soft}\sim 1$ TeV.

\item No cosmological moduli problem as the mass of the moduli are larger than $10$ TeV.

\item Good inflationary model: BI.
\end{itemize}
We also point out that there are two shortcomings:
\begin{enumerate}
\item No correct size for the QCD axion decay constant.

\item No correct size for the Majorana mass term for right-handed neutrinos.
\end{enumerate}
However these two seem not to be unsurmountable problems given that they might very well find viable solutions
looking at open string modes as QCD axion candidates \cite{OpenAxion} or considering non-perturbative effects for the
generation of neutrino masses \cite{Cvetic}.

For all these reasons the quiver locus seems more promising than the geometric regime. In this paper
we shall investigate if this is still the case once we focus on the study of reheating of the MSSM \textit{dof} after the end of inflation.

\subsection{Blow-up Inflation}
\label{BupInfl}

In this case the inflaton is the size of a blow-up mode \cite{kahlerinfl}. The volume looks like:
\be
\vo=\alpha\left(\tau_1^{3/2}-\gamma_2\tau_2^{3/2}-\gamma_3\tau_3^{3/2}-\gamma_4\tau_4^{3/2}\right),
\ee
while the tree-level and non-perturbative superpotential takes the form:
\be
W=W_0+A_2 e^{-a_2 T_2}+A_3 e^{-a_3 T_3},
\ee
and the tree-level K\"{a}hler potential with the leading order $\alpha'$ correction reads:
\be
K=-2\ln\left(\vo+\frac{\xi}{2 g_s^{3/2}}\right),\text{ \ \ with \ \ }
\xi=\frac{\left(h_{1,2}-h_{1,1}\right)\zeta(3)}{(2\pi)^3}>0.
\label{Kpot}
\ee
The cycle $\tau_4$ is supporting the visible sector and so it does not get any
non-perturbative correction. After minimising the axion directions and assuming that $\tau_4$
is fixed by either $D$-terms or $g_s$ corrections at the quiver locus or in the geometric regime,
the supergravity $F$-term scalar potential is given by:
\be
 V = \sum_{i=2}^3\frac{8 \, a_{i}^{2}A_i^2}{3\alpha\gamma_i}
 \left( \frac{\sqrt{\tau_{i}}}{\mathcal{V}}
 \right) e^{-2a_{i}\tau_i}
 -4 \,\sum_{i=2}^3\,W_{0}a_{i} A_i \left( \frac{\tau _{i}}{\mathcal{V}^{2}}
 \right) \, e^{-a_{i}\tau_i}+\frac{3 \xi W_0^2}{4 g_s^{3/2}\mathcal{V}^{3}}.
 \label{ygfdo}
\ee
This potential completely stabilises $\tau_2$, $\tau_3$ and the
volume $\vo\simeq \alpha \tau_1^{3/2}$ at:
\be
a_i \langle \tau_i \rangle\,=\frac{1}{g_s}\left( \frac{\xi}{2
\alpha J} \right)^\frac23,\hskip0.7cm \langle
 \mathcal{V}\rangle = \left( \frac{ 3 \,\alpha \gamma_i }{4a_{i}A_i}
 \right) W_0 \, \sqrt{\langle \tau_i\rangle }
 \; e^{a_i \langle \tau_i
 \rangle },\text{ \ \ \ }\forall\, i=2,3,\label{minhea}
\ee
where $J=\sum_{i=2}^3\gamma_i/a_i^{3/2}$.
Now displacing $\tau_2$ far from its minimum, due to the rapid exponential suppression,
this blow-up mode experiences a very flat direction which is suitable for inflation.
The reason why we have added the other blow-up mode $\tau_3$, which sits
at its minimum while $\tau_2$ rolls down to the minimum, is to keep the volume minimum
stable during inflation.

In terms of the canonically normalised inflaton $\phi$, the inflationary potential looks like \cite{kahlerinfl}:
\be
V\simeq V_0 -\beta\left(\frac{\phi}{\vo}\right)^{4/3} e^{-a \vo^{2/3}\phi^{4/3}}.
\ee
This is a model of small field inflation, and so no detectable gravity waves are produced during inflation:
$r\equiv T/S \ll 1$. The spectral index turns out to be in good agreement with the observations:
$0.960<n_s<0.967$, and the requirement of generating enough density perturbations fixes $\vo \simeq 10^{6-7}$.
This value of the volume is also the one preferred at the quiver locus to get both TeV scale SUSY and
GUT theories. However for $\vo \simeq 10^{6-7}$ the models in the geometric regime do not yield TeV scale SUSY.
Hence this model of inflation seems to give a strong indication in favour of LVS
at the quiver locus.

\subsubsection{Potential problems with string loops}
\label{ProbBI}

As we have seen in section \ref{BupInfl}, the inflationary potential is generated by tiny non-perturbative
effects which make it naturally very flat. However potential dangerous problems can come from string loop corrections
to the K\"{a}hler potential \cite{stringloops}. As pointed out in \cite{GenAnalofLVS}, due to the extended no-scale structure,
these particular corrections are subleading in a large volume expansion with respect to the non-perturbative effects which
fix the inflationary cycle $\tau_2$ small. However this is true only in a region in field-space around the LVS minimum,
while once $\tau_2$ is displaced far from the minimum to drive inflation, the $g_s$ corrections quickly come to dominate
the non-perturbative ones due to the rapid exponential suppression. Then these corrections spoil the flatness of the potential
which is not suitable for inflation anymore. This can easily be seen from the form of the string loop corrections to the
scalar potential in terms of the canonically normalised inflaton $\phi\sim \vo^{-1/2} \tau_2^{3/4}$:
\be
\delta V_{(g_s)}^{1-loop}\sim \frac{1}{\sqrt{\tau_2}\,\vo^3}\sim\frac{1}{\phi^{2/3}\vo^{10/3}},
\ee
which induce a correction to the $\eta$-parameter of the form:
\be
\delta \eta \sim \frac{\partial^2 \left(\delta V_{(g_s)}^{1-loop}\right)}{\partial \phi^2}\frac{\vo^3 g_s^{3/2}}{\xi}
\sim \frac{g_s^{3/2}}{\phi^{8/3} \vo^{1/3}\xi}\sim \frac{g_s^{3/2}\vo}{\tau_2^2 \xi}\gg 1.
\ee
In principle, there are two different way-outs to avoid this potential problem with string loop corrections:
\begin{enumerate}
\item Do not wrap any $D7$-brane around the inflationary 4-cycle;

\item Fine-tune the coefficients of the $g_s$ corrections.
\end{enumerate}
Let us see these two cases in more detail.

\newpage
$\bullet \textbf{ No D7 wrapped around the inflaton cycle}$

\bigskip

If the inflaton 4-cycle is wrapped by an $ED3$, the source of the string loop corrections is absent and the blow-up
mode $\tau_2$ can safely drive inflation without any hidden sector located on it. However this is not possible
for the following reason:
denoting the inflaton 4-cycle as $\tau_2$ and the other blow-up mode which keeps the volume minimum stable during inflation as $\tau_3$,
the overall volume in the string frame scales as:
\be
\vo\sim e^{\frac{a_2 \tau_2}{g_s}}\sim e^{\frac{a_3\tau_3}{g_s}},
\ee
where $a_i=2\pi$, $i=2,3$, in the case of an $ED3$, while $a_i=2\pi/N$ in the case of an $SU(N)$ hidden sector that undergoes
gaugino condensation. In addition both $\tau_2$ and $\tau_3$ has to be fixed larger than unity
and $g_s\ll 1$ in order to trust the effective field theory.
This implies that an $ED3$ can never yield $\vo\simeq 10^{6-7}$ as it is needed to generate enough density perturbations.
In fact, in the limiting case for $g_s=0.1$ and $\tau_i=1$, we already have:
\be
\vo\sim e^{20\pi}\sim 10^{27},
\ee
and increasing $\tau_i$, the volume can just get larger. On the other hand, if $\tau_i$ is wrapped by $N=10$ $D7$-branes,
for $g_s=0.1$ and $\tau_i=2.6$, we obtain:
\be
\vo\sim e^{\frac{20\pi\tau_i}{N}}\sim e^{16.3}\sim 10^7.
\ee
Hence, in order not to fix any cycle supporting non-perturbative effects at a size smaller than the string scale,
both $\tau_2$ and $\tau_3$ has to support a hidden sector that undergoes gaugino condensation.
Thus we realise that hidden sectors are always present in these models, both in the geometric regime and at the
quiver locus, and the only way to avoid problems with string loop corrections is to fine-tune their coefficient small.
In addition the Hubble parameter during inflation $H_{inf}$ has always to be smaller than the gaugino condensation scale
$\Lambda$ \cite{Oleg}, and so one might be worried that at the beginning of inflation at large $\tau_2$,
$\Lambda$ could turn out to be very low. However, in section \ref{HiddenSector}, we shall show that this is never the case throughout all the
inflationary dynamics.

As we have already pointed out, the cycle supporting the visible sector, which we called $\tau_4$,
cannot receive non-perturbative corrections since an instanton wrapped around $\tau_4$ will
generically have chiral intersections with the visible sector. We expect that similar conclusions apply also
to the case when the non-perturbative superpotential is generated via gaugino condensation since the presence
of chiral matter generically forbids the emergence of such a condensate.
Therefore these considerations seem to imply that the visible sector cannot be wrapped around $\tau_2$
resulting in an effective decoupling of the inflaton from the visible sector which can be very dangerous for reheating.
However we could still envisage a
quite generic situation where the visible sector $D7$-stack wraps a combination of the cycles $\tau_2$ and
$\tau_4$ such that the chiral intersections are with $\tau_4$ but not with $\tau_2$. In this set-up, one
could still have non-perturbative corrections dependent on $\tau_2$ and, at the same time, the visible
sector wrapped around $\tau_2$. Therefore in our study of reheating we shall consider two possible brane set-ups
with the inflaton 4-cycle wrapped by both the visible sector and a hidden sector undergoing gaugino condensation
or just by the hidden sector.

\bigskip

$\bullet \textbf{ Fine-tuning the coefficient of the string loops}$

\bigskip

The inflationary potential (\ref{ygfdo}) takes the schematic form:
\begin{equation}
V_{(np+\alpha')}=\frac{\lambda_1\sqrt{\tau_2}
e^{-2 a_2\tau_2}}{\mathcal{V}} -\frac{\lambda_2\tau_2
e^{-a_2\tau_2}}{\mathcal{V}^2}+\frac{\lambda_3\sqrt{\tau_3}
e^{-2 a_3\tau_3}}{\mathcal{V}} -\frac{\lambda_4\tau_3
e^{-a_3\tau_3}}{\mathcal{V}^{2}}
+\frac{\lambda_5}{g_s^{3/2}\mathcal{V}^{3}}. \label{Unp}
\end{equation}
Our aim is then to use $\tau_2$ as the inflationary direction, but we
have now to worry about the string loop corrections which could
spoil the flatness of this potential in the inflationary region,
since they take the form \cite{stringloops}:
\begin{equation}
\delta V_{(g_{s})}^{1-loop}\simeq \frac{\left(\mathcal{C}^{KK}_2
g_s\right)^2}{\mathcal{V}^3\sqrt{\tau_2}}, \label{Ugs}
\end{equation}
where we are assuming that the blow-up mode $\tau_2$ does not intersect
with any other 4-cycle. Hence no winding corrections can be generated but only
KK corrections coming from the exchange of Kaluza-Klein closed string modes
between the $D7$-brane wrapping $\tau_2$ and any other $D7$ or $D3$ brane
present in the compactification.

Given that inflation takes place in the region of field space such
that $a_2\tau_2>2\ln\vo$, the large exponential suppression in
(\ref{Unp}) tends to render the non-perturbative generated
inflationary potential smaller than the loop generated potential
(\ref{Ugs}) even though this is not the case close to the global
minimum. Hence we need to fine tune $\mathcal{C}^{KK}_2\ll 1$ (by
fine-tuning the complex structure moduli), and in order to
understand the amount of fine-tuning needed to avoid this problem,
we have to make sure that $V_{(g_s)}<V_{(np)}$ for initial values
of $\tau_2$ that give rise to around 60 e-foldings of inflation.
The number of e-foldings is given by \cite{kahlerinfl}:
\begin{equation}
N_e\simeq\frac{1}{\vo^2}\int_{2\ln\vo}^{n\ln\vo}\frac{e^{a_2\tau_2}}{\sqrt{\tau_2}
\left(a_2\tau_2-1\right)}d\tau_2, \label{NE}
\end{equation}
while the ratio between (\ref{Unp}) and (\ref{Ugs}) looks like
(considering just the $\tau_2$-dependent bit of each expression):
\begin{equation}
R\equiv\frac{V_{(g_s)}}{V_{(np)}}\simeq \frac{\left(\mathcal{C}^{KK}_2
g_s\right)^2}{\vo}\frac{e^{a_2\tau_2}}{\lambda_2\tau_2^{3/2}
},\text{ \ \ with \ \ }\lambda_2\simeq 10. \label{R}
\end{equation}
Then we need to impose that $R\ll 1$ at $a_2\tau_2=n\ln\vo$ with
$n$ such that $N_e=60$. Expressing the volume as $\vo=10^x$,
$\mathcal{C}^{KK}_2\simeq 10^{-y}$ and fixing the string coupling such
that $g_s=0.1$,\footnote{We prefer to focus all the fine-tuning on
$\mathcal{C}^{KK}_2$ preserving $g_s$ not too small since the volume of a
4-cycle in the string frame is related to the same quantity in the
Einstein frame as $\tau_s=g_s\tau_E$, and so a very small $g_s$
might lead $\tau_s\ll 1$ in the regime where we do not trust
anymore the supergravity approximation.} this corresponds to
imposing:
\begin{equation}
R\simeq \frac{10^{\left(n-1\right)x-2y-3}}{x^{3/2}}\ll 1.
\label{Rnew}
\end{equation}
For each value of $x$, we can work out $n$ numerically from
(\ref{NE}) and then, substituting these two values in
(\ref{Rnew}), we derive the amount of fine-tuning $y$. The results
are summarised in Table 1.

\begin{figure}[ht]
\begin{center}
\begin{tabular}{|c|c|c|c|c|}
$x$ & $n$ & $y$ & $R$ \\
\hline \hline
4 & 3.03 & 2.5 & 0.16\\
\hline
6 & 2.72 & 3.5 & 0.14 \\
\hline
8 & 2.56 & 4.5 & 0.13 \\
\hline
10 & 2.46 & 5.5 & 0.12 \\
\hline
12 & 2.39 & 6.5 & 0.11 \\
\hline
14 & 2.34 & 7.5 & 0.11 \\
\end{tabular} \\\smallskip
{\bf Table {1}:} Estimation of the fine-tuning of
$\mathcal{C}^{KK}_2=10^{-y}$ needed to render the string loops
subleading with respect to the inflationary potential. Here
$\vo=10^x$ and $n$ is such that $N_e\simeq 60$.
\end{center}
\end{figure}

\subsection{Fibre Inflation}
\label{FibInfl}

In this case the inflaton is the size of a $K3$ fibre over a $\mathbb{C}P^1$ base \cite{fiberinfl}. The volume looks like:
\be
\vo=\alpha\left(\sqrt{\tau_1}\tau_2-\gamma_3\tau_3^{3/2}-\gamma_4\tau_4^{3/2}\right),
\ee
while the superpotential takes the form\footnote{Even in this case, in order not to have $\langle\tau_3\rangle\ll 1$,
$\tau_3$ has to be wrapped by a stack of $D7$-branes supporting a hidden sector
where gaugino condensation takes place.}:
\be
W=W_0+A_3 e^{-a_3 T_3},
\ee
and the K\"{a}hler potential has the same form as in (\ref{Kpot}).
The cycle $\tau_4$ is supporting the visible sector and so it does not get any
non-perturbative correction. After minimising the axion directions and assuming that $\tau_4$
is fixed by either $D$-terms or $g_s$ corrections at the quiver locus or in the geometric regime,
the supergravity $F$-term scalar potential reads:
\be
 V = \frac{8 \, a_3^{2}A_3^2}{3\alpha\gamma_3}
 \left( \frac{\sqrt{\tau_3}}{\mathcal{V}}
 \right) e^{-2a_3\tau_3}
 -4 \,W_{0}a_3 A_3 \left( \frac{\tau_3}{\mathcal{V}^{2}}
 \right) \, e^{-a_3\tau_3}+\frac{3 \xi W_0^2}{4 g_s^{3/2}\mathcal{V}^{3}}.
 \label{PotK3}
\ee
This potential completely stabilises $\tau_3$ and the
volume $\vo\simeq \alpha \sqrt{\tau_1}\tau_2$ at:
\be
\langle \tau_3 \rangle\,=\frac{1}{g_s}\left( \frac{\xi}{2
\alpha \gamma_3} \right)^\frac23,\hskip0.7cm \langle
 \mathcal{V}\rangle = \left( \frac{ 3 \,\alpha \gamma_3 }{4a_3 A_3}
 \right) W_0 \, \sqrt{\langle \tau_3\rangle }
 \; e^{a_3 \langle \tau_3
 \rangle }.
 \label{LVmin}
\ee
The direction in the ($\tau_1,\tau_2$)-plane orthogonal to the overall volume is still flat.
It can be stabilised by the inclusion of string loop corrections to the K\"{a}hler potential
which arise once we wrap a stack of $D7$ branes around $\tau_1$ and another stack around $\tau_2$.
These two stacks can correspond to either a hidden or a visible sector. These $g_s$-corrections turn out to be
subleading with respect to (\ref{PotK3}), due to the extended no-scale structure \cite{GenAnalofLVS},
and so they do not destroy the large volume minimum (\ref{LVmin})
but they can lift the remaining flat direction. The loop generated potential looks like \cite{fiberinfl}:
\be
\delta V_{(g_s)}=\left(\frac{A}{\tau_1^2}-\frac{B}{\vo\sqrt{\tau_1}}+\frac{C\tau_1}{\vo^2}\right)
\frac{W_0^2}{\vo^2}
\label{Vloop}
\ee
and this fixes $\tau_1$ at:
\be
\langle\tau_1\rangle = c\, \vo^{2/3},\text{ \ \ where \ \ }
c\simeq \frac{\left(g_s \mathcal{C}_1\right)^{4/3}}{\mathcal{C}_{12}^{2/3}}.
\ee
As shown in appendix \ref{appendice}, the direction $\chi$ fixed by the string loops, turns out to be naturally lighter than
the volume mode which sets the scale of the potential: $V\sim M_P^2 m_{\vo}^2$. Therefore
this field is a natural candidate to drive inflation given that the $\eta$-parameter looks like:
\be
\eta=M_P^2\frac{V_{\chi\chi}}{V}\simeq\frac{m_{\chi}^2}{m_{\vo}^2}\simeq\frac{\vo^{-10/3}}{\vo^{-3}}
\simeq \frac{1}{\vo^{1/3}}\ll 1.
\ee
Now working in the $(\vo, \tau_1)$-plane and displacing $\tau_1$ far from its minimum,
due to the fact that this mode is naturally lighter than $H$ by a factor $\vo^{-1/3}\ll 1$,
the K3 fibre experiences a very flat direction which is suitable for inflation.
In terms of the canonically normalised inflaton $\phi=(\sqrt{3}/2)\ln\tau_1$,
the inflationary potential looks like \cite{fiberinfl}:
\be
V=\frac{\beta}{\vo^{10/3}}\left(3-4\, e^{-\phi/\sqrt{3}}+e^{-4\phi/\sqrt{3}}+R\, e^{2\phi/\sqrt{3}}\right),
\label{FIPOT}
\ee
where
\be
R\simeq g_s^4 \ll 1.
\ee
The potential in the inflationary region is dominated by the up-lifting bit plus the
first negative exponential in (\ref{FIPOT}), and so it can be very well be approximated as:
\be
V\simeq \frac{\beta}{\vo^{10/3}} \left(3-4 e^{-\phi/\sqrt{3}}\right),
\ee
which is a typical large-field inflaton potential that can give rise to observable gravity waves
due to a trans-Planckian motion of $\phi$ in field space: $\Delta \phi \gtrsim M_P$. It is also interesting
to notice that all the adjustable parameters enter only in the prefactor rendering this
inflationary model very predictive. The slow-roll parameters $\epsilon$ and $\eta$ are naturally
smaller than unity for large $\phi$. In addition both the tensor-to-scalar ratio $r$ and the
spectral index $n_s$ are functions of just the number of e-foldings $N_e$, and they can be expressed one
in terms of the other via the interesting relation \cite{fiberinfl}:
\be
r\simeq 6\left(n_s-1\right)^2.
\ee
The requirement of generating enough density perturbations fixes $\vo \simeq 10^{3-4}$, corresponding to a
GUT-scale $M_s$. Then for different values of the reheating temperature $T_{RH}$, one can obtain a different
number of e-foldings $N_e$, which, in turn, fixes the predictions for the cosmological observables
$r$ and $n_s$. This prediction does not depend strongly on $T_{RH}$, and one generically obtains
the promising outcome \cite{fiberinfl}:
\be
n_s=0.97,\text{ \ \ and \ \ }r=0.005,
\ee
with a tensor-to-scalar ratio potentially detectable by future cosmological observations.

Let us make a few final remarks:
\begin{itemize}
\item As in the case of BI, also in FI, both in the geometric regime and at the
quiver locus, it is impossible to build a model without the presence of a hidden sector.
In fact, in FI a stack of $D7$-branes wrapping the two large cycles $\tau_1$ and $\tau_2$
has to be present in order to generate the string loop corrections that fix
the K3 fibre and provide the inflationary potential. In addition the blow-up cycle $\tau_3$ has still
to support a hidden sector that undergoes gaugino condensation in order to stabilise the overall volume
at a phenomenologically viable value.

\item In section \ref{ProbBI} we noticed that $g_s$ corrections in general would destroy BI,
whereas for FI we have seen that the inflationary potential is loop-generated.
So in one case, string loops destroy inflation whereas in the other case they give rise to a nice potential
naturally suitable to drive inflation. The reason of this different behaviour of the same kind of corrections is the different
topology of the inflaton 4-cycle, which in the first case is the size of a blow-up mode resolving a point-like singularity whereas
in the second case, it is the volume of a K3-fibre. The different topology is reflected in a different volume-scaling of the
elements of the inverse K\"{a}hler metric which enter into the $F$-term supergravity scalar potential.

\item FI requires a rather high string-scale which would seem
to be in disagreement with its intermediate value that was needed to obtain TeV scale SUSY for
the models in the geometric regime. Hence this would seem to be another indication in favour of LVS at the
quiver locus. However it has to be said that the potential of FI is able to give inflation
for all scales and the only issue which sets $M_s\sim M_{GUT}$, is the requirement of generating enough density perturbations.
Therefore one could try to let the inflaton just drive inflation and generate the density fluctuations
via another curvaton-like or modulating field. In this way, it might be possible either to lower the
inflationary scale or to generate large non-Gaussianities in the CMB spectrum \cite{BCGQTZ}, but one should worry about excess isocurvature
perturbations as the inflaton and curvaton decay products might not thermalise before BBN \cite{RM}.
\end{itemize}


\subsection{Moduli dynamics after inflation and reheating}

At the end of inflation, the inflaton field which acts like a homogeneous condensate oscillates coherently
around its minimum. During this time there could be perturbative and non-perturbative ways of creating
particles which are coupled to the inflaton~\cite{RM, Reheat-rev}. There could be various distinct ways of exciting non-perturbative
particles in general:

\begin{enumerate}
\item Tachyonic preheating: if there exists a scalar potential where the effective mass of the inflaton becomes $m^2_{eff}<0$,
then there arises an instability~\cite{Tachyon-Lev} for the modes $k\leq m_{eff}$. The occupation number of the inflaton quanta
grows exponentially, $n_k\sim \exp(2m_{eff}t_{\ast})$, where $t_\ast $ is a typical time scale during which the growth of the momentum
mode $k$ takes place. For a typical self coupling of the inflaton, $\lambda$, the growth factor can be at most $n_k\sim \pi^2/\lambda$.
Note that for smaller $\lambda$ the occupation number can be larger. However this alone does not drain the inflaton energy density into visible or hidden \textit{dof} since this process simply excites the self quanta of the inflaton~\cite{Tachyon-Lev}. The inflaton still needs to decay perturbatively into
the observable and hidden \textit{dof} in order to reach full thermalisation.

\item Instant reheating: if the inflaton couples to visible sector \textit{dof}, it is possible to excite them just within few
oscillations. This happens typically for very large Yukawa or gauge couplings in a renormalisable interaction. Note that by virtue of the inflaton coupling to visible sector \textit{dof}, they obtain time dependent masses during the inflaton oscillations. However this mass vanishes if the inflaton passes through a zero VEV. It is then more effective to excite the zero modes of the visible sector \textit{dof} from the vacuum fluctuations. In a typical scenario, a zero VEV would correspond to a point of {\it enhanced gauge symmetry}, where the gauge bosons are effectively massless~\cite{MSSM-inf}. Particle production happens due to the violation of the adiabaticity condition: $\dot \omega_k \gg \omega_k^2$, where $\omega_k$ is the frequency of the excited quanta with momentum $k$~\cite{Instant-Felder,Kofman}. However in our case the instant preheating is rather unlikely, as the inflaton couplings to hidden and visible sectors are governed by non-renormalisable interactions. Since the couplings are so weak, the excitations of the visible sector \textit{dof} do not exhaust the inflaton energy density.

\item Parametric resonance: in this case the particle creation mechanism is similar to the instant reheating scenario, albeit particle creation takes place in each oscillation whenever the inflaton passes through its minimum. If the inflaton couples to ordinary matter and gauge fields, it can then excite these \textit{dof} from the time dependent induced mass term~\cite{Kofman, parametric}. In both instant and parametric resonance the particle creation mechanism is based on the violation of the adiabaticity condition. In the later case particle creation happens continuously over the oscillations. In both cases the visible sector \textit{dof} can be excited if the couplings are large. However the decay products have a non-thermal spectra and a complete thermalisation of all the \textit{dof} happens at a much larger time scale which is comparable to that of the perturbative decay of the inflaton, i.e. $\Gamma_{infl}^{-1}$ (see~\cite{Reheat-rev, Micha}). Furthermore notice that in order to excite all the MSSM and hidden \textit{dof} the inflaton has to decay completely, which happens only perturbatively.

\item Fragmentation: if the inflaton possesses a global or gauged $U(1)$ symmetry then it is possible to fragment the inflaton condensate to form inflatonic Q-balls~\cite{MSSM-rev,Q} provided the effective potential near the minimum grows slower than $\phi^2$ by virtue of quantum corrections. These inflatonic
non-topological solitons, known as Q-balls,  decay into lighter fermions from their surface and in some cases can be absolutely stable if the energy of the Q-ball becomes smaller than the lightest baryons in the theory~\cite{MSSM-rev}. The stable Q-balls can become a dark matter component of the Universe~\cite{Kusenko-dark}.

\item Reheating and thermalisation: given that all the above mentioned effects are typically scatterings of the inflaton to excite new quanta, preheating cannot exhaust all the inflaton energy density. Preheating will end when the backreaction of the excited quanta becomes important. The inflaton then must decay perturbatively. The decay products of the inflaton and the excited quanta due to preheating must thermalise to produce a thermal bath of  MSSM \textit{dof}.
    Typically thermalisation of all the \textit{dof} take considerably longer time than one would expect, especially in the context of the MSSM~\cite{AM1,AM2,AM3}. One of the main reasons is that the MSSM contains $F$-and $D$-flat directions (parametrised by a complex scalar monomial field), which are made up of MSSM squarks and sleptons \cite{MSSM-rev}. During inflation these flat directions tend to take large VEVs by virtue of random quantum fluctuations imparted on light scalars with TeV scale mass due to a large Hubble friction term. Once these flat directions take a large VEV during inflation, their VEV slowly rolls down towards their minimum. Since these flat directions are charged under the MSSM gauge group, during their dynamical evolution after inflation they produce VEV dependent masses to gauge bosons and gauginos similar to the Higgs mechanism. Heavy gauge bosons and gauginos tend to slow down any $2\leftrightarrow 2$ or $2\leftrightarrow 3$ interactions required to obtain {\it kinetic} and {\it chemical} equilibrium, which are essential ingredients for acquiring local thermodynamical equilibrium~\cite{AM1}. These flat directions have conserved global numbers, like baryonic, leptonic or some combination of both, which enable them to live long enough and decay perturbatively close to their TeV mass scale~\cite{AM3}. One of the important consequences of the finite VEV of the flat direction is that they can kinematically block preheating all together~\cite{Reheat-rev,AM2}.

The bottom line is that the thermalisation time scale $\Gamma_{therm}^{-1}$ in the presence of MSSM flat directions in the visible sector can be longer than the time scale of the perturbative decay of the inflaton,
$\Gamma_{infl}^{-1}$: $\Gamma_{infl}^{-1}\leq \Gamma_{therm}^{-1}\simeq H_{therm}^{-1}$. The final reheating temperature will then be given by~\cite{AM1}:
\be
T^{therm}_{RH}\simeq (\Gamma_{therm} M_{P})^{1/2}\,.
\ee
Since $\Gamma_{therm}$ depends on the nature of the MSSM flat directions and their dynamics, we will leave this detailed study for future investigation.
In a rather model independent scenario, it is expected that a complete thermalisation will take place when the flat directions completely evaporate.
This happens at temperatures fairly close to the TeV scale~\cite{AM1,AM2,AM3}.

For the purpose of illustrating the current issues at hand we shall work under the approximation of sudden thermalisation of the decayed particles, which are assumed to be relativistic at the time of decay. Thus we shall obtain a maximal value for the reheating temperature which we shall denote as $T_{RH}^{max}$. This implicitly assumes also that the perturbative decay channels are the main source of energy transfer from the moduli sector. Given the couplings of the moduli to hidden and visible sectors to be so weak, preheating effects are expected to be largely suppressed. By equating the inflaton decay rate to the Hubble parameter for radiation dominance by assuming $ \Gamma_{infl}\simeq H$, we obtain:
\be
T_{RH}^{therm}\leq T_{RH}^{max} \simeq (\Gamma_{infl} M_{P})^{1/2}\,.
\ee
%
%
Note that since in our case the visible and hidden sectors do not have any direct coupling, the two sectors are not expected to thermalise before BBN. For all practical purposes the hidden sector will become like a dark sector. Thus there will be two separate thermal baths formed with two distinct temperatures, $T_{vis}$ and $T_{hid}$ both of them of the order $T_{RH}^{max}$, and \textit{dof}, $g_*^{vis}$ and $g_*^{hid}$, which we assume to be relativistic:
\be\label{evolution-both}
H^2 \simeq g_*^{vis}\frac{T_{vis}^4}{M_P^2}+g_*^{hid}\frac{T_{hid}^4}{M_P^2}\,.
\ee
\end{enumerate}

In this paper we will do the first preliminary check regarding the inflaton decay rates to visible and hidden \textit{dof}.
It is then important to know what fraction of the inflaton energy density gets diverted into the hidden sector. Since in our case a K\"{a}hler modulus plays the r\^{o}le of the inflaton within the 4D effective field theory, it is crucial to derive the inflaton couplings to visible and hidden sector \textit{dof} which are localised in particular regions of the Calabi-Yau compactification. This derivation is performed in detail in appendix \ref{appendice}.
We also stress that in sections \ref{ProbBI} and \ref{FibInfl} we showed that it is impossible
to avoid the presence of hidden sectors. Therefore a full understanding of reheating
requires a non-trivial interplay between local and global aspects of string compactifications but it can also
set severe constraints on which internal 4-cycle can be wrapped by what kind of branes.

There could be three potential problems which may arise in the study of reheating:
\begin{itemize}
\item The inflaton decay rate to visible sector \textit{dof}
is suppressed with respect to its decay rate to hidden sector \textit{dof} by some power of the overall volume:
\be
\frac{\Gamma_{infl\to vis. sec.}}{\Gamma_{infl\to hid. sec.}}\sim \mathcal{O}\left(\frac{1}{\vo^p}\right)\ll 1\text{, \ \ for \ \ }p>0.
\ee
In this case the inflaton dumps all its energy into hidden sector,
instead of visible sector, \textit{dof}, and so it is impossible to reheat the visible sector.
Notice that the reheating of the visible sector cannot proceed via the decay of the hidden sector particles
to the visible ones. In fact there is no direct coupling between these particles since they correspond to
two open string sectors localised in different regions of the Calabi-Yau. The only indirect gravitational
coupling via a graviton or a modulus exchange is definitely too weak to give rise to a viable reheating of
the visible sector due to the decay of hidden to visible \textit{dof}.

\item The inflaton decay rate to visible and hidden sector \textit{dof} has the same volume scaling:
\be
\frac{\Gamma_{infl\to vis. sec.}}{\Gamma_{infl\to hid. sec.}}\sim \mathcal{O}(1).
\ee
In this case the inflaton decays at the same time to hidden and visible sector particles
releasing the same amount of entropy to both sectors. This might generically yield an overproduction of dark matter
particles given by the stable hidden superpartners which interact with the visible \textit{dof}
only gravitationally and get an $\mathcal{O}(M_{soft})$ mass due to SUSY-breaking effects.
These dark matter particles can be produced either non-thermally directly by the inflaton decay or
by thermal effects. In the last case,
two different thermal baths get formed with temperatures $T_{vis}$ and $T_{hid}$, with $T_{vis}\neq T_{hid}$,
and the hidden sector dark matter particles would be created thermally if $T_{hid}>M_{soft}$.

\item The inflaton decay rate to hidden sector \textit{dof}
is subleading with respect to its decay rate to visible sector \textit{dof}:
\be
\frac{\Gamma_{infl\to hid. sec.}}{\Gamma_{infl\to vis. sec.}}\sim \mathcal{O}\left(\frac{1}{\vo^p}\right)\ll 1\text{, \ \ for \ \ }p>0.
\ee
In this case the inflaton can reheat the visible sector but the reheating temperature can turn out to be
lower than the BBN temperature $T_{BBN}\sim $1 MeV: $T_{RH}^{max}<T_{BBN}$, due to an effective decoupling of the inflaton
from the visible sector because of the geometrical separation of the two corresponding 4-cycles within the Calabi-Yau.
\end{itemize}


\section{Hidden sector dynamics}
\label{HiddenSector}

As we have seen in the previous section, in order not to have an overall volume which is too large for VEVs of the moduli
larger than the string scale, one is forced to
consider non-perturbative effects coming from gaugino condensation. Thus we need a hidden stack of $D7$-branes supporting
a supersymmetric field theory that undergoes gaugino condensation.

Let us focus on a 4D $N=1$ $SU(N_c)$ gauge theory with
$N_f$ chiral superfields in the fundamental representation, $Q^i_a$, with $a=1,...,N_c$ and $i=1,...,N_f$,
and $N_f$ chiral superfields in the antifundamental representation, $\overline{Q}^a_i$ (for a review see \cite{NPReview}). The global anomaly-free symmetry is
$SU(N_f)_L\times SU(N_f)_R\times U(1)_B\times U(1)_R$. The unique gauge-invariant chiral superfield is $M_{ij}=Q_i\cdot\overline{Q}_j$
and it can be thought of as a meson superfield whose scalar component is a colour-singlet combination of squarks.
This theory gives rise to gaugino condensation for $N_f<(N_c-1)$, and so we shall focus only on this case\footnote{For $N_f=(N_c-1)$,
the non-perturbative superpotential is generated by gauge instantons and so we are not interested in this case. In addition no
non-perturbative superpotential is generated for $N_f>(N_c-1)$.}. Let us analyse the two subcases $N_f=0$ and $N_f\neq 0$ separately.


\subsection{$N_f=0$}

For $N_f=0$, we are in the case of a pure supersymmetric Yang-Mills theory with no matter.
Defining the fundamental coupling constant $g^2$ at the UV scale $\Lambda_{UV}$,
the effective running coupling constant is given by:
\be
\frac{1}{g^2}(\mu)=\frac{1}{g^2}+\frac{\beta}{16\pi^2}\ln\left(\frac{\Lambda_{UV}^2}{\mu^2}\right).
\label{running}
\ee
This theory confines in the IR due to strong dynamics at the scale $\Lambda$ where the expression (\ref{running}) formally
diverges:
\be
\Lambda^{\beta}=\Lambda_{UV}^{\beta}e^{-8\pi^2/g^2},
\ee
where the coefficient of the 1-loop $\beta$-function is given by $\beta=3 N_c$. In addition, at the scale
$\Lambda$ the theory develops a mass gap since glueballs ($gg$), $\langle F_{\mu\nu}F^{\mu\nu}\rangle\sim \Lambda^4$,
gluinoball-mesons ($\tilde{g}\tilde{g}$), $\langle\lambda\lambda\rangle\sim \Lambda^3$
and `glueballinos' ($g\tilde{g}$), all get a mass of the order $\Lambda$. The resulting non-perturbative
superpotential looks like:
\be
W_{np}=\Lambda^3=\Lambda_{UV}^3 e^{-8\pi^2/(N_c g^2)}.
\label{p1}
\ee
Notice that these condensates
break the original $\mathbb{Z}_{2N}$ global symmetry but since this is a discrete symmetry no massless Goldstone boson arises.

In string compactifications, the fundamental gauge coupling is given by the modulus $\tau$ supporting
this pure $N=1$ SYM theory, $g^{-2}=\tau/(4\pi)$,
and so we can integrate out all the condensates and remain with a holomorphic non-perturbative superpotential
for $\tau=$Re$(T)$ below the scale $\Lambda$:
\be
W_{np}=M_P^3 e^{-2\pi T/N_c}.
\label{p2}
\ee
Let us estimate the order of magnitude of $\Lambda$. There are two ways to do it: either from (\ref{p1}) or from the expression
of the normalised superpotential $e^{K/2}W_{np} = \Lambda^3$ in terms of (\ref{p2}). From (\ref{p1}) we obtain:
\be
\Lambda=\Lambda_{UV}e^{-a\tau/3}\text{, \ \ with \ \ }a\equiv 2\pi/N_c.
\ee
There is a subtlety now pointed out in \cite{Pedro}. It might be that in the case
when the cycle is invariant under the orientifold then the running of the coupling starts from
$M_s$ while when the cycle is not invariant under the orientifold the UV scale is higher: $\Lambda_{UV}=M_s R \simeq M_s \vo^{1/6}$.
In the first case we would obtain:
\be
\Lambda\simeq\frac{M_P}{\vo^{1/2}}e^{-a\tau/3},
\label{firstcase}
\ee
while in the second case we find:
\be
\Lambda\simeq\frac{M_P}{\vo^{1/3}}e^{-a\tau/3}.
\label{secondcase}
\ee
On the other hand, from (\ref{p2}) we realise that:
\be
\Lambda=e^{K/6}W_{np}^{1/3}=e^{K/6}M_P e^{-a \text{Re}(T)/3}\simeq \frac{M_P}{\vo^{1/3}}e^{-a \text{Re}(T)/3}.
\label{Compare}
\ee
Comparing (\ref{Compare}) with (\ref{secondcase}), we see that in the second case we have Re$(T)=\tau$,
but comparing (\ref{Compare}) with (\ref{firstcase}) we realise that
in the first case we need to redefine the modulus as Re$(T)=\tau+\ln(\vo^{1/2})/a$ \cite{Pedro}.
Given that in both cases $e^{-a\tau}\simeq \vo^{-1}$, we end up with $\Lambda\simeq M_P\vo^{-5/6}$ for the first case,
and $\Lambda\simeq M_P\vo^{-2/3}$ for the second.

The scale of IR strong dynamics $\Lambda$ has to be no lower than the Hubble parameter
during inflation $H_{inf}$: $H_{inf}\leq \Lambda$ \cite{Oleg}.
This is indeed the case for FI:
\be
H_{inf}\simeq \frac{V^{1/2}}{M_P}\simeq \frac{M_P}{\vo^{5/3}},
\ee
and for BI when $\tau$ sits at its VEV:
\be
H_{inf}\simeq \frac{V^{1/2}}{M_P}\simeq \frac{M_P}{\vo^{3/2}}.
\label{Hinf}
\ee
However, as we pointed out in section \ref{ProbBI}, the non-perturbative inflationary potential for BI has to be
generated by gaugino condensation and not by stringy instantons. Therefore in this case,
given that the inflaton $\tau$ varies during inflation, also $\Lambda$ varies while $H_{inf}$ remains constant.
Thus we need to check that $H_{inf}<\Lambda$ throughout all inflationary dynamics and not just when $\tau$ relaxes at its minimum.
Focusing on the case when the 4-cycle is invariant under the orientifold, we have to make sure that
the expression (\ref{firstcase}) for $\Lambda$ evaluated at the value of $\tau$ such that you get 60 e-foldings of inflation,
is not lower than $H_{inf}$ given by (\ref{Hinf}).
As can be seen from Table 1, taking $a\tau=n \ln\vo$, small variations of $n$ give large modifications of the number of e-foldings.
In particular, $N_e=60$ is obtained for $n\simeq 2.7$ if $\vo\simeq 10^6$. Then we obtain:
\be
\Lambda\simeq\frac{M_P}{\vo^{1/2}}e^{-a\tau/3}\simeq\frac{M_P}{\vo^{1/2}}e^{-(\ln\vo)^{9/10}}\simeq\frac{M_P}{\vo^{7/5}}
\text{ \ }\Rightarrow\text{ \ } \frac{H_{inf}}{\Lambda}\simeq\frac{1}{\vo^{1/10}}\leq \mathcal{O}(1),
\ee
and so we conclude that throughout all inflationary dynamics we always have $H_{inf}\leq \Lambda$.

Another very interesting observation is that the inflaton mass $m_{inf}$ after inflation for both BI and FI is
lower than $\Lambda$ since we have:
\be
m_{inf}\simeq \frac{M_P}{\vo}\text{ \ for BI, \ \ \ } m_{inf}\simeq\frac{M_P}{\vo^{5/3}}\text{ \ for FI.}
\ee
Thus the inflaton decay to hidden sector \textit{dof} is {\it kinematically forbidden}!
In addition the Kaluza-Klein scale is given by:
\be
M_{KK}\simeq \frac{M_s}{R} \simeq \frac{M_s}{\tau_4^{1/4}} \simeq \frac{M_P}{\sqrt{\vo}\tau_4^{1/4}},
\ee
and so we end up with the following final hierarchy among scales:
\be
m_{inf}<H_{inf}\leq\Lambda<M_{KK}<M_s.
\ee
Finally we need also to check that the gaugino condensate does not get destabilised by thermal effects as it would be the
case for reheating temperatures in the hidden sector $T_{hid}>\Lambda$. However we can check that this is never the case. The
largest  temperature in the hidden sector is given by $T_{hid}\simeq(\Gamma_{inf}M_P)^{1/2}$,
where $\Gamma_{inf}= m_{inf}^3\vo^x/M_P^2$ is the inflaton decay rate.
Given that the strongest possible inflaton coupling is $1/M_s$ we
infer that $x\leq 1$, and so it suffices to check that $T_{hid}<\Lambda$ for $x=1$. However in this case we can conclude that we always have $T_{hid}<\Lambda$ since:
\be
\frac{T_{hid}}{\Lambda}\simeq \frac{1}{\vo^{1/6}}<1\text{ \ for BI, \ \ \ } \frac{T_{hid}}{\Lambda}\simeq\frac{1}{\vo^{7/6}}<1\text{ \ for FI.}
\ee
Let us see now how an $N=1$ $SU(N_c)$ gauge theory with $N_f=0$ can be realised in type IIB string compactifications. All the cycles which in our scenarios are stabilised non-perturbatively are local blow-up modes, and so they are rigid cycles which do not intersect with any other cycle. Thus these cycles are perfect to generate no matter either from possible intersections with other cycles or from deformations of the cycle. Given that these cycles are \textit{non-Spin}, one has to turn on half-integer world-volume fluxes to cancel the Freed-Witten anomalies \cite{FW}. These fluxes could generate chiral matter but since these cycles do not intersect with any other cycle, one can always choose the $B$-field to compensate these Freed-Witten fluxes and have a total world-volume flux which is zero\footnote{Let us illustrate this claim with a simple example. Let $\Sigma_1$ and $\Sigma_2$
be two blow-up 4-cycles which do not intersect with each other, and let $D_1$ and $D_2$ be the corresponding dual 2-forms.
Then the half-integer Freed-Witten fluxes $F_1$ and $F_2$ will look like:
\be
F_1=\frac{f_1}{2}\cdot D_1+0\cdot D_2\text{ \ \ and \ \ }F_2=0\cdot D_1+\frac{f_2}{2}\cdot D_2,
\ee
with $f_1$ and $f_2$ integer coefficients. The $B$-field can be adjusted to take the form $B=F_1+F_2$,
so that the total world-volume flux $(B-F_i)$ on $\Sigma_i$, $i=1,2$, is vanishing.
Notice that if the two cycles intersected, then $B$ would not be able to compensate both half-integer fluxes at the same time.}.

Let us see now what happens if one wraps a stack of $D7$-branes around a local blow-up 4-cycle which is invariant under the orientifold so that the $D7$-branes are on top of the $O7$-planes\footnote{We shall not consider the case when the cycle is not invariant under the orientifold due to the similarity with instantons which contribute to the superpotential only if the cycle wrapped by the instanton is invariant under the orientifold, and the fact that it would be difficult to satisfy tadpole cancelation.}. In order to cancel the tadpole, 4 $D7$-branes have to sit on top of each $O7$-plane. The gauge theory on each stack of $N$ $D7$-branes is $SO(2N)$ and so the gauge group is $SO(8)^n$ for $n$ $O7$-planes. In addition one can turn on a world-volume Abelian gauge flux $F_2$ on this 4-cycle: $F_2= T_0 F_2^{(0)}+ T_i F_2^{(i)}$, where $T_0$ is the generator of the diagonal $U(1)_d\subset SO(2N)$ while $T_i$ are the traceless Abelian elements of $SU(N)$. It is then important to consider what happens to $F_2$:
\begin{enumerate}
\item $F_2=0$ and so we obtain a pure $N=1$ $SO(8)^n$ gauge theory that undergoes gaugino condensation and develops a mass-gap.

\item Only a diagonal flux is turned on: $F_2=T_0 F_2^{(0)}$. This flux breaks the gauge group $SO(2N)\to SU(N)\times U(1)_d$ and the $U(1)_d$
acquires a string-scale mass via the Green-Schwarz mechanism. No chiral matter is generated and the theory below $M_s$ is still $N=1$ pure SYM.
However the non-zero gauge flux generates a Fayet-Iliopoulos term $\xi$ associated to $U(1)_d$ which cannot be cancelled by any scalar field.
Hence in order to cancel the $D$-terms, one has to impose $\xi=0$ which, due to the $\tau$-dependence of $\xi$, implies that the cycle supporting the hidden sector is forced to shrink to zero size \cite{quiver}. This is what we do not want to have both for the inflaton and for the cycle keeping the volume minimum stable during inflation, and so we need to discard this case.

\item Only some of the $D7$-branes get magnetised, say just one of them: $F_2=T_0 F_2^{(0)}+ T_1 F_2^{(1)}$. This flux breaks the gauge group $SO(2N)\to SU(N)\times U(1)_d\to SU(N-1)\times U(1)_1\times U(1)_d$. The new $U(1)_1$ gets massive via the Green-Schwarz mechanism only if $F_2^{(1)}$ is turned on on a 2-cycle which is also a 2-cycle of the Calabi-Yau. In addition, chiral matter in the fundamental representation
    gets originated and so one has to check that $N_f<(N_c-1)$ in order to have still gaugino condensation.

    Notice that in this case one can prevent the cycle to shrink to zero size by cancelling the Fayet-Iliopoulos term against the VEV of a scalar field $\Phi$ charged under the anomalous $U(1)$ which becomes massive: $\langle \Phi \rangle \sim \sqrt{\xi} \sim M_s$. If then the superpotential contains a term of the form:
    \be
    W=\lambda \Phi Q_i\cdot\overline{Q}_j,
    \label{Yuk}
    \ee
    the VEV of $\Phi$ generates a mass for the matter fields of the order $M_s > \Lambda$ (for $\lambda\sim 1$). Therefore the matter fields can be safely integrated out obtaining a pure $N=1$ SYM theory that develops a mass gap \cite{Choi}.
\end{enumerate}


\subsection{$N_f < (N_c-1)$}

For $N_f < (N_c-1)$, the original $SU(N_c)$ gauge theory has $2N_fN_c$ complex scalars. Then $D$-terms break the gauge group to $SU(N_c-N_f)$ and the resulting theory consists of $(N_c-N_f)^2-1$ gauge bosons and gauginos plus $N_f^2$ scalar moduli $M_{ij}$ \cite{NPReview}. The non-perturbative superpotential for $M_{ij}$ is generated by gaugino condensation. This is the famous Affleck-Dine-Seiberg superpotential whose form can be constrained using symmetries \cite{ADS}:
\be
W_{np}\simeq\left(\frac{\Lambda^{3N_c-N_f}}{\det(M)}\right)^{\frac{1}{N_c-N_f}}.
\label{WADS}
\ee
Notice that for $N_f=0$ we get the previous expression (\ref{p1}): $W_{np}\simeq\Lambda^3$. The states $gg$, $\lambda\lambda$ and $g\lambda$ get a mass of the order $\Lambda$ but the superpotential (\ref{WADS}) gives a run-away for $M_{ij}$. Therefore there is no stable minimum and the system is driven to $\langle M_{ij}\rangle\to\infty$. Notice that the $F$-term of $M_{ij}$ gives the quark condensate:
$F_{M_{ij}}\propto\langle\psi_{Q_i}\cdot\psi_{\overline{Q}_j}\rangle$. In addition $F_{M_{ij}}\sim \langle M_{ij}\rangle^p$ with $p<0$, hence $\langle M_{ij}\rangle\to\infty$ implies $\langle\psi_{Q_i}\cdot\psi_{\overline{Q}_j}\rangle\to 0$. Thus the theory does not show any spontaneous breaking of the original chiral symmetry $SU(N_f)_L\times SU(N_f)_R$.

However now we are working in supergravity where, due to the $e^K$ prefactor of the scalar potential, $M_{ij}$ can be stabilised at a finite VEV. In addition in flux compactifications the background fluxes break supersymmetry inducing a non-zero
$F$-term for the modulus supporting the hidden sector\footnote{Notice that there are constructions where gaugino condensation can restore supersymmetry but this is never the case for LVS.}. This generates a SUSY-breaking mass term for the scalars:
$\Delta\mathcal{L}=-m_0^2\left(|Q|^2+|\overline{Q}|^2\right)$, that is the main contribution that yields a minimum at finite
$\langle M_{ij}\rangle=t_{\star}\delta_{ij}$, with $t_{\star}\sim \left(\frac{\Lambda^{k_1}}{m_0}\right)^{k_2}$, $k_1=\frac{5N_c-3N_f}{N_c-N_f}$ and $k_2=\frac{N_c-N_f}{2N_c-N_f}$ \cite{Peskin}. Expanding around this VEV all the fermions get a mass of the order $m_0$. Also some scalars get a mass of the order $m_0$ but some of them are left massless. These are the Goldstone bosons of the spontaneously broken chiral symmetry due to the presence of mass terms:
$SU(N_f)_L\times SU(N_f)_R\to SU(N_f)_V$. Notice that now $\langle\psi_{Q_i}\cdot\psi_{\overline{Q}_j}\rangle\neq 0$.

This discussion and the previous one for the case $N_f=0$ suggest that an $N=1$ $SU(N_c)$ gauge theory with $N_f\neq 0$ can be realised in type IIB string compactifications in several ways. Focusing on local blow-up modes invariant under the orientifold, one has to magnetise only some of the $D7$-branes
wrapping this cycle checking that $N_f$ chiral flavours are generated with $N_f<(N_c-1)$. Then one has to make sure that the cycle does not shrink to zero size by cancelling the Fayet-Iliopoulos term associated to anomalous $U(1)$ factors against the VEV of a scalar, but without generating a mass term of the order $M_s$ for the matter fields due to couplings like (\ref{Yuk}). Then the theory undergoes gaugino condensation, and in the presence of chiral symmetry, its spontaneous breaking will always give rise to massless Goldstone bosons except for the case of a low energy Higgs-like mechanism.


\subsection{Constraint on hidden sector configurations}

We have seen that the simple requirement of having a hidden sector that undergoes gaugino condensation supported by a rigid orientifold-invariant 4-cycle which does not shrink at the quiver locus and does not intersect with any other cycle, does not restrict the possible particle content and mass spectrum of the hidden sector. Let us summarise the various options which we have found:
\begin{enumerate}
\item The hidden sector consists just of a pure $N=1$ SYM theory that develops a mass-gap. All particles acquire a mass of the order $\Lambda$ and are heavier than the inflaton.

\item The hidden sector consists of a pure $N=1$ SYM theory that develops a mass-gap plus a massless $U(1)$. The mass-spectrum below $\Lambda$ consists of massless hidden photons and hidden photini which get an $\mathcal{O}(M_{soft})$ mass due to SUSY-breaking effects.

\item The hidden sector consists of an $N=1$ $SU(N_c)$ theory with $N_f<(N_c-1)$ flavours. Furthermore there might be an additional massless $U(1)$.
The condensates made of gauge bosons and gauginos get a mass of the order $\Lambda$.
In the presence of spontaneous chiral symmetry breaking some matter condensates are massless while the other get a mass of the order $M_{soft}$.
If chiral symmetry is explicitly broken by a low energy Higgs-like mechanism, all the matter fields will get a $\delta m \ll M_{soft}$ contribution to their masses.

Therefore there are two options for the matter spectrum:
\begin{enumerate}
\item The hidden photons are massless while the hidden photini, matter fermionic and scalar condensates acquire a mass of the order $M_{soft}$. In addition there are very light (or even massless) pion-like mesons.

\item There are matter fermionic and scalar condensates with mass of the order $M_{soft}$ and
very light (or even massless) pion-like mesons, but all the hidden gauge bosons and gauginos are heavier than the inflaton.
\end{enumerate}
\end{enumerate}


\section{Inflaton couplings to hidden and visible sectors}


In order to investigate what fraction of energy density is transferred to the hidden and visible sectors, we have to work out the moduli mass spectrum and the inflaton coupling to visible and hidden sector \textit{dof}. This can be done starting to derive
the canonical normalisation. The first step is to find a global minimum and then
expand each modulus around its VEV:
\be
\tau_i=\langle\tau_i\rangle+\delta \tau_i, \text{ \ \ }\forall i=1,...,h_{1,1},
\ee
ending up with:
\be
\mathcal{L}=K_{ij}\partial_{\mu}\left(\delta\tau_i\right)\partial^{\mu}\left(\delta\tau_j\right)
-\langle V \rangle -\frac{1}{2}V_{ij}\delta\tau_i\delta\tau_j+\mathcal{O}(\delta\tau^3).
\label{Lncn}
\ee
Writing the original moduli $\delta\tau_i$ in terms of the canonically normalised fields around the
minimum $\delta\phi_i$ as:
\be
\delta\tau_i=\frac{1}{\sqrt{2}}C_{ij}\delta\phi_j,
\label{CN}
\ee
the Lagrangian (\ref{Lncn}) takes the canonical form:
\be
\mathcal{L}=\frac{1}{2}\sum_{i=1}^{h_{1,1}}\partial_{\mu}\left(\delta\phi_i\right)
\partial^{\mu}\left(\delta\phi_i\right)-\langle V \rangle - \frac{m_i^2}{2}\sum_{i=1}^{h_{1,1}}\delta\phi_i^2,
\ee
only if:
\be
C_{ai}K_{ij}C_{jb}=\delta_{ab}\text{ \ \ and \ \ }\frac{1}{2}C_{ai}V_{ij}C_{jb}=m_a^2\delta_{ab}.
\ee
The two previous relations are satisfied if $C_{ja}$ and $m_a^2$ are, respectively, the eigenvectors and
the eigenvalues of the mass-squared matrix $\left(M^2\right)_{ij}\equiv \frac{1}{2}\left(K^{-1}\right)_{ik}V_{kj}$.

In this way we have been able to derive the moduli mass-spectrum.
Then the inflaton coupling to visible and hidden sector \textit{dof} can be explicitly
worked out by knowing the moduli dependence of the kinetic and mass terms of open string modes.
Subsequently the moduli are expanded around their VEVs and then expressed in terms of the
canonically normalised fields using (\ref{CN}). Following this procedure, in appendix \ref{appendice},
we derived the moduli canonical normalisation, mass spectrum and couplings to all particles in the model
for the case of BI and FI both in the geometric regime and at the quiver locus.

Let us outline here just the derivation of the moduli coupling to the massless
gauge bosons of the field theory
living on a stack of $D7$-branes wrapping a 4-cycle whose volume
is given by $\tau$ (which can be any of our moduli). We focus just on this case since this is
the simplest derivation and, above all, in appendix \ref{appendice} we show that
the strongest moduli decay rates are to visible or hidden gauge bosons.

These couplings can be worked out from the moduli dependence of the tree-level gauge kinetic
function \cite{astro}, which in the case when $\tau$ is fixed in the geometric regime, is given by $4\pi g^{-2}=\tau$.
The kinetic terms read:
\begin{equation}
\mathcal{L}_{gauge}=-\frac{\tau}{M_P}F_{\mu\nu}F^{\mu\nu}.
\end{equation}
We then expand $\tau$ around its minimum and go to the canonically
normalised field strength $G_{\mu\nu}$ defined as $G_{\mu\nu}=2\sqrt{\tau}F_{\mu\nu}$ and obtain:
\begin{equation}
\mathcal{L}_{gauge}=-\frac{1}{4}G_{\mu\nu}G^{\mu\nu}-\frac{\delta\tau}
{4 M_P\langle\tau\rangle}G_{\mu\nu}G^{\mu\nu}. \label{FmunuFmunu}
\end{equation}
Now by substituting one of the expressions (\ref{CN}) for $\delta\tau$ in (\ref{FmunuFmunu}), we obtain the
moduli couplings to the corresponding gauge bosons.

We notice that the expression of the gauge kinetic function changes in the case when $\tau$ shrinks down
at the singularity. In fact in this case the tree-level bit depends on the axio-dilaton $S=e^{-\phi}+i C_0$ and the K\"{a}hler
modulus enters only at 1-loop in the presence of a non vanishing world-volume flux $F$ \cite{quiver}:
\be
4\pi g^{-2}=\text{Re}(S)+h(F)\tau.
\ee
Hence in this case, in order to derive the moduli couplings to the the corresponding gauge bosons, one has to work out
the moduli mixing with the axio-dilaton $S$ (see appendix \ref{appendice}).

Let us illustrate these claims in the explicit examples of BI and FI both in the
geometric regime and at the quiver locus.


\subsection{Blow-up inflation}
\label{BI}

Let us focus on the inflationary model described in section \ref{BupInfl}. We shall consider both the
case when the cycle $\tau_4$ supporting the visible sector is stabilised in the geometric regime or
at the quiver locus.

\subsubsection{Geometric Regime}
\label{GeomReg}

The canonical normalisation reads (see appendix \ref{appendice}):
\begin{eqnarray}
\delta \tau _{1} &\sim&\mathcal{O}(\mathcal{V}^{2/3})\delta \phi
_{1}+\sum_{i=2}^{4}\mathcal{O}(\mathcal{V}^{1/6})\delta \phi_i\sim
\mathcal{O}(\mathcal{V}^{2/3})\delta \phi
_{1}, \\
\delta \tau _{2} &\sim&\mathcal{O}(1)\delta \phi _{1}+\mathcal{O}(\mathcal{V}
^{1/2})\delta \phi _{2}+\sum_{i=3}^{4}\mathcal{O}(\mathcal{V}^{-1/2})\delta
\phi_i\sim \mathcal{O}(\mathcal{V}
^{1/2})\delta \phi _{2},
\label{cn2} \\
\delta \tau _{3} &\sim&\mathcal{O}(1)\delta \phi _{1}+\mathcal{O}(\mathcal{V}
^{1/2})\delta \phi _{3}+\sum_{i=2,4}\mathcal{O}(\mathcal{V}^{-1/2})\delta
\phi_i\sim \mathcal{O}(\mathcal{V}
^{1/2})\delta \phi_3, \\
\delta \tau _{4} &\sim&\mathcal{O}(1)\delta \phi _{1}+\mathcal{O}(\mathcal{V}
^{1/2})\delta \phi _{4}+\sum_{i=2}^{3}\mathcal{O}(\mathcal{V}^{-1/2})\delta
\phi_i\sim\mathcal{O}(\mathcal{V}
^{1/2})\delta \phi_4,
\end{eqnarray}
and the moduli mass spectrum looks like:
\be
m^2_1\simeq\frac{M_P^2}{\vo^3\ln\vo},\text{ \ \ and \ \ }m_i^2\simeq\frac{(2\ln\vo)^2}{\vo^2}M_P^2,\text{ \ }
\forall i=2,3,4.
\ee
We point out that the volume-scaling of the
canonical normalisation of the small blow-up modes $\delta\tau_i$, $i=2,3,4$, can
be understood from a geometric point of view. In fact, each canonically normalised
field $\delta\phi_i$, $i=2,3,4$, is mostly given by the corresponding blow-up mode with a
power of $\vo^{1/2}$, then the next mixing in a large volume expansion is with $\delta\tau_1$
which corresponds to the overall volume mode. Finally the mixing with all the other blow-up modes
is further suppressed by a power of $\vo^{-1/2}$. This suppression with respect to the mixing with the
overall volume mode $\delta\tau_1$, is reflecting the geometrical separation in the internal space
between the different localised singularities resolved by each different blow-up mode.

As we have seen in section \ref{ProbBI}, we need to fine-tune the coefficient of the
string loop corrections in order to prevent them from spoiling the flatness of the inflationary potential.
In addition, we shall consider two situations:
\begin{enumerate}
\item Inflaton 4-cycle $\tau_2$ wrapped just by the hidden sector $D7$ stack;

\item Inflaton 4-cycle $\tau_2$ wrapped by both the visible and the hidden sector $D7$ stack. More precisely,
the hidden sector $D7$-branes are wrapped around $\tau_2$ while the visible sector is wrapped around a combination of
$\tau_2$ and $\tau_4$ with chiral intersections only on $\tau_4$.
\end{enumerate}

\medskip
\textit{Inflaton 4-cycle not wrapped by the visible sector}
\medskip

The brane set-up is (assuming that the tadpole-cancelation condition can be satisfied by an appropriate choice
of background fluxes):
\begin{itemize}
\item $\tau_1$ is not wrapped by any brane;

\item $\tau_2$ is wrapped by a hidden sector $D7$-stack that undergoes gaugino-condensation;

\item $\tau_3$ is wrapped by a hidden sector $D7$-stack that undergoes gaugino-condensation;

\item $\tau_4$ is wrapped by the visible sector stack of $D7$-branes.
\end{itemize}

\begin{figure}[ht]
\begin{center}
\epsfig{file=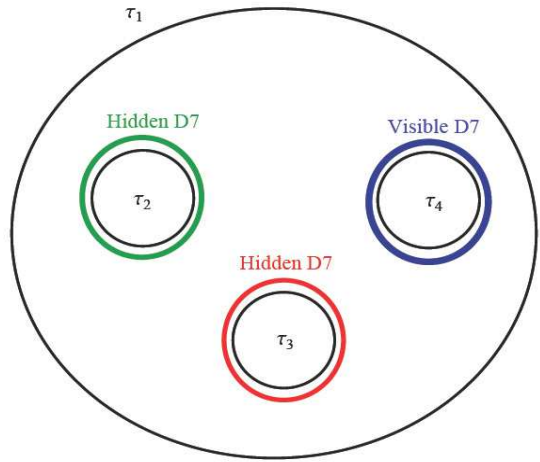, height=63mm,width=72mm} \caption{Brane set-up for BI
in the geometric regime with the inflaton $\tau_2$ wrapped by a hidden sector.} \label{Fig2}
\end{center}
\end{figure}

\newpage
The order of magnitude of the moduli couplings to hidden gauge bosons living on $\tau_2$ and $\tau_3$,
and to visible gauge bosons living on $\tau_4$ reads (see appendix \ref{appendice} for a detailed derivation):
\begin{figure}[ht]
\begin{center}
\begin{tabular}{c||c|c|c|c}
  & $\delta\phi_1$ & $\delta\phi_2$
  & $\delta\phi_3$ & $\delta\phi_4$ \\
  \hline\hline
  \\ & & & & \vspace{-0.9cm}\\
  $(F_{\mu \nu}^{(2)} F^{\mu \nu}_{(2)})$
  & $\frac{1}{M_P}$
  & $\frac{\vo^{1/2}}{M_P}$
  & $\frac{1}{\vo^{1/2}M_P}$
  & $\frac{1}{\vo^{1/2}M_P}$ \\
  \hline
  \\ & & & & \vspace{-0.9cm}\\
  $(F_{\mu \nu}^{(3)} F^{\mu \nu}_{(3)})$
  & $\frac{1}{M_P}$
  & $\frac{1}{\vo^{1/2}M_P}$
  & $\frac{\vo^{1/2}}{M_P}$
  & $\frac{1}{\vo^{1/2}M_P}$ \\
  \hline
  \\ & & & & \vspace{-0.9cm}\\
  $(F_{\mu \nu}^{(4)} F^{\mu \nu}_{(4)})$
  & $\frac{1}{M_P}$
  & $\frac{1}{\vo^{1/2}M_P}$
  & $\frac{1}{\vo^{1/2}M_P}$
  & $\frac{\vo^{1/2}}{M_P}$
\end{tabular}
\end{center}
\end{figure}

As studied in \cite{Preheating}, at the end of inflation, due to the steepness of the potential,
the inflaton $\tau_2$ stops oscillating just after two or three oscillations due to an extremely efficient
non-perturbative particle production of $\delta\tau_2$ fluctuations.
From the canonical normalisation (\ref{cn2}), we then realise that our Universe will mostly be filled with $\delta\phi_2$-particles
plus some $\delta\phi_1$ and fewer $\delta\phi_3$ and $\delta\phi_4$-particles.
Therefore the energy density of the Universe is dominated by $\delta\phi_2$ whose decay will be responsible for reheating.

Then denoting as $g$ the visible gauge bosons and with $X_2$ and $X_3$ the hidden ones,
we immediately realise that the coupling of $\delta\phi_2$ to $gg$ is suppressed by an inverse
power of the overall volume with respect to its coupling to $X_2 X_2$. Hence the inflaton will dump
all its energy to hidden, instead of visible, sector \textit{dof}. More precisely the first particles to decay are:
\be
\delta\phi_2 \to X_2 X_2\text{, \ \ \ \ \ }\delta\phi_3 \to X_3 X_3\text{, \ \ \ \ \ }\delta\phi_4 \to g g.
\ee
Later on also $\delta\phi_1$ decays to visible and hidden \textit{dof} out of thermal equilibrium.

The only way to avoid the inflaton energy dumping to hidden sector \textit{dof} is to forbid the decay of $\delta\phi_2$ to any hidden particle.
In section \ref{HiddenSector} we have studied the dynamics of the hidden sector gauge theory and the relative mass spectrum.
Moreover we focused on the simplest case of a rigid blow-up cycle invariant under the orientifold, not intersecting with any other cycle and not
shrinking at the quiver locus. In this case we have found that the mass spectrum is strictly dependent on the way the $D7$-branes get magnetised,
finding some cases where all hidden sector particles get a mass $m$ greater then the inflaton mass $m_2$: $m\geq \Lambda > m_2$.
This is the case when no gauge flux is turned on, or $F_2\neq 0$ but all matter fields get a string scale mass due to their Yukawa interaction
with a scalar that takes an $\mathcal{O}(M_s)$ VEV via $D$-terms \cite{Choi}.

We shall therefore consider these cases where the decay of $\delta\phi_2$ to $X_2 X_2$ is kinematically forbidden. As pointed out before,
we stress again that the coupling of the hidden \textit{dof} to the visible ones is too weak to reheat the visible sector via the decay of hidden
to visible \textit{dof}. Then the first particles to decay are $\delta\phi_3$ and $\delta\phi_4$ but their decay does not give rise to any reheating since the energy density of the Universe is dominated by $\delta\phi_2$. Reheating takes place only later on when $\delta\phi_2$ decays to visible gauge bosons
with total decay rate (see appendix \ref{appendice}):
\be
\Gamma_{\delta\phi_2\to gg}^{TOT}\simeq (\ln\vo)^3\frac{M_P}{\vo^4}.
\ee
The maximal reheating temperature for the visible sector in the approximation of sudden thermalisation can be worked out as follows:
\be
\frac{4\left(\Gamma_{\delta\phi_2\to gg}^{TOT}\right)^2}{3}=3 H^2 =\left(\frac{g_* ^{vis}\pi^2}{30}\right) \frac{\left(T_{RH}^{max}\right)^4}{M_P^2}
\,\,\,\,\,\,\Rightarrow \,\,\,\,\,\,T_{RH}^{max}=\left(\frac{40}{g_*^{vis} \pi^2}\right)^{1/4}\sqrt{\Gamma_{\delta\phi_2\to gg}^{TOT} M_P},
\label{defTRH}
\ee
where $g_*^{vis}\simeq 200$ is the total number of relativistic \textit{dof} in the thermal bath. Hence we end up with:
\be
T_{RH}^{max}\simeq \left(\frac{(\ln\vo)^{3/2}}{2}\right)\frac{M_P}{\vo^2}.
\label{TRH1}
\ee
We notice that for $\vo\sim 10^{6 - 7}$ (as needed to match the COBE normalisation
for the density fluctuations in BI), we would obtain $T_{RH}^{max}\simeq 10^{6- 8}$ GeV for the ideal case when all
the \textit{dof} are relativistic. This
is the largest temperature a visible sector can have in this particular scenario.


However, given that $\delta\phi_2$ couples to $gg$ and $X_3 X_3$ with the same $(1/(M_P\sqrt{\vo}))$-strength, the inflaton
would decay at the same time to visible and hidden sector \textit{dof} living on $\tau_3$. This might give rise to the problem
of hidden sector dark matter overproduction due to the fact that $\delta\phi_2$ releases the same
amount of entropy to both sectors. Let us see in detail the behaviour with respect to this problem
of each realisation of the hidden sector outlined in section \ref{HiddenSector}:

\begin{enumerate}
\item Pure SYM theory: the inflaton decay to hidden sector \textit{dof} is kinematically forbidden, and so reheating is viable.

\item Pure SYM theory plus a massless $U(1)$: the mass spectrum below $\Lambda$ consists of massless photons plus photini
with $\mathcal{O}(M_{soft})$ mass. From the derivation of the moduli decay rates in appendix \ref{appendice}, we find
that the branching ratios read\footnote{Here the visible sector particles are denoted as $\psi$ (for fermions),
$H$ (for Higgs), $\lambda$ (for gauginos), $\varphi$ (for SUSY scalars) and $\tilde{H}$ (for Higgsinos) whereas
$\lambda_{hid}^{(3)}$ denotes hidden gauginos on $\tau_3$.}:
\begin{gather}
BR(\delta\phi_2\to gg)\simeq BR(\delta\phi_2\to \overline{\varphi}\varphi)\simeq 42\%, \notag \\
BR(\delta\phi_2\to g\overline{\psi}\psi)\simeq BR(\delta\phi_2\to \overline{H}H)\simeq
BR(\delta\phi_2\to X_3 X_3)
\simeq BR(\delta\phi_2\to \lambda_{hid}^{(3)}\lambda_{hid}^{(3)})\simeq 4\%, \notag \\
BR(\delta\phi_2\to\text{other visible \textit{dof}})\simeq BR(\delta\phi_2\to\delta\phi_1 \delta\phi_1)\simeq 0\%. \notag
\end{gather}
Therefore we might overproduce dark matter since about 4 over 100 inflaton decays yield hidden gauginos. We point out that this dark matter
would be produced directly from the inflaton decay given that hidden photons and photini cannot form a thermal bath.
In addition in the presence of $N_G^{hid}>1$ additional massless $U(1)$ factor the situation becomes even worse since:
\be
\frac{BR(\delta\phi_2\to gg)}{BR(\delta\phi_2\to X_3 X_3)}\simeq \frac{10}{N_G^{hid}}.
\ee
Hence we would prefer to discard this case.

\item Pure SYM theory plus matter but no massless $U(1)$: the mass spectrum below $\Lambda$
consists of massless pions plus massive condensates with $\mathcal{O}(M_{soft})$ mass.
From the results of appendix \ref{appendice}, we find that the branching ratios read\footnote{The hidden
\textit{dof} on $\tau_3$ are denoted as $\psi_{hid}^{(3)}$ (for massive fermionic condensates),
$\varphi_{hid}^{(3)}$ (for massive scalar condensates) and
$\pi_3$ (for massless pion-like mesons).}:
\begin{itemize}
\item For $N_f=2$
\begin{gather}
BR(\delta\phi_2\to gg)\simeq BR(\delta\phi_2\to \overline{\varphi}{\varphi})\simeq 40\%,\text{ \ \ \ }
BR(\delta\phi_2\to \overline{\psi}_{hid}^{(3)}\psi_{hid}^{(3)})\simeq 8\%, \notag \\
BR(\delta\phi_2\to g\overline{\psi}\psi)\simeq BR(\delta\phi_2\to \overline{H}H)\simeq 4\%, \notag \\
BR(\delta\phi_2\to\pi_3\pi_3)\simeq BR(\delta\phi_2\to\varphi_{hid}^{(3)}\varphi_{hid}^{(3)})\simeq 2\%, \notag \\
BR(\delta\phi_2\to\text{other visible \textit{dof}})\simeq BR(\delta\phi_2\to \delta\phi_1\delta\phi_1)\simeq 0\%. \notag
\end{gather}

\item For $N_f=5$
\begin{gather}
BR(\delta\phi_2\to \overline{\psi}_{hid}^{(3)}\psi_{hid}^{(3)})\simeq 32\%,
\text{ \ \ \ }BR(\delta\phi_2\to gg)\simeq BR(\delta\phi_2\to \overline{\varphi}{\varphi})\simeq 25\%, \notag \\
BR(\delta\phi_2\to\pi_3\pi_3)\simeq BR(\delta\phi_2\to\varphi_{hid}^{(3)}\varphi_{hid}^{(3)})\simeq 6\%, \notag \\
BR(\delta\phi_2\to g\overline{\psi}\psi)\simeq BR(\delta\phi_2\to \overline{H}H)\simeq 3\%, \notag \\
BR(\delta\phi_2\to\text{other visible \textit{dof}})\simeq BR(\delta\phi_2\to \delta\phi_1\delta\phi_1)\simeq 0\%. \notag
\end{gather}
\end{itemize}
Due to the large branching ratio of the inflaton decay to hidden massive fermionic condensates,
hidden dark matter would be overproduced, and so we need to discard this case.
On top of this, these particles will form a thermal bath due to their residual
QCD-like interactions, and massless hidden pions could then be produced not just directly from the inflaton decay
but also thermally.

Given that the hidden sector particles interact with the
visible ones only gravitationally, two thermal baths will get formed with two different temperatures $T_{vis}$ and $T_{hid}$, with
$T_{vis}\neq T_{hid}$ but both of them of the same order of magnitude as $T_{RH}^{max}$ given by (\ref{TRH1}).
Then both the thermal baths will contribute to the expansion of the Universe given by eq.~(\ref{evolution-both}),
where the number of hidden \textit{dof} is given by:  $g_*^{hid}=N_f^2$,~\footnote{The number of massless Goldstone bosons is given by the dimension of the
broken global symmetry group which is dim$(U(N_f))=N_f^2$ (the $SU(N_f)$ factor comes from chiral
symmetry breaking whereas the $U(1)$ factor comes from the breaking of the original $R$-symmetry).}
yielding too many new relativistic \textit{dof} in contrast with
the severe bounds from BBN, which allows {\it only} one relativistic species, i.e. $N_f=1$. Hence this is another reason why we need to discard this case.

\item Pure SYM theory plus matter and a massless $U(1)$: this case has to be discarded due to the reasons outlined above.
\end{enumerate}
We have seen that the only way to avoid the overproduction of hidden dark matter living on $\tau_3$ from the inflaton decay,
is to forbid the decay of $\delta\phi_2$ to any hidden particle.
Hence, like in the case of $\tau_2$, also $\tau_3$ has to support a pure $N=1$ SYM theory.
We finally stress that the requirement of having a viable reheating
sets severe constraints on the particular brane constructions realising the hidden sectors.
For example, no hidden photon can be present in the effective field theory even for a brane wrapping the large cycle $\tau_1$
since $\delta\phi_2$ would also couple to those photons as $1/(M_P \sqrt{\vo})$.

\bigskip
\textit{Inflaton 4-cycle wrapped by the visible sector}
\bigskip

The brane set-up is (assuming that the tadpole-cancelation condition can be satisfied by an appropriate choice
of background fluxes):
\begin{itemize}
\item $\tau_1$ is not wrapped by any brane;

\item $\tau_2$ is wrapped by a hidden sector $D7$-stack that undergoes gaugino-condensation
plus the visible sector $D7$-stack which has chiral intersections with $\tau_4$ but not with $\tau_2$;

\item $\tau_3$ is wrapped by a hidden sector $D7$-stack that undergoes gaugino-condensation;

\item $\tau_4$ is wrapped by the visible sector stack of $D7$-branes.
\end{itemize}
The moduli couplings to hidden gauge bosons living on $\tau_2$ and $\tau_3$,
and to visible gauge bosons living on $\tau_4$ scale as:
\begin{figure}[ht]
\begin{center}
\begin{tabular}{c||c|c|c|c}
  & $\delta\phi_1$ & $\delta\phi_2$
  & $\delta\phi_3$ & $\delta\phi_4$ \\
  \hline\hline
  \\ & & & & \vspace{-0.9cm}\\
  $(F_{\mu \nu}^{(2)} F^{\mu \nu}_{(2)})$
  & $\frac{1}{M_P}$
  & $\frac{\vo^{1/2}}{M_P}$
  & $\frac{1}{\vo^{1/2}M_P}$
  & $\frac{1}{\vo^{1/2}M_P}$ \\
  \hline
  \\ & & & & \vspace{-0.9cm}\\
  $(F_{\mu \nu}^{(3)} F^{\mu \nu}_{(3)})$
  & $\frac{1}{M_P}$
  & $\frac{1}{\vo^{1/2}M_P}$
  & $\frac{\vo^{1/2}}{M_P}$
  & $\frac{1}{\vo^{1/2}M_P}$ \\
  \hline
  \\ & & & & \vspace{-0.9cm}\\
  $(F_{\mu \nu}^{(4)} F^{\mu \nu}_{(4)})$
  & $\frac{1}{M_P}$
  & $\frac{\vo^{1/2}}{M_P}$
  & $\frac{1}{\vo^{1/2}M_P}$
  & $\frac{\vo^{1/2}}{M_P}$
\end{tabular}
\end{center}
\end{figure}

\begin{figure}[ht]
\begin{center}
\epsfig{file=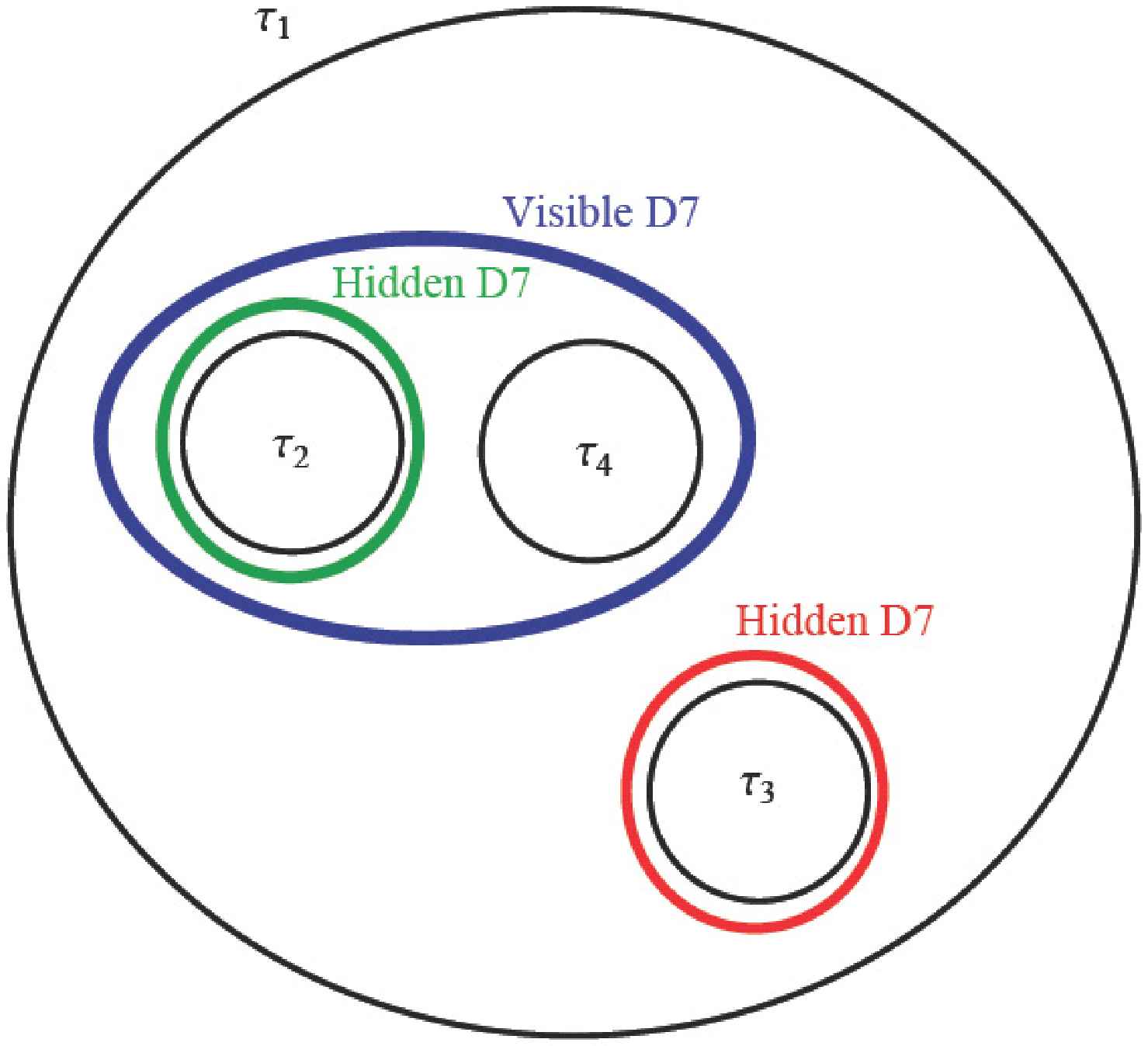, height=70mm,width=80mm} \caption{Brane set-up for BI
in the geometric regime with the inflaton $\tau_2$ wrapped by the visible sector.} \label{Fig1}
\end{center}
\end{figure}
At the end of inflation there will be non-perturbative production of $\delta\tau_2$
excitations given that the field $\tau_2$ is the inflaton. From the canonical normalisation
(\ref{cn2}), we then realise that our Universe will mostly be filled with $\delta\phi_2$-particles
plus some $\delta\phi_1$ and fewer $\delta\phi_3$ and $\delta\phi_4$-particles.
The inflaton $\delta\phi_2$ couples to visible gauge bosons $g$ and hidden gauge bosons $X_2$ with the
same $1/M_s$-strength while it couples much more weakly (as $1/(M_P\sqrt{\vo})$)
to the hidden gauge bosons living on $\tau_3$. Therefore $\delta\phi_2$ will decay at the same time
to visible and hidden sector \textit{dof} with the possible overproduction of dark matter particles
from the hidden sector. More precisely the first particles to decay are:
\be
\delta\phi_2\to gg\text{ \ or \ }X_2 X_2\text{, \ \ \ \ \ }\delta\phi_3 \to X_3 X_3
\text{, \ \ \ \ \ }\delta\phi_4 \to g g.
\ee
Only the decay of $\delta\phi_2$ gives rise to reheating since the energy density of the Universe is dominated by this particle.
The maximal reheating temperature of the visible sector turns out to be (see appendix \ref{appendice}):
\be
\Gamma_{\delta\phi_2\to gg}^{TOT}\simeq 0.1(\ln\vo)^3\frac{M_P}{\vo^2}
\,\,\Rightarrow\,\, T_{RH}^{max}\simeq 0.1(\ln\vo)^{3/2}\frac{M_P}{\vo}.
\label{TRH2}
\ee
We notice that for $\vo\simeq 10^{6- 7}$, we would obtain a rather high reheating temperature: $T_{RH}^{max}\simeq 10^{12- 13}$ GeV,
that would allow any mechanism for the generation of the matter-antimatter asymmetry. Furthermore there is no gravitino
problem since for $\vo\simeq 10^7$ the gravitino would be heavier than $10$ TeV: $m_{3/2}\simeq M_P / \vo \simeq T_{RH}^{max}\simeq 10^{11}$ GeV.
Later on $\delta\phi_1$ decays out of thermal equilibrium to visible and hidden \textit{dof}.

Like in the case when the inflaton 4-cycle is not wrapped by the visible sector, we are again facing the same problem:
dark matter overproduction due to the fact that $\delta\phi_2$ releases the same
amount of entropy to the visible and the hidden sector living on $\tau_2$.
Let us see in detail the behaviour of each possible realisation of the hidden sector outlined in section \ref{HiddenSector}:
\begin{itemize}
\item Pure SYM theory: the inflaton decay to hidden sector \textit{dof} is kinematically forbidden, and so reheating is viable.

\item Pure SYM theory plus a massless $U(1)$: the mass spectrum below $\Lambda$
consists of massless photons plus photini with $\mathcal{O}(M_{soft})$ mass.
From the results of appendix \ref{appendice}, we find that the branching ratios read:
\begin{gather}
BR(\delta\phi_2\to gg)\simeq 50\%,\text{ \ \ }BR(\delta\phi_2\to \overline{\varphi}\varphi)\simeq 25\%, \notag \\
BR(\delta\phi_2\to g\overline{\psi}\psi)\simeq BR(\delta\phi_2\to g \lambda\lambda)\simeq BR(\delta\phi_2\to X_2 X_2)
\simeq BR(\delta\phi_2\to \lambda_{hid}^{(2)}\lambda_{hid}^{(2)})\simeq 5\% \notag \\
BR(\delta\phi_2\to \overline{H}H)\simeq BR(\delta\phi_2\to g\tilde{H}\tilde{H})\simeq 2.5\%,
\text{ \ \ }BR(\delta\phi_2\to\delta\phi_1\delta\phi_1)\simeq 0\%. \notag
\end{gather}
Therefore we might overproduce dark matter since about 5 over 100 inflaton decays yield hidden gauginos. We point out that this dark matter
would be produced directly from the inflaton decay given that hidden photons and photini cannot form a thermal bath. In addition in the presence
of more than one additional massless $U(1)$ factor the situation becomes even worse hence we would prefer to discard this case.

\item Pure SYM theory plus matter but no massless $U(1)$: the mass spectrum below $\Lambda$ consists of massless pions plus massive condensates with $\mathcal{O}(M_{soft})$ mass. The branching ratios read (see appendix \ref{appendice}):
\begin{itemize}
\item For $N_f=2$
\begin{gather}
BR(\delta\phi_2\to gg)\simeq 50\%,\text{ \ \ \ \ \ \ }BR(\delta\phi_2\to \overline{\varphi}\varphi)\simeq 25\%, \notag \\
BR(\delta\phi_2\to\overline{\psi}_{hid}^{(2)}\psi_{hid}^{(2)})\simeq 10\%,
\text{ \ \ }BR(\delta\phi_2\to g\overline{\psi}\psi)\simeq BR(\delta\phi_2 \to g\lambda\lambda) \simeq 5\%,\notag \\
BR(\delta\phi_2\to\overline{H}H)\simeq BR(\delta\phi_2\to g\tilde{H}\tilde{H}) \simeq 2.5\%, \text{ \ }
BR(\delta\phi_2\to\text{other hidden \textit{dof}})\simeq 0\%. \notag
\end{gather}

\item For $N_f=5$
\begin{gather}
BR(\delta\phi_2\to\overline{\psi}_{hid}^{(2)}\psi_{hid}^{(2)})\simeq 40\%,\text{ \ \ \ }BR(\delta\phi_2\to gg)\simeq 32\%,\text{ \ \ \ }
BR(\delta\phi_2\to \overline{\varphi}\varphi)\simeq 16\%, \notag \\
BR(\delta\phi_2\to g\overline{\psi}\psi)\simeq BR(\delta\phi_2 \to g\lambda\lambda) \simeq 3\%,\notag \\
BR(\delta\phi_2\to\overline{H}H)\simeq BR(\delta\phi_2\to g\tilde{H}\tilde{H}) \simeq 2\%, \notag \\
BR(\delta\phi_2\to\pi_2\pi_2)\simeq BR(\delta\phi_2\to \varphi_{hid}^{(2)}\varphi_{hid}^{(2)}) \simeq 1\%,\text{ \ \ }
BR(\delta\phi_2\to\delta\phi_1\delta\phi_1)\simeq 0\%. \notag
\end{gather}

\end{itemize}
We immediately realise that we need to discard this case since we would overproduce hidden dark matter.

\item Pure SYM theory plus matter and a massless $U(1)$: this case has to be discarded due to the reasons outlined above.
\end{itemize}
We finally stress that we have seen also in this case how the requirement of having a viable reheating sets severe constraints
on the hidden sector model building. Again the only clearly viable scenario is the one where the hidden sector on $\tau_2$
consists of just a pure SYM theory. On the other hand, in this case there is no constraint on the hidden sector wrapping $\tau_3$
or on another possible hidden sector wrapping $\tau_1$. Thus this model would allow the presence of hidden photons living on
both a small and a large 4-cycle.


\subsubsection{Quiver Locus}
\label{Quiv}

In this case the inflaton 4-cycle can never be wrapped by the visible sector since this is realised by $D3$-branes at
the $\tau_4$-singularity\footnote{The authors of \cite{Preheating} considered the very speculative case of the inflaton shrinking
below the string scale at the end of inflation assuming that it is possible to derive such a model of inflation.
The inflaton's energy should be transferred to closed strings which are created by the merging of two winding modes
which appear in the theory when the cycle shrinks below the string scale.
Afterwards these excited closed strings decay to KK gravitons
which in turn decay to SM \textit{dof}, in a way somehow similar to what happens at the end of brane inflation.
However the realisation of inflation and the transition from the geometric to the singular regime is not very clear, and so we shall
not consider this case.}. Therefore we need to focus only on the case of the inflaton 4-cycle wrapped by just the
hidden sector undergoing gaugino condensation\footnote{It is understood that we assume that the coefficients of the string loop
corrections are fine-tuned small in order not to spoil the flatness of the inflationary potential.}. The resulting brane set-up is:

\begin{itemize}
\item $\tau_1$ is not wrapped by any brane;

\item $\tau_2$ is wrapped by a hidden sector $D7$-stack that undergoes gaugino-condensation;

\item $\tau_3$ is wrapped by a hidden sector $D7$-stack that undergoes gaugino-condensation;

\item $\tau_4$ is shrunk down at the quiver locus
and the visible sector is built via $D3$-branes at the singularity.
\end{itemize}

\begin{figure}[ht]
\begin{center}
\epsfig{file=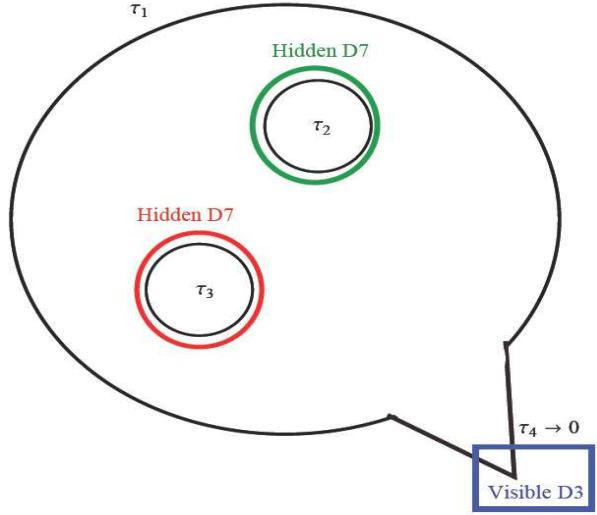, height=70mm,width=80mm} \caption{Brane set-up for BI
at the quiver locus.} \label{Fig3}
\end{center}
\end{figure}

In this case the gauge kinetic function for the visible sector takes the form \cite{quiver}:
\be
4\pi g^{-2}=\text{Re}(S)+h(F)\tau_4,
\ee
hence in order to work out the coupling of the inflaton $\tau_2$ to the visible gauge bosons,
we have to derive its mixing with the real part of the axio-dilaton $s=$Re$(S)$
and the other K\"{a}hler modulus $\tau_4$.
The effective field theory at the quiver locus admits a K\"{a}hler potential which can be
expanded around the singularity as (with $\hat{\xi}\equiv\xi/g_s^{3/2}=\xi/s^{3/2}$) \cite{quiver}:
\be
K=-2\ln\left(\vo'+\frac{\hat{\xi}}{2}\right)+\lambda \frac{\tau_4^2}{\vo'}-\ln\left(2 s\right),
\label{Kquivero}
\ee
with:
\be
\vo'=\alpha\left(\tau_1^{3/2}-\gamma_2\tau_2^{3/2}-\gamma_3\tau_3^{3/2}\right).
\ee
The particular form of the K\"{a}hler potential (\ref{Kquivero}) and $\langle\tau_4\rangle=0$
imply that at leading order there is no
mixing between $\tau_2$ and $\tau_4$. In addition, at tree level, the K\"{a}hler potential
takes a factorised form, and so $\tau_2$ does not mix with the axio-dilaton either. However $\tau_2$ mixes with $s$
due to $\alpha'$ corrections to $K$ leading to the following general form of the canonical normalisation
(see appendix \ref{appendice} for the detailed derivation):
\begin{eqnarray}
\delta \tau _{1} &\sim&\mathcal{O}(\mathcal{V}^{2/3})\delta \phi
_{1}+\sum_{i=2}^{3}\mathcal{O}(\mathcal{V}^{1/6})\delta \phi _{i}+\mathcal{O}(\vo^{1/6})\delta\phi_s\sim
\mathcal{O}(\mathcal{V}^{2/3})\delta \phi
_{1}, \\
\delta \tau _{2} &\sim&\mathcal{O}(1)\delta \phi _{1}+\mathcal{O}(\mathcal{V}
^{1/2})\delta \phi _{2}+\mathcal{O}(\mathcal{V}^{-1/2})\delta
\phi _3+\mathcal{O}(\vo^{-1/2})\delta\phi_s\sim \mathcal{O}(\mathcal{V}
^{1/2})\delta \phi _{2},
\label{cn2quiver} \\
\delta \tau _{3} &\sim&\mathcal{O}(1)\delta \phi _{1}+\mathcal{O}(\mathcal{V}
^{1/2})\delta \phi _{3}+\mathcal{O}(\mathcal{V}^{-1/2})\delta
\phi _2+\mathcal{O}(\vo^{-1/2})\delta\phi_s\sim \mathcal{O}(\mathcal{V}
^{1/2})\delta \phi_3, \\
\delta \tau_4 &\sim&\mathcal{O}(\mathcal{V}
^{1/2})\delta \phi_4 \\
\delta s &\sim&\mathcal{O}(\vo^{-1/2})\delta \phi _{1}+\sum_{j=2}^3\mathcal{O}(\mathcal{V}
^{-1})\delta \phi_j+\mathcal{O}(1)\delta\phi_s \sim \mathcal{O}(1)\delta\phi_s.
\end{eqnarray}
Then the volume scaling of the moduli couplings to hidden gauge bosons living on $\tau_2$ and $\tau_3$,
and to visible gauge bosons at the $\tau_4$-singularity reads:
\begin{figure}[ht]
\begin{center}
\begin{tabular}{c||c|c|c|c|c}
  & $\delta\phi_1$ & $\delta\phi_2$
  & $\delta\phi_3$ & $\delta\phi_4$ & $\delta\phi_s$ \\
  \hline\hline
  \\ & & & & \vspace{-0.9cm}\\
  $(F_{\mu \nu}^{(2)} F^{\mu \nu}_{(2)})$
  & $\frac{1}{M_P}$
  & $\frac{\vo^{1/2}}{M_P}$
  & $\frac{1}{\vo^{1/2}M_P}$
  & -
  & $\frac{1}{\vo^{1/2}M_P}$ \\
  \hline
  \\ & & & & \vspace{-0.9cm}\\
  $(F_{\mu \nu}^{(3)} F^{\mu \nu}_{(3)})$
  & $\frac{1}{M_P}$
  & $\frac{1}{\vo^{1/2}M_P}$
  & $\frac{\vo^{1/2}}{M_P}$
  & -
  & $\frac{1}{\vo^{1/2}M_P}$ \\
  \hline
  \\ & & & & \vspace{-0.9cm}\\
  $(F_{\mu \nu}^{(4)} F^{\mu \nu}_{(4)})$
  & $\frac{1}{\vo^{1/2}M_P}$
  & $\frac{1}{\vo M_P}$
  & $\frac{1}{\vo M_P}$
  & $\frac{\vo^{1/2}}{M_P}$
  & $\frac{1}{M_P}$
\end{tabular}
\end{center}
\end{figure}

At the end of inflation there will be non-perturbative production of $\delta\tau_2$
excitations given that the field $\tau_2$ is the inflaton. From the canonical normalisation
(\ref{cn2quiver}), we then realise that our Universe will mostly be filled with $\delta\phi_2$-particles
plus some $\delta\phi_1$ and fewer $\delta\phi_3$ and $\delta\phi_s$-particles.

This model is affected by the problem of the dumping of the inflaton energy to hidden, instead of visible, sector \textit{dof}.
In fact the coupling of $\delta\phi_2$ to $(F_{\mu \nu}^{(2)} F^{\mu \nu}_{(2)})$ is stronger than the coupling
to $(F_{\mu \nu}^{(3)} F^{\mu \nu}_{(3)})$ which, in turn, is stronger than the coupling to the visible $(F_{\mu \nu}^{(4)} F^{\mu \nu}_{(4)})$.
The explanation of this different behaviour is the geometric separation between $\tau_2$ and $\tau_3$ and, on top of that, the
sequestering of the visible sector at the $\tau_4$-singularity.

Thus the first particles to decay are $\delta\phi_2 \to X_2 X_2$ and $\delta\phi_3 \to X_3 X_3$. As we have seen in section \ref{GeomReg},
the only way to solve this problem is to forbid the decay of $\delta\phi_2$ to any hidden particle. This requirement forces us to consider
brane constructions on both $\tau_2$ and $\tau_3$ that yield just a pure SYM theory with no matter that develops a mass gap, so that
the decay of $\delta\phi_2$ to $X_2 X_2$ or $X_3 X_3$ is kinematically forbidden.
Then the first particle to decay is $\delta\phi_s\to gg$ but without giving rise to any reheating
since the energy density of the Universe is dominated by the inflaton field
$\delta\phi_2$ which has not decayed yet. Reheating takes place only later on when $\delta\phi_2$ decays to $gg$
producing the following maximal reheating temperature for the visible sector:
\be
\Gamma_{\delta\phi_2\to gg}^{TOT}\simeq (\ln\vo)^3\frac{M_P}{\vo^5}
\,\,\Rightarrow\,\, T_{RH}^{max}\simeq \left(\frac{(\ln\vo)^{3/2}}{2}\right)\frac{M_P}{\vo^{5/2}}.
\label{TRH3}
\ee
We notice that for $\vo\simeq 10^6$, we would obtain $T_{RH}^{max}\simeq 5\cdot 10^4$ GeV but for $\vo\simeq 10^7$, we would find $T_{RH}^{max}\simeq 10^2$ GeV.
However this reheating temperature is quite higher than $T_{BBN}\simeq 1$ MeV, and so
it would not give rise to any problem if the matter-antimatter asymmetry could be realised in a non-thermal/thermal way. Note that baryogenesis
is still possible at very low scales if R-parity violating interactions are introduced~\cite{KMS}.
We finally point out that this model does not allow the presence of any hidden photon
since the coupling of $\delta\phi_2$ to possible hidden photons living on the large cycle $\tau_1$
would scale as $1/(M_P \sqrt{\vo})$, and so it would be stronger than the coupling to visible \textit{dof} on $\tau_4$.


\subsection{Fibre Inflation}
\label{FI}

Let us focus on the inflationary model described in section \ref{FibInfl} considering both the
case when the cycle $\tau_4$ supporting the visible sector is stabilised in the geometric regime or
at the quiver locus.

\subsubsection{Geometric Regime}

The canonical normalisation reads (see appendix \ref{appendice}):
\begin{eqnarray}
\delta \tau_1 &\sim&\sum_{i=1}^2\mathcal{O}(\mathcal{V}^{2/3})\delta \phi_i
+\sum_{j=3}^4\mathcal{O}(\mathcal{V}^{1/6})\delta \phi_j\sim
\sum_{i=1}^2\mathcal{O}(\mathcal{V}^{2/3})\delta \phi_i, \label{CANnorm1} \\
\delta \tau_2 &\sim&\sum_{i=1}^2\mathcal{O}(\mathcal{V}^{2/3})\delta \phi_i
+\sum_{j=3}^4\mathcal{O}(\mathcal{V}^{1/6})\delta \phi_j\sim
\sum_{i=1}^2\mathcal{O}(\mathcal{V}^{2/3})\delta \phi_i, \label{CANnorm2} \\
\delta \tau_3 &\sim&\mathcal{O}(c^{-1/2}\mathcal{V}
^{-1/3})\delta \phi_1+\mathcal{O}(1)\delta \phi_2+\mathcal{O}(\mathcal{V}
^{1/2})\delta \phi_3+\mathcal{O}(\mathcal{V}^{-1/2})\delta
\phi_4\sim \mathcal{O}(\mathcal{V}
^{1/2})\delta \phi_3, \\
\delta \tau_4 &\sim&\mathcal{O}(c^{-1/2}\mathcal{V}
^{-1/3})\delta \phi_1+\mathcal{O}(1)\delta \phi_2+\mathcal{O}(\mathcal{V}^{-1/2})\delta
\phi_3+\mathcal{O}(\mathcal{V}
^{1/2})\delta \phi _4\sim \mathcal{O}(\mathcal{V}
^{1/2})\delta \phi_4.
\end{eqnarray}
and the moduli mass spectrum looks like (with $\langle\tau_1\rangle=c\,\vo^{2/3}$):
\be
m^2_1\simeq\frac{M_P^2}{c^{1/2}\vo^{10/3}},\text{ \ \ }m^2_2\simeq\frac{M_P^2}{\vo^3\ln\vo},
\text{ \ \ and \ \ }m_i^2\simeq\frac{(2\ln\vo)^2}{\vo^2}M_P^2,\text{ \ }
\forall i=3,4.
\label{MASSES}
\ee
As shown in appendix \ref{appendice}, inverting the exact form of (\ref{CANnorm1}) and (\ref{CANnorm2}), it turns out that $\delta\phi_2$
is a particular combination of $\delta\tau_1$ and $\delta\tau_2$ which corresponds to the overall volume, whereas
$\delta\phi_1$ is a direction orthogonal to this one which is fixed only at subleading order by string loops.
This is why, as can be seen from (\ref{MASSES}), $\delta\phi_1$ is lighter than $\delta\phi_2$, and plays the r\^{o}le
of the inflaton. We also stress that the canonical normalisation for a K3 fibration and a Swiss-cheese Calabi-Yau
has the same volume scaling once we identify $\delta\phi_2$ as the large overall volume mode. The brane set-up is:
\begin{itemize}
\item $\tau_1$ is wrapped by a hidden sector $D7$-stack which generates string loops;

\item $\tau_2$ is wrapped by a hidden sector $D7$-stack which generates string loops;

\item $\tau_3$ is wrapped by a hidden sector $D7$-stack that undergoes gaugino-condensation;

\item $\tau_4$ is wrapped by the visible sector stack of $D7$-branes.
\end{itemize}

\begin{figure}[ht]
\begin{center}
\epsfig{file=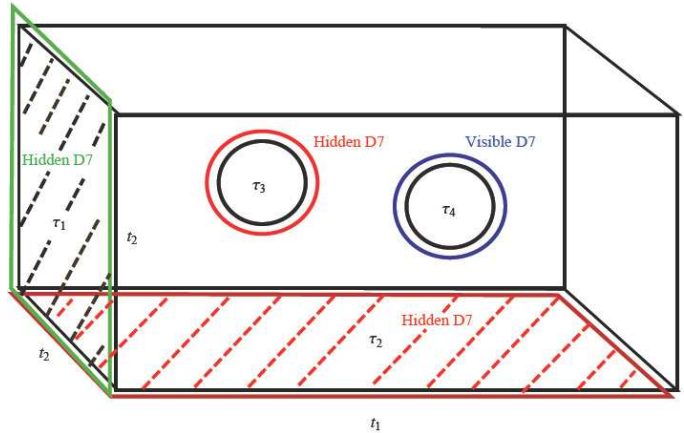, height=58mm,width=90mm} \caption{Brane set-up for FI
in the geometric regime.} \label{Fig4}
\end{center}
\end{figure}

The volume scaling of the moduli couplings to visible gauge bosons living on $\tau_4$,
and to hidden gauge bosons living on $\tau_1$,  $\tau_2$ and $\tau_3$ reads:
\newline
\begin{figure}[ht]
\begin{center}
\begin{tabular}{c||c|c|c|c}
  & $\delta\phi_1$ & $\delta\phi_2$
  & $\delta\phi_3$ & $\delta\phi_4$ \\
  \hline\hline
  \\ & & & & \vspace{-0.9cm}\\
  $(F_{\mu \nu}^{(1,2)} F^{\mu \nu}_{(1,2)})$
  & $\frac{1}{M_P}$
  & $\frac{1}{M_P}$
  & $\frac{1}{\vo^{1/2}M_P}$
  & $\frac{1}{\vo^{1/2}M_P}$ \\
  \hline
  \\ & & & & \vspace{-0.9cm}\\
  $(F_{\mu \nu}^{(3)} F^{\mu \nu}_{(3)})$
  & $\frac{c^{-1/2}}{\vo^{1/3} M_P}$
  & $\frac{1}{M_P}$
  & $\frac{\vo^{1/2}}{M_P}$
  & $\frac{1}{\vo^{1/2}M_P}$ \\
  \hline
  \\ & & & & \vspace{-0.9cm}\\
  $(F_{\mu \nu}^{(4)} F^{\mu \nu}_{(4)})$
  & $\frac{c^{-1/2}}{\vo^{1/3} M_P}$
  & $\frac{1}{M_P}$
  & $\frac{1}{\vo^{1/2}M_P}$
  & $\frac{\vo^{1/2}}{M_P}$
\end{tabular}
\end{center}
\end{figure}

As we have pointed out before, the inflaton $\chi$
is a particular combination of $\tau_1$ and $\tau_2$ orthogonal to the overall volume
($a$ and $b$ are $\mathcal{O}(1)$ constants):
\be
\chi = a\ln\tau_1+b\ln\tau_2,
\ee
and so at the end of inflation there will be non-perturbative production of $\delta\chi$
excitations. Given that the shape of the inflationary potential for FI is much less steep
than the one for BI, we expect a different behaviour for preheating
without such a violent non-perturbative production
of inflaton inhomogeneities. It might even be that in this case all the dynamics is purely perturbative.
The canonical normalisation around the minimum for $\chi$ is given by:
\be
\delta\chi\sim \mathcal{O}(1)\delta\phi_1+\mathcal{O}(c^{-1/2}\vo^{-1/3})\delta\phi_2
+\sum_{i=3}^4 \mathcal{O}(\vo^{-1/2})\delta\phi_i,
\label{cnchi}
\ee
where the subleading mixing with $\delta\phi_2$ is introduced by the string loop corrections to the K\"{a}hler potential (\ref{Vloop}).
From the canonical normalisation eq.~(\ref{cnchi}), we then realise that our Universe will mostly be filled with $\delta\phi_1$-particles
plus some $\delta\phi_2$ and fewer $\delta\phi_3$ and $\delta\phi_4$-particles.

Then denoting as $g$ the visible gauge bosons and with $X_1$, $X_2$ and $X_3$ the hidden sector ones,
the first particles to decay are $\delta\phi_3$ and $\delta\phi_4$: $\delta\phi_3 \to X_3 X_3$ and $\delta\phi_4 \to gg$.
However these decays do not give rise to any reheating since the energy density of the Universe is dominated by $\delta\phi_1$.
This is also the case when the $\delta\phi_2$ particles decay to all the gauge bosons in the theory at the same time.
Finally also $\delta\phi_1$ decays but no reheating takes place in the visible sector since the inflaton coupling to the hidden
gauge bosons $X_1$ and $X_2$ is stronger than the one to visible gauge bosons and $X_3$ since (see appendix \ref{appendice}):
\be
\frac{\Gamma _{\delta \phi_1\rightarrow gg
}}{\Gamma _{\delta \phi_1\rightarrow X_1 X_1}}\sim \left(\frac{0.01}{(\ln\vo)^2}\right)\frac{1}{c\vo^{2/3}}
=\left(\frac{0.01}{(\ln\vo)^2}\right)\frac{1}{\langle \tau_1\rangle }\ll 1.
\ee
Thus reheating turns out to be a problem in this model where the inflaton dumps all its energy to
hidden, instead of visible, \textit{dof}. There are two possible solutions to this problem:

\begin{enumerate}
\item Forbid the inflaton decay to hidden sector \textit{dof}.

If all the hidden sector \textit{dof} are heavier than
the inflaton, then the inflaton decay to these particles is kinematically forbidden. In section \ref{BI}, we have seen
that this can indeed be the case if the hidden sector consists of a pure $N=1$ SYM theory that develops
a mass gap. Nonetheless, in this case, the hidden sector supported by $\tau_1$ and $\tau_2$
is only responsible to source the string loop corrections that generate the inflationary potential,
and so it does not have to undergo gaugino condensation.
Hence it might have a much more complicated gauge group structure and particle content with several particles
that would naturally be lighter than the inflaton. However there is nothing against having non-perturbative effects supported on
$\tau_1$ and $\tau_2$ since these two cycles are stabilised large, and so in a regime where the non-perturbative corrections are
subleading with respect to the string loops and do not modify the moduli stabilisation picture. We shall therefore assume that
it is indeed possible to render all the hidden sector \textit{dof} on $\tau_1$ and $\tau_2$ heavier than the inflaton, for example
considering a pure SYM theory living on these cycles.

We have then to consider a pure SYM theory also on $\tau_3$ otherwise,
as discussed in section \ref{BI}, too much hidden sector dark matter would be produced by the simultaneous inflaton decay to visible and
hidden \textit{dof} living on $\tau_3$. Under all these constraints on the hidden sector model building, reheating can take place
leading to the formation of a visible sector thermal bath with maximal temperature:
\be
\Gamma_{\delta\phi_1\to gg}^{TOT}\simeq \left(\frac{5\cdot 10^{-4}}{\ln\vo}\right)\frac{M_P}{c^{7/4}\vo^{17/3}}
\,\,\Rightarrow\,\, T_{RH}^{max}\simeq \left(\frac{0.01}{(\ln\vo)^{1/2}}\right)\frac{M_P}{c^{7/8}\vo^{17/6}}.
\label{TRH4}
\ee
In the case of FI the requirement of matching the COBE
normalisation for the density fluctuations sets $\vo\simeq 10^{3- 4}$, and using the same parameters of
\cite{fiberinfl}, $c\simeq 5\cdot 10^{-2}$. For these values we would obtain a maximal reheating temperature of the
order $T_{RH}^{max}\simeq 5\cdot 10^{5- 8}$ GeV. We finally stress that this model does not allow the presence of any light hidden photon.

\item Locate the visible sector on $\tau_1$.

Another possible way-out is to change our initial
brane set-up assuming that the visible sector is located on $\tau_1$
without any need to wrap $D7$-branes around $\tau_4$. The first problem
that one encounters with this set-up, is that the K3 fiber is not a rigid cycle and
so one has to worry about how to fix the $D7$-brane deformation
moduli that would give rise to unwanted matter in the adjoint
representation. Here we shall assume that these moduli can be
fixed by the use of background fluxes. The second problem is the
order of magnitude of the visible sector gauge coupling:
\be
\alpha_1=\frac{g_1^2}{4\pi}=\frac{1}{\langle\tau_1\rangle}=\frac{1}{c\,\vo^{2/3}},
\label{avis}
\ee
which for large volume would clearly become too small. However
in the case of FI the volume is not too large:
$\vo\simeq 10^4$. In addition at the end of inflation the K3 fibre sits
at a small size $\langle\tau_1\rangle\sim\mathcal{O}(10)$ (for $c\simeq 5\cdot 10^{-2}$) which reproduces
the correct order of magnitude of $\alpha_1$. Hence we shall
assume that it is indeed possible to wrap the visible sector around $\tau_1$.
Notice that in this case the stack of $D7$-branes wrapped around $\tau_2$ would intersect
with the visible sector stack wrapped around $\tau_1$. Hence there would be gauge interactions
among the particles living on $\tau_1$ and $\tau_2$. Thus $\tau_2$ is not
supporting a hidden sector but a different gauge group of the visible sector.
The gauge coupling of this additional force is given by (with $\vo\simeq \alpha\sqrt{\tau_1}\tau_2$):
\be
\alpha_2=\frac{g_2^2}{4\pi}=\frac{1}{\tau_2}=\frac{\alpha \sqrt{\tau_1}}{\vo}=\frac{\alpha \,c^{1/2}}{\vo^{2/3}}
\,\,\,\Leftrightarrow\,\,\,\frac{\alpha_1}{\alpha_2}=\frac{1}{\alpha\, c^{3/2}},
\ee
and so for $c\simeq 5\cdot 10^{-2}$ and $\alpha\simeq 0.1$ (as for typical Calabi-Yau examples \cite{ExplicitK3}),
the ratio $\alpha_1/\alpha_2\simeq 10^3$
takes the same order of magnitude of the ratio between the electromagnetic and the weak gauge coupling:
$\alpha_{EM}/\alpha_W \simeq 5\cdot 10^3$. Notice that the presence of at least two different visible sector
gauge groups does not allow to build a GUT theory with a unified gauge group unless the gauge bosons on $\tau_2$
decouple from the effective field theory getting an $\mathcal{O}(M_s)$ mass.

Therefore the inflaton couples to visible gauge bosons
living on $\tau_1$ and $\tau_2$ with the same $1/M_P$ strength. In addition, in the presence of
hidden massless pions living on $\tau_3$, $\delta\phi_1$ would also couple to these particles
as $1/M_P$. Hence the inflaton decay would yield too many new relativistic \textit{dof}
or it would overproduce hidden dark matter if the hidden pions on $\tau_3$ get massive via a
low-energy Higgs-like mechanism. On the other hand, $\delta\phi_1$ would couple to hidden photons
on $\tau_3$ as $1/(M_P \vo^{1/3})$, so more weakly than its coupling to visible gauge bosons.
Therefore the presence of an additional massless $U(1)$ factor on $\tau_3$ would not induce any problem.
Assuming that the hidden sector on $\tau_3$ is a pure SYM theory
with possibly the presence of an additional massless $U(1)$ factor,
the visible sector can be safely reheated to the temperature:
\be
\Gamma_{\delta\phi_1\to gg}^{TOT}\simeq 0.05 \frac{M_P}{c^{3/4}\vo^5}
\,\,\Rightarrow\,\, T_{RH}^{max}\simeq 0.1\frac{M_P}{c^{3/8}\vo^{5/2}}.
\label{TRH5}
\ee
For $\vo\simeq 10^{3- 4}$ and $c\simeq 5\cdot 10^{-2}$, we would obtain $T_{RH}^{max}\simeq 10^{8- 10}$ GeV.
We finally stress that this model would allow the presence of hidden photons living on the small cycle $\tau_3$.

\end{enumerate}


\subsubsection{Quiver Locus}

In this case the visible sector 4-cycle is shrunk down at the singularity and the resulting
brane set-up is:
\begin{itemize}
\item $\tau_1$ is wrapped by a hidden sector $D7$-stack which generates string loops;

\item $\tau_2$ is wrapped by a hidden sector $D7$-stack which generates string loops;

\item $\tau_3$ is wrapped by a hidden sector $D7$-stack that undergoes gaugino-condensation;

\item $\tau_4$ is shrunk down at the quiver locus
and the visible sector is built via D3-branes at the singularity.
\end{itemize}

\begin{figure}[ht]
\begin{center}
\epsfig{file=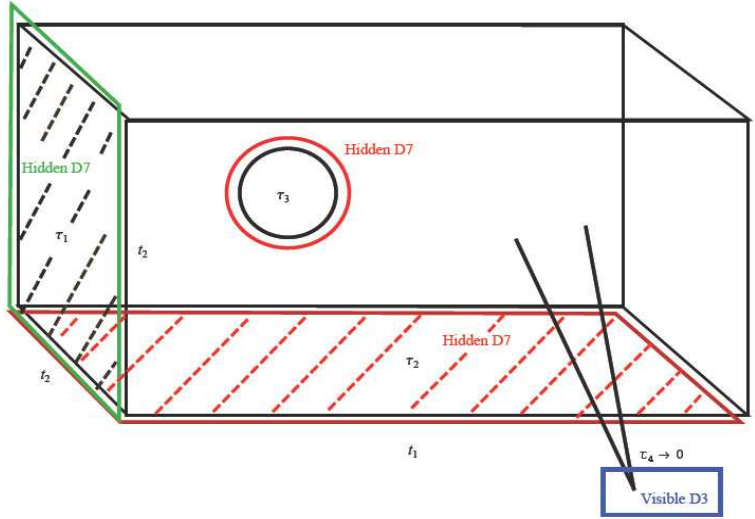, height=70mm,width=100mm} \caption{Brane set-up for FI
at the quiver locus.} \label{Fig5}
\end{center}
\end{figure}

As we have seen in section \ref{Quiv}, the particular form of the K\"{a}hler potential and $\langle\tau_4\rangle=0$
imply that at leading order there is no mixing between $\tau_4$ and the other moduli, leading to the following
canonical normalisation:
\begin{eqnarray}
\delta \tau_1 &\sim&\sum_{j=1}^2\mathcal{O}(\mathcal{V}^{2/3})\delta \phi_j
+\mathcal{O}(\mathcal{V}^{1/6})\delta \phi_3+\mathcal{O}(\vo^{1/6})\delta\phi_s\sim
\sum_{j=1}^2\mathcal{O}(\mathcal{V}^{2/3})\delta \phi_j, \\
\delta \tau_2 &\sim&\sum_{j=1}^2\mathcal{O}(\mathcal{V}^{2/3})\delta \phi_j
+\mathcal{O}(\mathcal{V}^{1/6})\delta \phi_3+\mathcal{O}(\vo^{1/6})\delta\phi_s\sim
\sum_{j=1}^2\mathcal{O}(\mathcal{V}^{2/3})\delta \phi_j,
\label{cn2quiverNew} \\
\delta \tau_3 &\sim&\mathcal{O}(c^{-1/2}\vo^{-1/3})\delta \phi_1+\mathcal{O}(1)\delta \phi_2+\mathcal{O}(\mathcal{V}
^{1/2})\delta \phi _{3}+\mathcal{O}(\vo^{-1/2})\delta\phi_s\sim \mathcal{O}(\mathcal{V}
^{1/2})\delta \phi_3, \\
\delta \tau _{4} &\sim&\mathcal{O}(\mathcal{V}
^{1/2})\delta \phi _{4} \\
\delta S &\sim&\mathcal{O}(c^{-1/2}\vo^{-5/6})\delta \phi_1+\mathcal{O}(\vo^{-1/2})\delta \phi_2+\mathcal{O}(\mathcal{V}
^{-1})\delta \phi_3+\mathcal{O}(1)\delta\phi_s\sim \mathcal{O}(1)\delta\phi_s.
\end{eqnarray}
Then the moduli couplings to visible gauge bosons at the $\tau_4$-singularity,
and to hidden gauge bosons living on $\tau_1$, $\tau_2$ and $\tau_3$ scale as:
\begin{figure}[ht]
\begin{center}
\begin{tabular}{c||c|c|c|c|c}
  & $\delta\phi_1$ & $\delta\phi_2$
  & $\delta\phi_3$ & $\delta\phi_4$ & $\delta\phi_s$ \\
  \hline\hline
  \\ & & & & \vspace{-0.9cm}\\
  $(F_{\mu \nu}^{(1,2)} F^{\mu \nu}_{(1,2)})$
  & $\frac{1}{M_P}$
  & $\frac{1}{M_P}$
  & $\frac{1}{\vo^{1/2}M_P}$
  & -
  & $\frac{1}{\vo^{1/2}M_P}$ \\
  \hline
  \\ & & & & \vspace{-0.9cm}\\
  $(F_{\mu \nu}^{(3)} F^{\mu \nu}_{(3)})$
  & $\frac{c^{-1/2}}{\vo^{1/3} M_P}$
  & $\frac{1}{M_P}$
  & $\frac{\vo^{1/2}}{M_P}$
  & -
  & $\frac{1}{\vo^{1/2}M_P}$ \\
  \hline
  \\ & & & & \vspace{-0.9cm}\\
  $(F_{\mu \nu}^{(4)} F^{\mu \nu}_{(4)})$
  & $\frac{c^{-1/2}}{\vo^{5/6}M_P}$
  & $\frac{1}{\vo^{1/2} M_P}$
  & $\frac{1}{\vo M_P}$
  & $\frac{\vo^{1/2}}{M_P}$
  & $\frac{1}{M_P}$
\end{tabular}
\end{center}
\end{figure}

At the end of inflation there will be non-perturbative production of $\delta\chi$
excitations. The canonical normalisation around the minimum for $\chi$ is given by:
\be
\delta\chi\sim \mathcal{O}(1)\delta\phi_1+\mathcal{O}(c^{-1/2}\vo^{-1/3})\delta\phi_2
+\mathcal{O}(\vo^{-1/2})\delta\phi_3+\mathcal{O}(\vo^{-1/2})\delta\phi_s,
\label{cnchiNew}
\ee
where the subleading mixing with $\delta\phi_2$ is introduced by the string loop corrections to the K\"{a}hler potential (\ref{Vloop}).
From the canonical normalisation
(\ref{cnchiNew}), we then realise that our Universe will mostly be filled with $\delta\phi_1$-particles
plus some $\delta\phi_2$ and fewer $\delta\phi_3$ and $\delta\phi_s$-particles.

Then denoting as $g$ the visible sector gauge bosons and with $X_1$, $X_2$ and $X_3$ the hidden sector ones,
the first decay is $\delta\phi_3\to X_3 X_3$ and later on $\delta\phi_s\to gg$.
However these decays do not give rise to any reheating since the energy density of the Universe is dominated by $\delta\phi_1$
which has not decayed yet. This is also the case when the $\delta\phi_2$ particles decay to all the hidden sector gauge bosons
in the theory at the same time.

Later on the inflaton dumps all its energy to hidden, instead of
visible, gauge bosons due to the fact that the coupling of $\delta\phi_1$ to $(F_{\mu \nu}^{(1,2)} F^{\mu \nu}_{(1,2)})$
is stronger than the coupling to $(F_{\mu \nu}^{(3)} F^{\mu \nu}_{(3)})$ which, in turn, is even stronger
than the coupling to the visible $(F_{\mu \nu}^{(4)} F^{\mu \nu}_{(4)})$.
Hence in this model the reheating of the visible sector turns out to be
a problem. We stress that we cannot try to overcome this problem as we did in the geometric regime case,
where we assumed a different set-up with the visible sector localised on $\tau_1$. In fact, in this case,
$\tau_1$ cannot shrink to zero size since the all volume of the Calabi-Yau would shrink to zero size.
However we can imagine a brane construction such that all the hidden sectors in the model consist of a pure SYM
theory that develops a mass gap. In this way all the hidden gauge bosons and gauginos condense and become heavier than
$\delta\phi_1$ forbidding the corresponding decays.

Under these assumptions, the maximal reheating temperature due to the $\delta\phi_1$ decay
to visible gauge bosons turns out to be:
\be
\Gamma_{\delta\phi_1\to gg}^{TOT}\simeq \left(\frac{5\cdot 10^{-4}}{\ln\vo}\right) \frac{M_P}{c^{7/4}\vo^{20/3}}
\,\,\Rightarrow\,\, T_{RH}^{max}\simeq \left(\frac{0.01}{(\ln\vo)^{1/2}}\right)\frac{M_P}{c^{7/8}\vo^{10/3}}.
\label{TRH6}
\ee
For $\vo\simeq 10^{3- 4}$ and $c\simeq 5\cdot 10^{-2}$, we would obtain $T_{RH}^{max}\simeq 10^{4- 7}$ GeV. As far as hidden photons are concerned,
we found that the requirement of having a viable reheating of the visible sector automatically forbids their presence in the effective field theory both for BI and for FI when the cycle supporting the visible sector is shrunk down at the singularity.


\section{Discussion and Conclusions}

Once the inflaton has completely decayed into visible sector \textit{dof}, the system evolves to a complete thermalisation and the thermal
bath generates a moduli-dependent finite temperature scalar potential\footnote{Note that within the MSSM, as it is the case for many scenarios,
the final reheating temperature can be as low as $1$ TeV.}. The behaviour of this potential for an arbitrary LVS has been studied in
\cite{LVSatFiniteT} where it was found that the finite temperature corrections do not
develop any new minimum, and so they cannot induce any phase transition, but they can still lead the system
to a dangerous decompactification limit. The computation of the decompactification temperature for LVS gives \cite{LVSatFiniteT}:
\be
T_{max}\simeq\frac{M_P}{\vo^{3/4}},
\label{Tmax}
\ee
and this is valid for all the different scenarios that we have studied here. This temperature sets the maximal
temperature of our Universe, and so one has to make sure that $T_{RH}^{max}<T_{max}$. It is reassuring to notice
that in all our cases the reheating temperature is always suppressed with respect to the
decompactification temperature (\ref{Tmax}) by inverse powers of the overall volume. Only in the case
of BI with the inflaton wrapped by the visible sector, the ratio
$R\equiv T_{RH}^{max}/T_{max}\simeq \vo^{-1/4}$ is not particularly small. Hence it is important to
compute the exact coefficient in from of the $\vo^{-1/4}$ factor. This has already been done in \cite{LVSatFiniteT}
and the final result is:
\be
R\equiv \frac{T_{RH}^{max}}{T_{max}}= c\frac{\left(\ln\vo\right)^{3/2}}{\vo^{1/4}}\text{ \ with \ }
c\simeq \frac{W_0\, e^{K_{cs}/2}}{2\langle\tau_2\rangle^{11/8}},
\ee
For $\vo\simeq 10^{6- 7}$, the previous ratio just depends on the $c$ coefficient: $R\simeq c$. However
for natural values $W_0\simeq K_{cs}\simeq \mathcal{O}(1)$ and $\langle\tau_2\rangle\simeq \mathcal{O}(10)$,
$c\simeq 0.1- 0.01$, and so we correctly have $R < 1$. In the fine-tuned cases when $R$ becomes dangerously
of the order one, larger values of the volume seem to be favoured. However for $\vo>10^7$, the inflaton fluctuations
are not able anymore to generate the right amount of density perturbations, which could be, on the other hand,
produced by some non-standard mechanism, like a curvaton or a modulating field \cite{BCGQTZ}.

\medskip

Let us now summarise the main results we have found in this paper:
\begin{itemize}
\item In any LVS hidden sectors are always present along with the visible sector.

\item The ubiquitous presence of a hidden sector wrapping the inflaton 4-cycle implies that BI needs fine-tuning whereas FI does not.

\item The hidden sector and the visible sector are not directly coupled, and furthermore the inflaton coupling to visible sector
\textit{dof} can never be made stronger than its coupling to hidden sector \textit{dof}.

\item Two generic problems arise:
\begin{enumerate}
\item Inflaton energy dumping to hidden, instead of visible, sector \textit{dof}. As such this
is not a problem provided the hidden sector \textit{dof} are relativistic and there exists a curvaton mechanism which is responsible for creating the
perturbations and the visible sector matter. Such a scenario was considered in past~\cite{EKM}, where the MSSM Higgs would dominate the
energy density during its oscillations and would create all the visible \textit{dof}. It would also be interesting to investigate if this problem could be solved by building a curvaton model along the lines of \cite{BCGQTZ} where the curvaton dominates the energy density.

\item Overproduction of hidden sector dark matter if there exists a massive long lived species, otherwise there would be too many new relativistic \textit{dof}, which would spoil the BBN predictions.
\end{enumerate}

\item A possible solution could be to render the inflaton decay to hidden sector \textit{dof} kinematically forbidden.
This sets severe constraints on hidden sector model building leaving as the most promising possibility
just a pure $N=1$ SYM theory. This is always the case for each hidden sector at the quiver locus and in the geometric regime
if the inflaton is not wrapped by the visible sector. On the contrary, in the geometric regime case with the inflaton wrapped
by the visible sector, the requirement of having just a pure SYM theory for the hidden sector not wrapping the inflaton
4-cycle can be relaxed.

\item This constraint seems to be in contrast with the generic presence of hidden photons whose study received a lot of attention recently \cite{Mark}. More precisely, the appearance of hidden photons living on small or large cycles seems to be compatible with reheating only for models in the
geometric regime with the inflaton 4-cycle wrapped by the visible sector. All the other cases seem to be incompatible
with the existence of hidden photons.
\end{itemize}

Assuming that these problems can be solved by an appropriate brane construction of the
hidden sector, let us summarise our results for the maximal reheating temperature of the visible sector:

\begin{figure}[ht]
\begin{center}
\begin{tabular}{c||c|c}
  & Geometric Regime & Quiver Locus \\
  \hline\hline
  \\ & & \vspace{-0.9cm}\\
  & Inflaton not wrapped by the visible sector:  \\
  Blow-up & $T_{RH}^{max}\sim M_P \vo^{-2}\sim 10^{6- 8}$ GeV for $\vo\sim 10^{6- 7}$
  & $T_{RH}^{max}\sim M_P \vo^{-5/2}\sim 10^{2- 4}$ GeV \\
  Inflation & Inflaton wrapped by the visible sector: & for $\vo\sim 10^{6- 7}$ \\
  & $T_{RH}^{max}\sim M_P \vo^{-1}\sim 10^{12- 13}$ GeV for $\vo \sim 10^{6- 7}$ \\
  \hline
  \\ & & \vspace{-0.9cm}\\
  & Inflaton not wrapped by the visible sector:  \\
  Fibre & $T_{RH}^{max}\sim M_P \vo^{-17/6}\sim 10^{6- 9}$ GeV for $\vo\sim 10^{3- 4}$
  & $T_{RH}^{max}\sim M_P \vo^{-10/3}\sim 10^{4- 7}$ GeV \\
  Inflation & Inflaton wrapped by the visible sector: & for $\vo\sim 10^{3- 4}$ \\
  & $T_{RH}^{max}\sim M_P \vo^{-5/2}\sim 10^{8- 10}$ GeV for $\vo\sim 10^{3- 4}$
\end{tabular}\\
\end{center}
\end{figure}

We stress that only the model of BI with the visible sector at the quiver locus gives rise to
a viable reheating with TeV scale SUSY at the same time. In all other cases $M_{soft}\gg 1$ TeV.
Let us now assume that the inflaton is only responsible to drive inflation but the
density fluctuations are generated in a non-standard way
by a second curvaton or modulating field \cite{BCGQTZ}. In this way we still obtain inflation
but the volume could be set so to obtain TeV scale SUSY. The resulting maximal reheating temperature in the visible sector would turn out to be:

\begin{enumerate}
\item BI in the geometric regime with the inflaton not wrapped by the visible sector:
for $\vo\simeq 10^{14}$ (as needed to obtain TeV scale SUSY in the geometric regime),
we would obtain $T_{RH}^{max}\simeq 10$ eV. This is in clear disagreement with the lower bound coming from
BBN: $T_{RH}^{max}>1$ MeV. Thus we infer that this model does not seem to be compatible with TeV scale SUSY.

\item BI in the geometric regime with the inflaton wrapped by the visible sector:
for $\vo\simeq 10^{14}$, we would obtain $T_{RH}^{max}\simeq 5\cdot 10^{5}$ GeV which is not in contrast with any observation.
Thus we infer that this model could be made compatible with TeV scale SUSY.

\item FI in the geometric regime with the inflaton not wrapped by the visible sector:
for $\vo\simeq 10^{14}$, and $c\simeq 5\cdot 10^{-2}$ we would obtain $T_{RH}^{max}\simeq 10^{-14}$ eV,
which is unrealistically small.
Thus we infer that this model does not seem to be compatible with TeV scale SUSY.

\item FI in the geometric regime with the inflaton wrapped by the visible sector:
for $\vo\simeq 10^{14}$, and $c\simeq 5\cdot 10^{-2}$ we would obtain $T_{RH}^{max}\simeq 10^{-8}$ eV,
which is unrealistically small. Thus we infer that this model does not seem to be compatible with TeV scale SUSY.
On top of this, we would obtain an
unrealistically small value for the visible sector gauge coupling:
$\alpha^{-1}_{vis}=\langle\tau_1\rangle=c\,\vo^{2/3}\simeq 10^8$.

\item FI at the quiver locus :
for $\vo\simeq 10^6$ (as needed to obtain TeV scale SUSY at the quiver locus),
we would obtain $T_{RH}^{max}\simeq 1$ MeV which is of the same order of magnitude as $T_{BBN}$.
Thus we infer that this model could be made compatible with TeV scale SUSY only if $T_{RH}^{max}$ turns out to be
actually a bit higher than $T_{BBN}$ and the matter-antimatter asymmetry is generated in a non-thermal way.
\end{enumerate}

Thus, even though FI does not require any fine-tuning of the coefficient of the string loop corrections,
it seems very difficult to obtain a viable reheating for these models which is also compatible with TeV scale
SUSY. On the contrary, both BI at the quiver locus and in the geometric regime with the inflaton
wrapped by the visible sector, seems to be more promising to achieve inflation,
reheating and TeV scale SUSY at the same time.

We finally stress that our estimate of the reheating temperature was based on the approximation of sudden thermalisation
of the visible sector \textit{dof}. In reality it is more likely that the system slowly evolves
to thermal equilibrium with an actual reheating temperature $T_{RH}< T_{RH}^{max}$. However we shall leave the detailed study of the
thermalisation process for future investigation.

\section*{Acknowledgments}

We would like to thank Neil Barnaby, Wildfried Buchmueller, Cliff Burgess, Joe Conlon, Manuel Drees, Mark Goodsell,
Thomas Grimm, Tetsutaro Higaki, Sven Krippendorf, Oleg Lebedev, Fernando Quevedo, Gianmassimo Tasinato for helpful discussions and
explanations. We are also grateful to Fernando Quevedo and Cliff Burgess for commenting on the final draft of this paper.
The research of AM is partly supported by the
European Union through Marie Curie Research and Training Network
``UNIVERSENET'' (MRTN-CT-2006-035863).

\bigskip

\appendix

\section{Calculation of the moduli couplings}
\label{appendice}

In this appendix we shall derive for each case studied, the canonical normalisation
which is then needed to derive the moduli mass spectrum and their couplings to visible
and hidden sector \textit{dof}.

\subsection{Blow-up Inflation}

Without loss of generality but just in order to simplify our calculations, we shall
focus on compactifications on Calabi-Yau orientifolds
with three instead of four K\"{a}hler moduli. The overall volume looks like:
\be
\vo =\alpha\left(\tau_1^{3/2}-\gamma_2\tau_2^{3/2}-\gamma_3\tau_3^{3/2}\right).
\ee

\subsubsection{Geometric regime}
\label{ModCoupGR}

Let us first of all start with the canonical normalisation and then compute the moduli couplings to ordinary and
supersymmetric particles.
\newpage
\textbf{Canonical normalisation}
\newline
\newline
After axion minimisation, the scalar potential takes the form:
\begin{equation}
V=\frac{\lambda_2\sqrt{\tau_2}
e^{-2 a_2\tau_2}}{\mathcal{V}} +\frac{\lambda_3\sqrt{\tau_3}
e^{-2 a_3\tau_3}}{\mathcal{V}} -\frac{\mu_2\tau_2
e^{-a_2\tau_2}}{\mathcal{V}^{2}}-\frac{\mu_3\tau_3
e^{-a_3\tau_3}}{\mathcal{V}^2}
+\frac{\nu}{\mathcal{V}^{3}}, \label{V}
\end{equation}
where $\lambda_i=8 \, a_i^2 A_i^2/\left(3\alpha\gamma_i\right)\,$ $\forall i=2,3$,
$\mu_j=4 W_0 a_j A_j\,$ $\forall j=2,3$, and $\nu=3 \xi W_0^2/(4 g_s^{3/2})$.
The global minimum is located at:
\be
a_i \langle \tau_i \rangle= \left(\frac{4\nu}{J} \right)^{2/3},\hskip0.7cm \langle
 \mathcal{V}\rangle = \left(\frac{ \mu_i }{2\lambda_i}\right)
 \sqrt{\langle \tau_i\rangle }
 \; e^{a_i \langle \tau_i
 \rangle },\text{ \ \ \ }\forall\, i=2,3,\label{minimo}
\ee
where:
\be
J\equiv \sum_{i=2}^3\left(\frac{\mu_i^2}{\lambda_i a_i^{3/2}}\right).
\ee
Let us now work out the canonical normalisation around the minimum, starting expanding
each modulus around its VEV:
\be
\tau_i=\langle\tau_i\rangle+\delta \tau_i, \text{ \ \ }\forall i=1,2,3.
\ee
Then the Lagrangian takes the form:
\be
\mathcal{L}=K_{ij}\partial_{\mu}\left(\delta\tau_i\right)\partial^{\mu}\left(\delta\tau_j\right)
-\langle V \rangle -\frac{1}{2}V_{ij}\delta\tau_i\delta\tau_j+\mathcal{O}(\delta\tau^3),
\label{Lagrangian}
\ee
where the direct K\"{a}hler metric at leading order looks like:
\begin{equation}
\mathcal{K}=\frac{9}{8\langle\tau_1\rangle^2}\left(
\begin{array}{ccc}
2/3 & -\gamma _{2}\epsilon _{2} & -\gamma _{3}\epsilon _{3} \\
-\gamma _{2}\epsilon _{2} & \gamma _{2}\epsilon _{2}^{-1}/3 & \gamma
_{2}\gamma _{3}\epsilon _{2}\epsilon _{3} \\
-\gamma _{3}\epsilon _{3} & \gamma _{2}\gamma _{3}\epsilon _{2}\epsilon
_{3} & \gamma _{3}\epsilon_3^{-1}/3
\end{array}
\right) ,  \label{LaDiretta}
\end{equation}
with $\epsilon_i\equiv\sqrt{\langle\tau_i\rangle/\langle\tau_1\rangle}\ll 1\,$ $\forall i=2,3$.
Writing the original moduli $\delta\tau_i$ in terms of the canonically normalised fields around the
minimum $\delta\phi_i$ as:
\begin{equation}
\left(
\begin{array}{c}
\delta \tau _{1} \\
\delta \tau _{2} \\
\delta \tau _{3}
\end{array}
\right) =\left(
\begin{array}{c}
\\
\vec{v}_1 \\ \,
\end{array}
\right) \frac{\delta \phi _{1}}{\sqrt{2}}+\left(
\begin{array}{c}
\\
\vec{v}_2 \\ \,
\end{array}
\right) \frac{\delta \phi _{2}}{\sqrt{2}}+\left(
\begin{array}{c}
\\
\vec{v}_3 \\ \,
\end{array}
\right) \frac{\delta \phi _{3}}{\sqrt{2}},
\label{vettori}
\end{equation}
the Lagrangian (\ref{Lagrangian}) takes the canonical form:
\be
\mathcal{L}=\frac{1}{2}\sum_{i=1}^{3}\partial_{\mu}\left(\delta\phi_i\right)
\partial^{\mu}\left(\delta\phi_i\right)-\langle V \rangle - \frac{m_i^2}{2}\sum_{i=1}^{3}\delta\phi_i^2,
\ee
only if $\vec{v_i}$ and $m_i^2$ are, respectively, the eigenvectors and
the eigenvalues of the mass-squared matrix $\left(M^2\right)_{ij}\equiv \frac{1}{2}\left(K^{-1}\right)_{ik}V_{kj}$.
On top of this, the eigenvectors have to be normalised as $\vec{v_i}^T\cdot \mathcal{K}\cdot \vec{v_j}=\delta_{ij}$.

The inverse K\"{a}hler metric at leading order is given by:
\begin{equation}
\mathcal{K}^{-1}=4\langle\tau_1\rangle^2\left(
\begin{array}{ccc}
1/3 & \epsilon_2^2 & \epsilon_3^2 \\
\epsilon_2^2 & 2\epsilon_2/(3\gamma_2) &
\epsilon_2^2\epsilon_3^2 \\
\epsilon_3^2 & \epsilon_2^2\epsilon_3^2
& 2\epsilon_3/(3\gamma_3)
\end{array}
\right) ,  \label{LaInversa}
\end{equation}
whereas the Hessian of the scalar potential evaluated at the global minimum (\ref{minimo}), at leading order, looks like:
\begin{equation}
V_{ij}=\frac{\alpha^{-3}}{\langle\tau_1\rangle^{13/2}}\left(
\begin{array}{ccc}
c_1-c_2\langle\tau_2\rangle^{3/2}-c_3\langle\tau_3\rangle^{3/2} &
- 4 a_2 c_2\langle\tau_1\rangle\langle\tau_2\rangle^{3/2}/27 &
-4 a_3 c_3\langle\tau_1\rangle\langle\tau_3\rangle^{3/2}/27 \\
-4 a_2 c_2\langle\tau_1\rangle\langle\tau_2\rangle^{3/2}/27 &
8 a_2^2 c_2\langle\tau_1\rangle^2\langle\tau_2\rangle^{3/2}/81 & 0 \\
-4 a_3 c_3\langle\tau_1\rangle\langle\tau_3\rangle^{3/2}/27 & 0
& 8 a_3^2 c_3\langle\tau_1\rangle^2\langle\tau_3\rangle^{3/2}/81
\end{array}
\right) ,  \label{Vij}
\end{equation}
where:
\be
c_1 \equiv \frac{99\nu }{4},\text{ \ \ \ }c_2\equiv \frac{81\mu_2^2}{16\lambda_2},
\text{ \ \ \ }c_3\equiv \frac{81\mu_3^2}{16\lambda_3}.
\ee
The mass-squared matrix can be obtained multiplying (\ref{LaInversa}) by (\ref{Vij}). The leading order
result is (setting without loss of generality $\gamma_2=\gamma_3=A_2=A_3=W_0=1$):
\begin{equation}
\mathcal{M}^2=\frac{\alpha^{-3}}{\langle\tau_1\rangle^{9/2}}\left(
\begin{array}{ccc}
-9\left(a_2\langle\tau_2\rangle^{5/2}+a_3\langle\tau_3\rangle^{5/2}\right)(1-7\delta) &
6 a_2^2 \langle\tau_1\rangle\langle\tau_2\rangle^{5/2}(1-5\delta) &
6 a_3^2\langle\tau_1\rangle\langle\tau_3\rangle^{5/2}(1-5\delta) \\
-6 a_2 \sqrt{\langle\tau_1\rangle}\langle\tau_2\rangle^2(1-5\delta) &
4 a_2^2 \langle\tau_1\rangle^{3/2}\langle\tau_2\rangle^2(1-3\delta) &
6 a_3^2 \langle\tau_2\rangle \langle\tau_3\rangle^{5/2} \\
-6 a_3 \sqrt{\langle\tau_1\rangle}\langle\tau_3\rangle^2(1-5\delta) & 6 a_2^2 \langle\tau_3\rangle \langle\tau_2\rangle^{5/2}
& 4 a_3^2 \langle\tau_1\rangle^{3/2}\langle\tau_3\rangle^2(1-3\delta)
\end{array}
\right)  \notag
\end{equation}
where $\delta\equiv 1/(4 a_2 \langle\tau_2\rangle)=1/(4 a_3 \langle\tau_3\rangle)\simeq 1/(4\ln\vo)\ll 1$.
The two small blow-up modes $\tau_2$ and $\tau_3$ behave in the same way and so they will have the same mass
and will be heavier that the large overall volume mode $\tau_1$: $m_2\sim m_3 \gg m_1$.
Therefore we can work out the leading order volume scaling of the moduli mass spectrum noticing that:
\begin{eqnarray}
Tr[\mathcal{M}^{2}] &=&m_1^2+m_2^2+m_3^2\sim m_2^2\sim
m_3^2\sim \frac{a_2^2\langle\tau_2\rangle^2}{\alpha^3\langle\tau_1\rangle^3}\sim \frac{
a_3^2\langle\tau_3\rangle^2}{\alpha^3\langle\tau_1\rangle^3}\sim \left( \frac{\ln \mathcal{V}}{
\mathcal{V}}\right)^2 M_P^2,
\label{masse} \\
\frac{Det[\mathcal{M}^{2}]}{Tr[\mathcal{M}^2]^2} &\sim &\frac{
m_1^2 m_2^2 m_3^2}{m_2^2 m_3^2}\sim m_{1}^{2}\sim \frac{
(\langle\tau_2\rangle^{3/2}+\langle\tau_3\rangle^{3/2})}{a_2\langle\tau_2\rangle\alpha^3\langle\tau_1\rangle^{9/2}}
\sim \frac{(\langle\tau_2\rangle^{3/2}+\langle\tau_3\rangle^{3/2})}{a_3\langle\tau_3\rangle\alpha^3\langle\tau_1\rangle^{9/2}}
\sim \frac{M_P^2}{\mathcal{V}^3\ln \mathcal{V}}.
\label{MASSE}
\end{eqnarray}
Let us now derive the corresponding eigenvectors. For the eigenvalue $m_1^2$ we have
(for $\vec{v}_1=(x_1,y_1,z_1)$):
\begin{equation}
\mathcal{M}^{2}\vec{v}_{1}=m_{1}^{2}\vec{v}_{1}\,\Leftrightarrow\, \left\{
\begin{array}{c}
x_1\simeq a_s \langle\tau_1\rangle (y_1+z_1) \\
y_1=z_1 \\
z_1
\end{array},
\right.
\end{equation}
where without loss of generality we have set $a_2=a_3=a_s$ and $\langle\tau_2\rangle=\langle\tau_3\rangle=\langle\tau_s\rangle$.
On the other hand, for the eigenvalue $m_2^2$ we get
(for $\vec{v}_2=(x_2,y_2,z_2)$):
\begin{equation}
\mathcal{M}^{2}\vec{v}_2=m_2^2\vec{v}_2\,\Leftrightarrow\, \left\{
\begin{array}{c}
x_2\simeq \langle\tau_s\rangle^{1/2}\langle\tau_1\rangle^{-1/2} y_2 \\
y_2 \\
z_2\simeq \langle\tau_s\rangle^{3/2}\langle\tau_1\rangle^{-3/2} y_2\ll y_2
\end{array}.
\right.
\end{equation}
Finally for the eigenvalue $m_3^2$ we find
(for $\vec{v}_3=(x_3,y_3,z_3)$):
\begin{equation}
\mathcal{M}^{2}\vec{v}_3=m_3^2\vec{v}_3\,\Leftrightarrow\, \left\{
\begin{array}{c}
x_3\simeq \langle\tau_s\rangle^{1/2}\langle\tau_1\rangle^{-1/2} z_3 \\
y_3\simeq \langle\tau_s\rangle^{3/2}\langle\tau_1\rangle^{-3/2} z_3 \ll z_3 \\
z_3
\end{array}.
\right.
\end{equation}
The remaining components which have not been fixed yet, can be worked out
by properly normalising the eigenvectors as:
\begin{equation}
\left\{
\begin{array}{c}
\vec{v}_{1}\cdot \mathcal{K}\cdot \vec{v}_{1}=1\,\Leftrightarrow
\,z_{1}\simeq a_{s}^{-1}, \\
\vec{v}_{2}\cdot \mathcal{K}\cdot \vec{v}_{2}=1\,\Leftrightarrow
\,y_{2}\simeq \langle\tau _{s}\rangle^{1/4}\langle\tau _{1}\rangle^{3/4}, \\
\vec{v}_{3}\cdot \mathcal{K}\cdot \vec{v}_{3}=1\,\Leftrightarrow
\,z_{3}\simeq \langle\tau _{s}\rangle^{1/4}\langle\tau _{1}\rangle^{3/4}.
\end{array}
\right.
\end{equation}
Therefore the general form (\ref{vettori}) for the canonical normalisation takes the form:
\begin{equation}
\left(
\begin{array}{c}
\delta \tau _{1} \\
\delta \tau _{2} \\
\delta \tau _{3}
\end{array}
\right) =\left(
\begin{array}{c}
\langle\tau_1\rangle \\
a_s^{-1} \\
a_s^{-1} \,
\end{array}
\right) \frac{\delta \phi _{1}}{\sqrt{2}}+\left(
\begin{array}{c}
\langle\tau_1\rangle^{1/4}\langle\tau_s\rangle^{3/4}  \\
\langle\tau_1\rangle^{3/4}\langle\tau_s\rangle^{1/4} \\
\langle\tau_1\rangle^{-3/4}\langle\tau_s\rangle^{7/4} \,
\end{array}
\right) \frac{\delta \phi _{2}}{\sqrt{2}}+\left(
\begin{array}{c}
\langle\tau_1\rangle^{1/4}\langle\tau_s\rangle^{3/4} \\
\langle\tau_1\rangle^{-3/4}\langle\tau_s\rangle^{7/4} \\
\langle\tau_1\rangle^{3/4}\langle\tau_s\rangle^{1/4} \,
\end{array}
\right) \frac{\delta \phi _{3}}{\sqrt{2}},
\label{vettoriFinali}
\end{equation}
which in terms of factors of the overall volume scales as:
\begin{eqnarray}
\frac{\delta \tau_1}{\langle\tau_1\rangle} &\simeq&\mathcal{O}(1)\,\delta \phi_1
+\mathcal{O}(\epsilon)\,\delta \phi_2+\mathcal{O}(\epsilon)\,\delta \phi_3\simeq
\mathcal{O}(1)\,\delta \phi_1, \label{C1} \\
\frac{\delta \tau_2}{\langle\tau_2\rangle} &\simeq&\mathcal{O}\left(\frac{1}{\ln\vo}\right)\delta \phi_1
+\mathcal{O}\left(\frac{1}{\epsilon}\right)\delta \phi_2+\mathcal{O}(\epsilon)\delta \phi_3
\simeq \mathcal{O}\left(\frac{1}{\epsilon}\right)\delta \phi_2, \label{C2} \\
\frac{\delta \tau_3}{\langle\tau_3\rangle} &\simeq&\mathcal{O}\left(\frac{1}{\ln\vo}\right)\delta \phi_1
+\mathcal{O}(\epsilon)\delta \phi_2+\mathcal{O}\left(\frac{1}{\epsilon}\right)\delta \phi_3
\simeq \mathcal{O}\left(\frac{1}{\epsilon}\right)\delta \phi_3, \label{C3}
\end{eqnarray}
where\footnote{For $\tau_s\sim g_s^{-1}\sim \mathcal{O}(10)$ and
$\tau_1\sim (\vo/\alpha)^{2/3}\sim (10\vo)^{2/3}$ since for an arbitrary Calabi-Yau
$\alpha\sim \mathcal{O}(1/10)$.}:
\be
\epsilon\equiv \left(\frac{\langle\tau_s\rangle}{\langle\tau_1\rangle}\right)^{3/4}\simeq \frac{10^{1/4}}{\vo^{1/2}}\ll 1.
\ee
As we have already pointed out in the main part of the text, this result has a nice and clear
geometric understanding. In fact, from (\ref{C1}), we see that the overall volume mode is mostly given by $\delta\phi_1$
and then it mixes at subleading order with $\delta\phi_2$ and $\delta\phi_3$ in the same way since the
volume `sees' the two blow-up modes in the same way. On the contrary, from (\ref{C2}) and (\ref{C3}), we realise that
each blow-up mode is mostly given by $\delta\phi_2$, or $\delta\phi_3$ respectively, then it mixes with the
overall volume, and lastly there is an even more suppressed mixing with the other blow-up mode, reflecting the
geometric separation between the two blow-up modes which are resolving two singularities located in different
regions of the Calabi-Yau three-fold.
\newline
\newline
\textbf{Moduli couplings to visible sector particles}
\newline
\newline
Assuming that the visible MSSM-like sector is built via magnetised
$D7$-branes wrapping $\tau_3$,\footnote{We consider the visible sector localised
on a blow-up cycle which is fixed non-perturbatively even though this is not possible,
since this does not affect the derivation of the volume scaling of the moduli canonical normalisation.}
the 4D effective field theory is
completely specified by expanding $W$, $K$ and the gauge kinetic functions $f_i$ as
power series in the matter fields \cite{GeomRegime}:
\begin{eqnarray}
W &=&W_{mod}(\varphi )+\mu (\varphi
)H_{u}H_{d}+\frac{Y_{ijk}(\varphi
)}{6}C^{i}C^{j}C^{k}+..., \label{Wmodvis} \\
K &=&K_{mod}(\varphi ,\bar{\varphi} )+\tilde{K}_{i\bar{j}}(\varphi
,\bar{\varphi} )C^{i}C^{\bar{j}}+
\left[ Z(\varphi ,\bar{\varphi})H_{u}H_{d}+h.c.\right] +..., \label{Kmodvis} \\
f_{i} &=&\frac{T_3}{4\pi }+h_i(F)S, \label{gaugeKinfunc}
\end{eqnarray}
where $\varphi$ and the $C$'s denote respectively all the
moduli and matter fields. In addition $h_i(F)$ are 1-loop
functions of the world-volume
fluxes $F$ on different branes (the index $i$ runs over the three
MSSM gauge group factors). Finally the moduli scaling of the
K\"{a}hler potentials for matter fields
$\tilde{K}_{i\bar{j}}(\varphi ,\bar{\varphi})$ and $Z(\varphi
,\bar{\varphi})$, reads \cite{cremades}:
\begin{equation}
\tilde{K}_{i\bar{j}}(\varphi
,\bar{\varphi})\sim\frac{\tau_3^{1/3}}{\tau_1}k_{i\bar{j}}(U)\text{
\ \ and \ \ }Z(\varphi
,\bar{\varphi})\sim\frac{\tau_3^{1/3}}{\tau_1}z(U).
\label{chiralMetric}
\end{equation}
We notice that, due to the axionic shift symmetry of the imaginary part of the K\"{a}hler moduli,
$T=\tau+ i b$ with $b\to b +\epsilon$, the superpotential cannot depend on $b$. In addition,
the holomorphicity of $W$, forbids also any dependence on $\tau= (T+\bar{T})/2$.
Thus the superpotential $W$ can depend only on the complex structure moduli and the dilaton,
but not on the K\"{a}hler moduli. This implies that there cannot be any dimension 4 operator
involving K\"{a}hler moduli and ordinary MSSM particles. On the contrary, the K\"{a}hler moduli
will couple only via higher order operators once we consider the normalised superpotential
$e^{K/2}W$ and expand the K\"{a}hler potential around the moduli VEVs.

The moduli couplings to ordinary and supersymmetric particles at high temperatures
above the EW symmetry breaking scale, $T>M_{EW}$, where all the gauge bosons and matter fermions are still massless
and no mixing of Higgsinos with gauginos into
charginos and neutralinos has taken place yet, have already been derived in \cite{astro, LVSatFiniteT} for the case
of Swiss-cheese Calabi-Yau manifolds with just one small blow-up cycle. We shall now generalise those
results for the case of more than one blow-up mode.

\bigskip

$\bullet \textbf{ Couplings to visible gauge bosons}$

\bigskip

The moduli couplings to the gauge bosons $g$ arise from the
moduli dependence of the gauge kinetic function
(\ref{gaugeKinfunc}). Given that different values of MSSM gauge groups
are obtained by turning on different world-volume fluxes at 1-loop,
the coupling of $\tau_3$ to the gauge bosons is the same for $U(1)$, $SU(2)$ and $SU(3)$.
The relevant dimension 5 operator turns out to be:
\begin{equation}
\mathcal{L}\supset \frac{\delta\tau_3}
{4\langle\tau_3\rangle M_P}G_{\mu\nu}G^{\mu\nu},
\label{FmunuFmunu}
\end{equation}
where $G_{\mu\nu}$ is the canonically normalised field strength.
Now using the canonical normalisation (\ref{C3}) we end up with the following couplings:
\begin{gather}
\mathcal{L}_{\delta \phi_1 gg }\simeq\mathcal{O}\left( \frac{1}{4 \ln\vo}
\right) \frac{\delta \phi_1}{M_P}G_{\mu \nu }G^{\mu \nu },\text{ \ \ \ \ \ \ \ }
\mathcal{L}_{\delta \phi _{2}gg }\simeq \mathcal{O}\left( \frac{10^{1/4}}{4 \sqrt{
\vo}}\right) \frac{\delta\phi_2}{M_P}G_{\mu \nu }G^{\mu \nu }, \notag \\
\mathcal{L}_{\delta \phi _{3}gg }\simeq \mathcal{O}\left( \frac{\sqrt{\mathcal{
V}}}{4 \cdot 10^{1/4}}\right) \frac{\delta\phi_3}{M_P}G_{\mu \nu }G^{\mu \nu }.
\label{ImpModuliCoupl}
\end{gather}

$\bullet \textbf{ Couplings to visible matter fermions}$

\bigskip

The moduli couplings to an ordinary matter fermion $\psi$ can be worked out starting from
its moduli-dependent kinetic and mass terms, then expanding the moduli around their VEVs,
canonically normalising the $\psi$ kinetic terms, and finally substituting the canonical normalisation
for the moduli. In terms of the canonically unnormalised moduli, the relevant operator is:
\begin{equation}
\mathcal{L}\supset \left(\frac{1}{2 \langle\tau_1\rangle}\frac{\delta \tau_1}{M_P}
+\frac{1}{3 \langle\tau_3\rangle}\frac{\delta \tau_3}{M_P}
\right) \lambda_c \langle H_c \rangle \overline{\psi}_c\psi_c, \label{secc}
\end{equation}
where the physical Yukawa coupling is given by: $\lambda_c = \lambda (U) \langle\tau_3\rangle^{-1/2}$.
Notice that for $T>M_{EW}$ there is no direct modulus-fermion-fermion coupling since
the Higgs VEV is still located at zero: $\langle H_c\rangle=0$. Therefore we have to consider the
4-particle vertex given by:
\be
\mathcal{L}_{int}\sim \left(\frac{1}{\langle\tau_1\rangle}\frac{\delta \tau_1}{M_P}
-\frac{1}{3 \langle\tau_3\rangle}\frac{\delta \tau_3}{M_P}
\right) q_c \overline{\psi}_c\gamma^{\mu}A_{\mu}^c\psi_c,
\ee
where $q_c\equiv q /(2\langle\tau_3\rangle^{1/2})$. The final moduli couplings look like:
\begin{gather}
\mathcal{L}_{\delta \phi _{1}g \overline{\psi}\psi }\sim \mathcal{O}\left(1\right)
\frac{\delta \phi_1}{M_P}\overline{\psi}_{c}\gamma ^{\mu }A_{\mu }^{c}\psi
_{c},\text{ \ \ \ \ }\mathcal{L}_{\delta \phi _{2}g\overline{\psi}\psi
}\sim\mathcal{O} \left( \frac{2\cdot 10^{1/4}}{3 \sqrt{\vo}}\right) \frac{\delta \phi _{2}}{M_P}\overline{\psi}
_{c}\gamma ^{\mu }A_{\mu }^{c}\psi _{c}, \\
\mathcal{L}_{\delta \phi_3 g \overline{\psi}\psi }\sim \mathcal{O}\left( \frac{\sqrt{
\mathcal{V}}}{3\cdot 10^{1/4}}\right) \frac{\delta \phi _{3}}{M_P}\overline{\psi}_{c}\gamma ^{\mu
}A_{\mu }^c\psi_c.
\end{gather}
\newpage
$\bullet \textbf{ Couplings to visible Higgs fields}$

\bigskip

The form of moduli couplings to the canonically normalised Higgs field $H_c$ (either $H_{up}$ or $H_{down}$), can similarly be
worked out from the expansion of the Higgs kinetic term. The relevant dimension 5 operator
looks like\footnote{Another coupling can be derived from the expansion of the
SUSY breaking contribution to the Higgs mass given by $m_H \sim |F^3|/\tau_3 \sim M_P/\vo$ \cite{GeomRegime}. However this coupling gives rise to a
subleading contribution to the moduli decay rates and so we shall ignore it.}:
\be
\mathcal{L}\supset \left(\frac{\delta\tau_1}{\langle\tau_1\rangle M_P}-\frac{\delta\tau_3}{3\langle\tau_3\rangle M_P} \right)
\partial_{\mu}\overline{H}_c \partial^{\mu}H_c.
\label{uu}
\ee
Thus substituting in (\ref{uu}) the expressions (\ref{C1}) and (\ref{C3}), the moduli couplings to Higgs fields take the form:
\begin{gather}
\mathcal{L}_{\delta \phi_1 \overline{H}_c H_c }\sim \mathcal{O}\left(1\right)
\frac{\delta \phi _{1}}{M_P}\partial_{\mu}\overline{H}_c \partial^{\mu}H_c,
\text{ \ \ }\mathcal{L}_{\delta \phi _{2}\overline{H}_c H_c
}\sim\mathcal{O} \left( \frac{2\cdot 10^{1/4}}{3 \sqrt{\vo}}\right)
\frac{\delta \phi _{2}}{M_P}\partial_{\mu}\overline{H}_c \partial^{\mu}H_c,
\label{Higgs1} \\
\mathcal{L}_{\delta \phi_3\overline{H}_c H_c}\sim \mathcal{O}\left( \frac{\sqrt{
\mathcal{V}}}{3\cdot 10^{1/4}}\right) \frac{\delta \phi _{3}}{M_P}\partial_{\mu}\overline{H}_c \partial^{\mu}H_c.
\label{Higgs2}
\end{gather}

$\bullet \textbf{ Couplings to visible gauginos}$

\bigskip

The moduli couplings to gauginos can again be derived by focusing on their moduli-dependent kinetic
and soft mass terms. One first expands the moduli around their VEVs, then canonically normalises the
gauginos and finally substitutes the expressions for the canonically normalised moduli fields.
We recall that for a field theory living on a brane wrapping a generic small cycle $\tau_s$,
the gaugino mass is proportional to the non-vanishing $F$-term of $\tau_s$ defined as $F^s=e^{K/2}K^{s\bar{i}}D_{\bar{i}} W$ \cite{Brignole}.
It is important then to realise that this $F$-term can take the form:
\begin{enumerate}
\item $F^s\simeq -2 W_0\frac{\tau_s}{\vo}\,\Rightarrow\,$ at the minimum
$|\langle F^s\rangle|\simeq \frac{W_0}{\langle\vo\rangle}\langle\tau_s\rangle$ if $\tau_s$ is stabilised by $D$-terms or perturbative effects \cite{GenAnalofLVS};

\item $F^s= -2 W_0 \frac{\tau_s}{\vo} + \left(\frac{8 a_s A_s}{3\alpha\gamma_s}\right) \sqrt{\tau_s}\,
 e^{-a_s \tau_s}\,\Rightarrow\,$ at the minimum $|\langle F^s\rangle|\simeq \frac{W_0}{\langle \vo \rangle}$ if $\tau_s$ is stabilised non-perturbatively \cite{GeomRegime}.
\end{enumerate}
The first case holds for the cycle $\tau_3$ supporting the visible sector while we will need to consider
the second case for the cycle $\tau_2$ supporting the hidden sector. Hence we realise that $F^2$ and $F^3$
have the same volume scaling but a different dependence on the small cycle. This difference becomes crucial
once the moduli are expanded around their VEVs in order to derive the moduli couplings,
leading to a different volume scaling in the two cases. In the case of the visible sector,
we need to start from the Lagrangian:
\be
\mathcal{L}= 4i\tau_3\lambda^{\dagger} \bar{\sigma}^{\mu}\partial_{\mu}\lambda+|F^3|
\left(\lambda\lambda+h.c.\right),
\label{erfo}
\ee
where $F^3$ takes the same form as in the first case above, and then expand the moduli around their VEVs.
In terms of the canonically normalised fields $\lambda_c = 2\langle\tau_3\rangle^{1/2} \lambda$, (\ref{erfo}) becomes:
\be
\mathcal{L}\sim \left(1+\frac{\delta\tau_3}{\langle\tau_3\rangle M_P}\right)
\left[\lambda^{\dagger}_c i \bar{\sigma}^{\mu}\partial_{\mu}\lambda_c
+\frac{M_{1/2}^{hid}}{2}\lambda_c\lambda_c\right]
-\frac{3}{4}\frac{\delta\tau_1}{\langle\tau_1\rangle}\frac{M_{1/2}}{M_P}
\left(\lambda_c\lambda_c+h.c.\right).
\label{erfgo}
\ee
with $M_{1/2}=|\langle F^3\rangle|/(2\langle\tau_3\rangle)\simeq M_P/\vo $.
Only the last term in (\ref{erfgo}) contributes to the moduli decay rates due to the equations of motion.
Hence by means of (\ref{C1}), we finally obtain the following couplings:
\begin{eqnarray}
\mathcal{L}_{\delta\phi_1 \lambda\lambda } &\sim &\mathcal{O}\left(
\frac{3}{4}\right) \frac{M_{1/2}}{M_P} \,\delta\phi_1 \lambda\lambda \sim \mathcal{O}\left(
\frac{3}{4\vo}\right) \delta\phi_1 \lambda\lambda,
\label{rr} \\
\mathcal{L}_{\delta\phi_i \lambda\lambda } &\sim &\mathcal{O}\left(
\frac{3\cdot 10^{1/4}}{4 \sqrt{\vo}}\right) \frac{M_{1/2}}{M_P}\, \delta\phi_i
\lambda\lambda \sim \mathcal{O}\left( \frac{3\cdot 10^{1/4}}{4 \vo^{3/2}}\right) \delta\phi_i \lambda\lambda ,\,\,\forall i=2,3.
\label{rrr}
\end{eqnarray}
It is interesting to notice that, contrary to the previous cases, there is no difference in the volume scaling
of the coupling of $\delta\phi_2$ and $\delta\phi_3$ to $\lambda\lambda$. Thus the moduli couplings to gauginos do not
reflect the geometric localisation of the visible sector on $\tau_3$ instead of $\tau_2$. The reason is that the moduli
couple to gauginos only via the gaugino mass term, $M_{1/2}\simeq M_P/\vo$, which depends only on the overall volume
without reflecting the locality of the visible sector $D7$-branes. However this locality becomes manifest if we look
at the coupling of $\delta\phi_2$ to $\lambda\lambda$ plus another particle, like a gauge boson $g$ or the light modulus $\delta\phi_1$:
\be
\mathcal{L}_{\delta \phi_3 g \lambda \lambda }\sim \mathcal{O}\left( \frac{\sqrt{
\vo}}{10^{1/4}}\right) \frac{\delta \phi _3}{M_P}\lambda_c^{\dagger}\overline{\sigma}_{\mu}A^{\mu}\lambda_c,\text{ \ \ \ }
\mathcal{L}_{\delta \phi_3\delta\phi_1\lambda \lambda }\sim \mathcal{O}\left( \frac{\sqrt{
\vo}}{10^{1/4}}\right) \frac{\delta \phi _3}{M_P}\delta\phi_1 \lambda_c\lambda_c.
\label{3bodyGaugino}
\ee
These couplings will give rise to 3-body decay rates which turn out to be larger than the 2-body decay rate coming from the coupling (\ref{rrr}).
\bigskip

$\bullet \textbf{ Couplings to visible SUSY scalars}$

\bigskip

The moduli couplings to squarks and sleptons collectively denoted as $\varphi$, are
worked out from the expansion of the their kinetic and mass terms exactly as we did for the moduli
couplings to the Higgs fields. Thus the results look like (\ref{Higgs1}) and (\ref{Higgs2}).
\bigskip

$\bullet \textbf{ Couplings to visible Higgsinos}$

\bigskip

The moduli couplings to Higgsinos can be derived in a way similar to the previous cases
obtaining the following relevant part of the supergravity Lagrangian:
\be
\mathcal{L} \supset \frac{3}{2} \frac{\delta \tau_1}{\langle \tau_1\rangle M_P}\,
m_{ \tilde{H}}\left(\tilde{H}_c\tilde{H}_c+h.c.\right),
\label{secco}
\ee
with the physical Higgsino mass of the same order of the soft SUSY masses:
$m_{\tilde{H}}\simeq M_{1/2}/3\simeq M_P/(3\vo)$ \cite{GeomRegime}.
Now substituting in (\ref{secco}) the expression (\ref{C1}) for the canonical normalisation of the
large modulus $\tau_1$, we obtain the following couplings:
\begin{eqnarray}
\mathcal{L}_{\delta\phi_1\tilde{H}_c\tilde{H}_c}&\sim&\mathcal{O}\left(\frac{3}{2}\right)\frac{m_{\tilde{H}}}{M_P}\,
\delta\phi_1\tilde{H}_c\tilde{H}_c
\sim\mathcal{O}\left(\frac{1}{2\mathcal{V}}\right)
\delta\phi_1\tilde{H}_c\tilde{H}_c, \\
\mathcal{L}_{\delta\phi_i\tilde{H}_c\tilde{H}_c}&\sim&\mathcal{O}\left(\frac{3\cdot 10^{1/4}}{2\sqrt{\vo}}\right)
\frac{m_{\tilde{H}}}{M_P}\,\delta\phi_i\tilde{H}_c\tilde{H}_c\sim\mathcal{O}\left(\frac{10^{1/4}}{2\vo^{3/2}}\right)
\delta\phi_i\tilde{H}_c\tilde{H}_c,\,\,\forall i=2,3.
\end{eqnarray}
Similarly to the case of the moduli couplings to gauginos, we find again that the moduli couplings to Higssinos
do not reflect the locality of the visible sector, which, on the other hand, becomes manifest once we consider
4-particles interactions like the ones in (\ref{3bodyGaugino}).
\bigskip

$\bullet \textbf{ Moduli self couplings}$

\bigskip

The moduli self interactions can be
obtained expanding the potential around the minimum as:
\begin{equation}
V=V(\langle \tau_i\rangle)+\frac{1}{2}\left. \frac{\partial ^{2}V}{\partial \tau
_{i}\partial \tau _{j}}\right\vert _{\min }\delta \tau _{i}\delta
\tau _{j}+\frac{1}{3!}\left. \frac{\partial ^{3}V}{\partial \tau
_{i}\partial \tau _{j}\partial \tau _{k}}\right\vert _{\min
}\delta \tau _{i}\delta \tau _{j}\delta \tau _{k}+....
\label{expandi}
\end{equation}
and then focusing on the trilinear terms. After substituting the
expressions (\ref{C1})-(\ref{C3}) for the moduli canonical normalisation, we
end up with ($\forall \,i,j,k=2,3$):
\begin{eqnarray}
\mathcal{L}_{111} &\simeq &\mathcal{O}\left(\frac{M_P}{\vo^3}\right)\delta \phi_1^3,\text{
\ \ \ \ \ \ \ \ }\mathcal{L}_{11i}\simeq \mathcal{O}\left(\frac{M_P}{\vo^{5/2}}\right)\delta \phi
_{1}^{2}\delta \phi _{i}, \\
\text{\ }\mathcal{L}_{1ij} &\simeq &\mathcal{O}\left(\frac{M_P}{\vo^{2}}\right)\delta \phi
_{1}\delta \phi _{i}\delta \phi _{j},\text{ \ \ \ \ \ \ \ \ \ }\mathcal{L}
_{ijk}\simeq \mathcal{O}\left(\frac{M_{P}}{\vo^{3/2}}\right)\delta \phi _{i}\delta \phi _{j}\delta
\phi _{k}.  \label{coupl}
\end{eqnarray}
\newline
\newline
\textbf{Moduli couplings to hidden sector particles}
\newline
\newline
We shall now assume that $\tau_2$ is wrapped by a stack of $D7$-branes
supporting a hidden $SU(N_c)$ gauge theory with $N_f<(N_c-1)$ fundamental flavours so that
this theory undergoes gaugino condensation. The total superpotential and K\"{a}hler potential
of the 4D effective field theory read:
\be
W_{tot}=W_{mod}+W_{vis}+W_{hid}\text{, \ \ \ }
K_{tot}=K_{mod}+K_{vis}+K_{hid},
\ee
where $(W_{mod}+W_{vis})$ is given by (\ref{Wmodvis}) and $(K_{mod}+K_{vis})$ by (\ref{Kmodvis}).
Denoting all the hidden matter fields as $\Phi$,
$W_{hid}$, $K_{hid}$ and the hidden sector gauge kinetic functions $f_{hid}^{(i)}$ look like \cite{cremades}:
\be
W_{hid} =\frac{Y_{ijk}^{hid}(U)}{6}\Phi^{i}\Phi^{j}\Phi^{k}+...,\text{ \ \ }
K_{hid} =\frac{\tau_2^{1/3}}{\tau_1}k_{i\bar{j}}^{hid}(U) \Phi^{i}\Phi^{\bar{j}}+...,\text{ \ \ }
f_{hid}^{(i)} =\frac{T_2}{4\pi }+h_{hid}^{(i)}(F)S. \label{HidgaugeKinfunc}
\ee
We shall now list the moduli couplings to all hidden sector \textit{dof} following the same procedure
used to derive the moduli couplings to visible sector particles.
\bigskip

$\bullet \textbf{ Couplings to hidden gauge bosons}$

\bigskip

The moduli couplings to hidden gauge bosons $X$ localised on $\tau_2$ can be worked out in a
manner similar to the couplings to visible gauge bosons with the only difference that
now the gauge kinetic function for the hidden sector
scales as $f_{hid}\sim \tau_2 + h_{hid}(F)s$. Therefore the relevant dimension 5 operators are:
\begin{gather}
\mathcal{L}_{\delta \phi _{1}XX}\sim \mathcal{O}\left( \frac{1}{4\ln\vo}\right) \frac{\delta
\phi _{1}}{M_P}G_{\mu \nu }^{hid}G_{hid}^{\mu \nu },\text{ \ \ \ \ \ \ \ \ }
\mathcal{L}_{\delta \phi _{2}XX}\sim \mathcal{O}\left( \frac{\sqrt{\vo}}{4\cdot 10^{1/4}}
\right) \frac{\delta \phi _{2}}{M_P}G_{\mu \nu }^{hid}G_{hid}^{\mu \nu }, \notag \\
\mathcal{L}_{\delta \phi _{3}XX}\sim \mathcal{O}\left( \frac{10^{1/4}}{4 \sqrt{\vo}}\right)
\frac{\delta \phi _{3}}{M_P}G_{\mu \nu }^{hid}G_{hid}^{\mu \nu }.
\label{HiddenModuliCoupl}
\end{gather}

$\bullet \textbf{ Couplings to hidden gauginos}$

\bigskip

In order to derive the moduli couplings to hidden gauginos, we need to start from the Lagrangian:
\be
\mathcal{L}= 4i\tau_2\lambda^{\dagger} \bar{\sigma}^{\mu}\partial_{\mu}\lambda+|F^2|
\left(\lambda\lambda+h.c.\right),
\label{erf}
\ee
where $F^2$ takes the same form as in the second case at page 42, and then expand the moduli around their VEVs.
Given that $F^2$ has a different scaling than $F^3$,
the moduli couplings to hidden gauginos do not have the same volume scaling as the
expressions (\ref{rr}) and (\ref{rrr}).
In fact, in terms of the canonically normalised fields
$\lambda_c = 2\langle\tau_2\rangle^{1/2} \lambda$, (\ref{erf}) becomes:
\be
\mathcal{L}\sim \left(1+\frac{\delta\tau_2}{\langle\tau_2\rangle M_P}\right)
\left[\lambda^{\dagger}_c i \bar{\sigma}^{\mu}\partial_{\mu}\lambda_c
+\frac{M_{1/2}^{hid}}{2}\lambda_c\lambda_c\right]
+\left[\mathcal{O}\left(\frac{\ln\vo}{M_P}\right)\frac{\delta\tau_1}{\langle\tau_1\rangle}
+\mathcal{O}\left(\frac{(\ln\vo)^2}{M_P}\right)\frac{\delta\tau_2}{\langle\tau_2\rangle}
\right]M_{1/2}^{hid} \lambda_c\lambda_c.
\label{erfg}
\ee
with $M_{1/2}^{hid}=|\langle F^2\rangle|/(2\langle\tau_2\rangle)\sim M_P/(\vo\ln\vo)\sim M_{1/2}^{vis}/\ln\vo$.
The moduli couplings to hidden gauginos can be read off from the last term in (\ref{erfg}):
\begin{gather}
\mathcal{L}_{int}\sim \left(\frac{M_{1/2}^{hid}}{M_{P}}\right)\left[\mathcal{O}\left(\ln\vo
\right) \delta \phi _{1}+\mathcal{O}\left( \frac{(\ln\vo)^2 \vo^{1/2}}{10^{1/4}}\right) \delta \phi _{2}+
\mathcal{O}\left( \frac{10^{1/4}(\ln\vo)^2}{\vo^{1/2}}\right) \delta \phi
_{3}\right]\lambda _{c}\lambda _{c},  \notag \\
\Longrightarrow \mathcal{L}_{int}\sim \mathcal{O}\left( \frac{1}{\vo}
\right) \delta \phi _{1}\lambda _{c}\lambda _{c}+\mathcal{O}\left( \frac{\ln\vo
}{10^{1/4}\vo^{1/2}}\right) \delta \phi _{2}\lambda _{c}\lambda _{c}+\mathcal{O}\left( \frac{
10^{1/4}\ln\vo}{\vo^{3/2}}\right) \delta \phi _{3}\lambda _{c}\lambda _{c}.
\end{gather}

$\bullet \textbf{ Couplings to hidden massless pion-like mesons}$

\bigskip
Let us compute the inflaton coupling to a massless pion-like Goldstone boson. The effective Lagrangian for the
canonically normalised field $\pi$ expanded around the moduli VEVs looks like:
\be
\mathcal{L} \sim \frac{1}{2}\frac{\partial_{\mu} \pi \partial^{\mu} \pi}{\left(1+\frac{\pi^2}{\Lambda^2}\right)^2}
\left(1+\frac{\delta\tau_2}{3\langle\tau_2\rangle M_P}-\frac{\delta\tau_1}{\langle\tau_1\rangle M_P}\right),
\ee
so that the interaction Lagrangian in terms of the canonically normalised moduli takes the form:
\be
\mathcal{L}_{int}\sim
\mathcal{O}\left(\frac{1}{2 M_P}\right)\delta\phi_1 \partial_{\mu}\pi\partial^{\mu}\pi
+\mathcal{O}\left(\frac{\vo^{1/2}}{6\cdot 10^{1/4}}\right)\frac{\delta\phi_2}{M_P}\partial_{\mu}\pi\partial^{\mu}\pi
+\mathcal{O}\left(\frac{10^{1/4}}{3\vo^{1/2}}\right)\frac{\delta\phi_3}{M_P}\partial_{\mu}\pi\partial^{\mu}\pi.
\label{Pions}
\ee

$\bullet \textbf{ Couplings to hidden fermionic condensates}$

\bigskip

The moduli couplings to hidden fermionic condensates $\psi_{hid}$ can be worked out again expanding
the kinetic and mass terms. We obtain:
\be
\mathcal{L}_{int}\sim \mathcal{O}\left( \frac{1}{2\sqrt{3}\vo}
\right) \delta \phi_1 \bar{\psi}_{hid}\psi_{hid}+\mathcal{O}\left( \frac{\ln\vo
}{\sqrt{3}\cdot 10^{1/4}\vo^{1/2}}\right) \delta \phi_2 \bar{\psi}_{hid}\psi_{hid}+\mathcal{O}\left( \frac{
10^{1/4}\ln\vo}{\sqrt{3}\vo^{3/2}}\right) \delta \phi_3 \bar{\psi}_{hid}\psi_{hid}. \notag
\ee

$\bullet \textbf{ Couplings to hidden massive scalar condensates}$

\bigskip

The moduli couplings to hidden massive scalar condensates $\varphi_{hid}$ can be worked out again expanding
the kinetic and mass terms. We obtain:
\begin{eqnarray}
\mathcal{L}_{int} &\sim &\mathcal{O}\left( \frac{1}{2}\right) \frac{\delta
\phi _{1}}{M_P}\partial _{\mu }\varphi_{hid}\partial ^{\mu }\varphi
_{hid}+\mathcal{O}\left( \frac{\sqrt{\vo}}{6\cdot 10^{1/4}}\right)
\frac{\delta \phi_2}{M_P}\partial _{\mu }\varphi_{hid}\partial ^{\mu
}\varphi _{hid} \notag \\
&&+\mathcal{O}\left(\frac{10^{1/4}}{3\sqrt{\vo}}\right)
\frac{\delta\phi_3}{M_P}\partial _{\mu }\varphi_{hid}\partial ^{\mu }\varphi
_{hid}.
\end{eqnarray}
\newline
\textbf{Moduli decay rates}
\newline
\newline
Having derived the moduli couplings to all ordinary and supersymmetric particles,
let us now compute the volume scaling of all the possible decay rates.
\bigskip

$\bullet \textbf{ Decays to visible gauge bosons}$

\bigskip
The moduli decay rates to visible gauge bosons look like:
\begin{eqnarray}
\Gamma_{\delta\phi_1 \to gg}&\simeq&\left[\frac{1}{64\pi (\ln\vo)^2}\right]\frac{m_1^3}
{M_P^2}\simeq \left[\frac{1}{64\pi(\ln\vo)^{7/2}}\right]\frac{M_P}{\vo^{9/2}}\simeq
\left[\frac{5\cdot 10^{-3}}{(\ln\vo)^{7/2}}\right]\frac{M_P}{\vo^{9/2}}, \\
\Gamma_{\delta\phi_2\to gg}&\simeq&\left(\frac{\sqrt{10}}{64\pi \vo}\right)\frac{m_2^3}
{M_P^2}\simeq\left[\frac{\sqrt{10}(\ln\vo)^3}{8\pi}\right]\frac{M_P}{\vo^4}\simeq
0.1\,(\ln\vo)^3\frac{M_P}{\vo^4}, \label{DecVisPhotons}\\
\Gamma_{\delta\phi_3\to gg}&\simeq&\left(\frac{\vo}{64\sqrt{10}\pi}\right)\frac{m_3^3}
{M_P^2}\simeq\left[\frac{(\ln\vo)^3}{8\sqrt{10}\pi}\right]\frac{M_P}{\vo^2}\simeq 0.01\,(\ln\vo)^3\frac{M_P}{\vo^2}.
\end{eqnarray}

$\bullet \textbf{ Decays to matter fermions}$

\bigskip

As we have already pointed out, for temperatures above the EW symmetry breaking scale, $T>M_{EW}$,
there is no direct modulus-fermion-fermion coupling and so there is no direct modulus decay to
two ordinary fermions. On the other hand, we need to consider a three-body modulus decay
to gauge boson-fermion-fermion obtaining\footnote{Notice that we could still have a two-body modulus decay to
fermion-fermion via a 1-loop process. However this process is very suppressed with respect to the three-body decay since
$\Gamma_{\delta\phi\to \overline{\psi}\psi}^{1-loop}\sim \alpha_{EM}^2\Gamma_{\delta\phi\to g g}
\sim 10^{-4} \Gamma_{\delta\phi\to g g}$.}:
\be
\Gamma_{\delta\phi_1 \to g\overline{\psi}\psi}\simeq 0.01 (\ln\vo)^2 \,\Gamma_{\delta\phi_1 \to gg},
\text{ \ \ }\Gamma_{\delta\phi_i \to g\overline{\psi}\psi}\simeq 0.01 \,\Gamma_{\delta\phi_i \to gg},
\text{ \ \ \ }\forall i=2,3.
\ee

$\bullet \textbf{ Decays to visible Higgs fields}$

\bigskip

Given that the SUSY breaking contribution to the Higgs mass $m_H\sim M_P \vo^{-1}$ is higher than
$m_1\sim M_P \vo^{-3/2}$, the decay of the large modulus $\delta\phi_1$ to two Higgs fields is kinematically forbidden.
On the other hand the small blow-up modes are heavier than $H$ by a $(\ln\vo)$ factor, as can be seen from (\ref{masse}):
$m_2\sim m_3 \sim M_P (\ln\vo) / \vo$. Hence both $\delta\phi_2$ and $\delta\phi_3$ can decay to two Higgs fields
with the following decay rates:
\begin{eqnarray}
\Gamma_{\delta\phi_2\to \overline{H}H}&\simeq&\left(\frac{\sqrt{10}}{144\pi \vo}\right)\frac{m_2^3}
{M_P^2}\simeq\left[\frac{\sqrt{10}(\ln\vo)^3}{18\pi}\right]\frac{M_P}{\vo^4}\simeq
0.05\,(\ln\vo)^3\frac{M_P}{\vo^4}, \\
\Gamma_{\delta\phi_3\to \overline{H}H}&\simeq&\left(\frac{\vo}{576\sqrt{10}\pi}\right)\frac{m_3^3}
{M_P^2}\simeq\left[\frac{(\ln\vo)^3}{72\sqrt{10}\pi}\right]\frac{M_P}{\vo^2}\simeq 0.001\,(\ln\vo)^3\frac{M_P}{\vo^2}.
\end{eqnarray}

$\bullet \textbf{ Decays to visible gauginos}$

\bigskip

For the same kinematical reasons pointed out in the case of the moduli decays to Higgs fields,
only the small blow-up modes can decay to visible gauginos with the following decay rates:
\begin{eqnarray}
\Gamma_{\delta\phi_2\to \lambda\lambda}&\simeq&\left(\frac{9\sqrt{10}}{128\pi \vo^3}\right)m_2
\simeq\left(\frac{9\sqrt{10}\ln\vo}{64\pi \vo^3}\right)\frac{M_P}{\vo^4}\simeq
0.1\,(\ln\vo)\frac{M_P}{\vo^4}, \\
\Gamma_{\delta\phi_3\to g\lambda\lambda}&\simeq&\left(\frac{\sqrt{10}\vo}{288\pi^3}\right)\frac{m_3^3}
{M_P^2}\simeq\left[\frac{\sqrt{10}(\ln\vo)^3}{36\pi^3}\right]\frac{M_P}{\vo^2}\simeq 10^{-3}\,(\ln\vo)^3\frac{M_P}{\vo^2}.
\end{eqnarray}

$\bullet \textbf{ Decays to visible SUSY scalars}$

\bigskip

Only the heaviest moduli $\delta\phi_2$ and $\delta\phi_3$ can decay to squarks and sleptons
since $m_1 \sim M_P \vo^{-3/2} < m_0\sim M_P \vo^{-1}$.
The decay rates of the small blow-up modes to SUSY scalars scale as $\Gamma_{\delta\phi_i\to \bar{\varphi} \varphi} \sim
\Gamma_{\delta\phi_i\to \overline{H} H}$, $\forall i=2,3$.
\bigskip

$\bullet \textbf{ Decays to visible Higgsinos}$

\bigskip

For the same kinematical reasons stressed above, only the small blow-up modes can decay
into two Higgsinos with decay rates that scale as:
\be
\Gamma_{\delta\phi_2\to \tilde{H}\tilde{H}}\simeq
0.05\,(\ln\vo)\frac{M_P}{\vo^4},\text{ \ \ \ \ }
\Gamma_{\delta\phi_3\to g\tilde{H}\tilde{H}}\simeq 10^{-4}\,(\ln\vo)^3\frac{M_P}{\vo^2}.
\ee
\newpage
$\bullet \textbf{ Decays to the lightest modulus}$

\bigskip

The only kinematically allowed moduli decays to other moduli are $\delta\phi_i\to\delta\phi_1+\delta\phi_1$, $\forall \,i=2,3$
with the corresponding decay rates:
\be
\Gamma_{\delta\phi_i\to\delta\phi_1\delta\phi_1}\simeq \left(\frac{1}{16\pi \vo^5}\right)\frac{M_P^2}{m_i}\simeq 10^{-2}\,(\ln\vo)^{-1}\frac{M_P}{\vo^4},\,\,\,\,\forall\,i=2,3.
\ee

$\bullet \textbf{ Decays to hidden gauge bosons}$

\bigskip

The moduli decay rates to hidden gauge bosons $X$ look like:
\begin{eqnarray}
\Gamma_{\delta\phi_1 \to XX}&\simeq&
\left[\frac{5\cdot 10^{-3}}{(\ln\vo)^{7/2}}\right]\frac{M_P}{\vo^{9/2}}, \\
\Gamma_{\delta\phi_2\to XX}&\simeq& 0.01\,(\ln\vo)^3\frac{M_P}{\vo^2}, \\
\Gamma_{\delta\phi_3\to XX}&\simeq&
0.1\,(\ln\vo)^3\frac{M_P}{\vo^4}.
\end{eqnarray}

$\bullet \textbf{ Decays to hidden gauginos}$

\bigskip

Given that $\delta\phi_1$ is lighter than the hidden gauginos, only $\delta\phi_2$ and $\delta\phi_3$ can decay to these particles
with corresponding decay rates:
\begin{eqnarray}
\Gamma_{\delta\phi_2\to \lambda_{hid}\lambda_{hid}}&\simeq& \left(\frac{(\ln\vo)^2}{8\sqrt{10}\pi\vo}\right)m_2\simeq
0.01 \,(\ln\vo)^3\frac{M_P}{\vo^2}, \\
\Gamma_{\delta\phi_3\to \lambda_{hid}\lambda_{hid}}&\simeq& \left(\frac{\sqrt{10}(\ln\vo)^2}{8\pi\vo^3}\right)m_3
\simeq 0.1\, (\ln\vo)^3\frac{M_P}{\vo^4}.
\end{eqnarray}

$\bullet \textbf{ Decays to hidden massless pion-like mesons}$

\bigskip

The moduli decay rates to hidden massless pions scale as:
\begin{eqnarray}
\Gamma_{\delta\phi_1 \to\pi\pi}&\simeq&\left(\frac{1}{256\pi}\right)\frac{m_1^3}
{M_P^2}\simeq \left[\frac{1}{256\pi(\ln\vo)^{3/2}}\right]\frac{M_P}{\vo^{9/2}}\simeq
10^{-3} (\ln\vo)^{-3/2}\frac{M_P}{\vo^{9/2}}, \\
\Gamma_{\delta\phi_2\to \pi\pi}&\simeq&\left(\frac{\vo}{2304\sqrt{10}\pi}\right)\frac{m_2^3}
{M_P^2}\simeq\left[\frac{(\ln\vo)^3}{288\sqrt{10}\pi}\right]\frac{M_P}{\vo^2}\simeq
10^{-4}\,(\ln\vo)^3\frac{M_P}{\vo^2}, \\
\Gamma_{\delta\phi_3\to \pi\pi}&\simeq&\left(\frac{\sqrt{10}}{576\pi\vo}\right)\frac{m_3^3}
{M_P^2}\simeq\left[\frac{\sqrt{10}(\ln\vo)^3}{72\pi}\right]\frac{M_P}{\vo^4}\simeq 0.01\,(\ln\vo)^3\frac{M_P}{\vo^4}.
\end{eqnarray}

$\bullet \textbf{ Decays to hidden fermionic condensates}$

\bigskip

Only the heaviest moduli $\delta\phi_2$ and $\delta\phi_3$ can decay to hidden fermionic condensates with corresponding
decay rates that scale as:
\be
\Gamma_{\delta\phi_2\to \overline{\psi}_{hid}\psi_{hid}}\simeq
5\cdot 10^{-3} \,(\ln\vo)^3\frac{M_P}{\vo^2}, \text{ \ \ \ \ }
\Gamma_{\delta\phi_3\to \overline{\psi}_{hid}\psi_{hid}}
\simeq 5\cdot 10^{-2} \, (\ln\vo)^3\frac{M_P}{\vo^4}.
\ee

$\bullet \textbf{ Decays to hidden massive scalar condensates}$

\bigskip

Given that $\delta\phi_1$ is lighter than the hidden massive scalar condensates,
only $\delta\phi_2$ and $\delta\phi_3$ can decay to these particles
with corresponding decay rates:
\be
\Gamma_{\delta\phi_2\to \varphi_{hid}\varphi_{hid}}\simeq 10^{-4}\,(\ln\vo)^3\frac{M_P}{\vo^2},
\text{ \ \ \ \ }
\Gamma_{\delta\phi_3\to \varphi_{hid}\varphi_{hid}}\simeq
0.01\,(\ln\vo)^3\frac{M_P}{\vo^4}.
\ee
\newpage
$\textbf{ Summary}$

\begin{itemize}
\item Light large modulus $\delta\phi_1$: it can decay only to
hidden and visible gauge bosons, massless fermions and hidden pions with the following
decay widths:
\begin{gather}
\Gamma _{\delta \phi _{1}\rightarrow gg}^{TOT}\simeq N_{G}^{vis}\Gamma
_{1}\simeq 10\Gamma _{1},\text{ \ \ \ }\Gamma _{\delta \phi _{1}\rightarrow g\overline{
\psi }\psi }^{TOT}\simeq \frac{N_F}{100}(\ln\vo)^2\Gamma _{1}\simeq (\ln\vo)^2\Gamma _{1},  \notag \\
\Gamma _{\delta \phi _{1}\rightarrow XX}^{TOT}\simeq N_{G}^{hid}\Gamma _{1},
\text{ \ \ \ \ }\Gamma _{\delta \phi _{1}\rightarrow \pi \pi }^{TOT}\simeq
5 N_f^2 (\ln\vo)^2\Gamma _{1},
\end{gather}
where:
\be
N_F\equiv N_{F}^{EM}+N_{F}^{NC}+2N_{F}^{CC}+8N_{F}^{QCD}=138,\text{ \ \ \ }
\Gamma_1\equiv \left(\frac{5\cdot 10^{-3}}{(\ln\vo)^{7/2}}\right)\frac{M_P}{\vo^{9/2}},
\ee
and $N_G^{hid}=(N_c^2-1)$ for a pure SYM theory or $N_G^{hid}=(N_c-N_f)^2-1$ in the presence
of hidden sector matter fields.

\item Heavy blow-up mode $\delta\phi_2$ supporting the hidden sector: it decays to visible \textit{dof} with decay widths:
\begin{gather}
\Gamma _{\delta \phi _2\rightarrow gg}^{TOT}\simeq \frac{N_{G}^{vis}}{10}\frac{\Gamma_0}
{\vo^2}\simeq \frac{\Gamma_0}{\vo^2},\text{ \ \ \ }\Gamma _{\delta \phi _2\rightarrow g\overline{
\psi }\psi }^{TOT}\simeq \frac{N_F}{1000}\frac{\Gamma_0}
{\vo^2}\simeq 0.1 \frac{\Gamma_0}{\vo^2},\text{ \ \ }
\Gamma _{\delta \phi _2\rightarrow \overline{H}H}^{TOT}\simeq 0.1\frac{\Gamma_0}{\vo^2}, \notag \\
\Gamma _{\delta \phi _2\rightarrow \lambda \lambda }^{TOT}\simeq
\frac{1}{(\ln\vo)^2}\frac{\Gamma_0}{\vo^2}, \text{ \ \ \ \ }
\Gamma_{\delta\phi_2 \to \overline{\varphi}\varphi}^{TOT}\simeq \frac{\Gamma_0}{\vo^2},\text{ \ \ \ \ }
\Gamma_{\delta\phi_2\to \tilde{H}\tilde{H}}^{TOT}\simeq \frac{0.1}{(\ln\vo)^2}\frac{\Gamma_0}{\vo^2}, \notag
\end{gather}
whereas the decay rates to hidden sector \textit{dof} look like:
\begin{gather}
\Gamma _{\delta \phi_2\rightarrow XX}^{TOT}\simeq \frac{N_{G}^{hid}}{100}\Gamma_0,
\text{ \ \ \ }\Gamma _{\delta \phi_2\rightarrow \lambda_{hid}\lambda_{hid}}^{TOT}
\simeq \frac{N_{G}^{hid}}{100}\Gamma_0,
\text{ \ \ }\Gamma_{\delta\phi_2\to\delta\phi_1\delta\phi_1}^{TOT}\simeq \frac{10^{-2}}{(\ln\vo)^4}\frac{\Gamma_0}{\vo^2} \notag \\
\Gamma _{\delta \phi_2\rightarrow \pi \pi }^{TOT}\simeq
10^{-4}\,N_f^2\Gamma_0,\text{ \ \ }
\Gamma_{\delta\phi_2\to\overline{\psi}_{hid}\psi_{hid}}^{TOT}\simeq 5\cdot 10^{-3}N_f^2\Gamma_0,
\text{ \ \ }\Gamma_{\delta\phi_2\to \varphi_{hid}\varphi_{hid}}^{TOT}\simeq 10^{-4}N_f^2\Gamma_0, \notag
\end{gather}
where:
\be
\Gamma_0\equiv (\ln\vo)^3\frac{M_P}{\vo^2}.
\ee

\item Heavy blow-up mode $\delta\phi_3$ supporting the visible sector: it decays to visible \textit{dof} with decay widths:
\begin{gather}
\Gamma _{\delta \phi _3\rightarrow gg}^{TOT}\simeq \frac{N_{G}^{vis}}{100}\Gamma_0\simeq 0.1 \Gamma_0,
\text{ \ \ \ }\Gamma _{\delta \phi_3\rightarrow g\overline{
\psi }\psi }^{TOT}\simeq 10^{-4} N_F \Gamma_0 \simeq 0.01 \Gamma_0,  \text{ \ \ \ }
\Gamma _{\delta \phi _3\rightarrow \overline{H}H}^{TOT}\simeq 5\cdot 10^{-3}\Gamma_0, \notag \\
\Gamma _{\delta \phi _3\rightarrow g\lambda \lambda }^{TOT}\simeq
0.01\Gamma_0, \text{ \ \ \ }
\Gamma_{\delta\phi_3 \to \overline{\varphi}\varphi}^{TOT}\simeq 0.05\Gamma_0,\text{ \ \ \ }
\Gamma_{\delta\phi_3\to g\tilde{H}\tilde{H}}^{TOT}\simeq 5\cdot 10^{-3}\Gamma_0, \notag
\end{gather}
whereas the decay rates to hidden sector \textit{dof} look like:
\begin{gather}
\Gamma _{\delta \phi_3\rightarrow XX}^{TOT}\simeq 0.1\,N_{G}^{hid}\frac{\Gamma_0}{\vo^2},
\text{ \ \ \ }\Gamma _{\delta \phi_2\rightarrow \lambda_{hid}\lambda_{hid}}^{TOT}
\simeq 0.1\,N_{G}^{hid}\frac{\Gamma_0}{\vo^2},
\text{ \ \ }\Gamma_{\delta\phi_3\to\delta\phi_1\delta\phi_1}^{TOT}\simeq \frac{10^{-2}}{(\ln\vo)^4}\frac{\Gamma_0}{\vo^2} \notag \\
\Gamma _{\delta \phi_3\rightarrow \pi \pi }^{TOT}\simeq
0.01\,N_f^2\frac{\Gamma_0}{\vo^2},\text{ \ \ }
\Gamma_{\delta\phi_3\to\overline{\psi}_{hid}\psi_{hid}}^{TOT}\simeq 0.05 N_f^2\frac{\Gamma_0}{\vo^2},
\text{ \ \ }\Gamma_{\delta\phi_3\to \varphi_{hid}\varphi_{hid}}^{TOT}\simeq 0.01 N_f^2\frac{\Gamma_0}{\vo^2}. \notag
\end{gather}
\end{itemize}
We notice that, except for $\delta\phi_1$, the strongest moduli decay rates are to visible or hidden gauge bosons and so, from now on,
we shall focus only on those couplings.

\subsubsection{Quiver locus}

Let us now focus on the case when the cycle supporting the visible sector shrinks at zero size: $\tau_3\to 0$.
We shall first of all start with the canonical normalisation and then compute the moduli couplings.
\newline
\newline
\textbf{Canonical normalisation}
\newline
\newline
Let us assume that the modulus $\tau_3$ shrinks at zero size. This can be achieved by using $D$-terms which
generically take the form:
\begin{equation}
V_D\sim g^2\left(\sum_{i}q_i\frac{\partial K}{\partial \varphi_i
}\varphi_i-\xi_3\right)^2, \label{Dterms}
\end{equation}
where:
\begin{equation}
\xi_3\sim \frac{1}{\mathcal{V}}\int_{CY}J\wedge (F_3-B)\wedge D_3,
\label{FIterm}
\end{equation}
is the Fayet-Iliopoulos term associated with the turning on of the
gauge flux $F_3$ on $\tau_3$ (we denote with $D_3$ the 2-form dual
to the 4-cycle whose volume is given by $\tau_3$). If $\tau_3$ is
a blow-up mode which does not intersect with any other cycle,
then $\xi_3\sim \sqrt{\tau_3}/\vo$.

Given that in a large volume expansion $V_D\sim \mathcal{V}^{-2}$
whereas $V_F \sim \mathcal{V}^{-3}$, a non-vanishing $V_D$ would
give a run-away behaviour for the volume mode. Hence we need to
impose $V_D=0$. If the sum in (\ref{Dterms}) comprises also a SM
singlet $\varphi_s$, then this requirement fixes,
$|\varphi_s|\sim \sqrt{\xi_3}$, and then $\tau_3$ can be fixed
by string loop corrections as pointed out in \cite{GenAnalofLVS}.
However if in (\ref{Dterms}) there are no SM singlets, then
$V_D=0$ implies $\xi_3=0$ forcing $\tau_3$ to shrink to zero size \cite{quiver}.

At the quiver locus the
tree-level gauge kinetic function is given by the axio-dilaton $S$ and not by
$\tau_3$, hence in order to work out the coupling of the inflaton
to visible sector \textit{dof}, one has to include also $S$ in the
canonical normalisation. More precisely, the complete description of the
effective field theory at the quiver locus is obtained by expanding around the
singularity \cite{quiver}:
\begin{eqnarray}
K &\sim&-2\ln\left[\alpha\left(\tau_1^{3/2}-\gamma_2\tau_2^{3/2}\right)+\frac{s^{3/2}\xi}{2 }\right]
+\lambda\frac{\tau_3^2}{\vo'}-\ln\left(2s\right), \label{Kquiver} \\
W &\sim&W_0+A_2 e^{-a_2 T_2},
\label{Wquiver} \\
f &\sim& S+ h T_3, \label{fquiver}
\end{eqnarray}
where $s=$Re$(S)$, $h$ is a 1-loop factor and $\vo'=\alpha\left(\tau_1^{3/2}-\gamma_2\tau_2^{3/2}\right)$.
The real part of the axio-dilaton gives the string coupling
$\langle s\rangle=g_s^{-1}$ and it is flux-stabilised at tree-level so that we are in the perturbative regime:
$\langle s \rangle\sim \mathcal{O}(10)$.
The direct K\"{a}hler metric at leading order and around the minimum $\langle \tau_3 \rangle =0$ looks like:
\begin{equation}
\mathcal{K}=\frac{9}{8 \langle\tau_1\rangle^2}\left(
\begin{array}{cccc}
2/3 & -\gamma_2\epsilon_2 & 0 & \xi\epsilon_s/(2\alpha) \\
-\gamma_2\epsilon_2 & \gamma_2\epsilon_2^{-1}/3 & 0 & -\gamma_2\xi\epsilon_2\epsilon_s/(2\alpha) \\
0 & 0 & 4\lambda\sqrt{\langle\tau_1\rangle}/(9\alpha) & 0 \\
\xi \epsilon_s/(2\alpha) & -\gamma_2\xi\epsilon_2\epsilon_s/(2\alpha) & 0 & 2 \epsilon_s^{-4}/9
\end{array}
\right),  \label{LaDirettaquiver}
\end{equation}
with $\epsilon_2=\sqrt{\langle\tau_2\rangle/\langle\tau_1\rangle}\ll 1\,$ and
$\epsilon_s=\sqrt{\langle s\rangle/\langle\tau_1\rangle}\ll 1\,$.
It is easy to realise that at leading order, $\mathcal{K}$ is diagonal and in addition $\tau_3$
does not mix with any other modulus. In fact the leading order kinetic Lagrangian reads:
\begin{gather}
\mathcal{L}_{kin}^{lead}=\frac{3}{4\langle\tau_1\rangle^2}\partial_{\mu}(\delta\tau_1)\partial^{\mu}(\delta\tau_1)
+\frac{3\gamma_2}{8\langle\tau_1\rangle^{3/2}\sqrt{\langle\tau_2\rangle}}\partial_{\mu}(\delta\tau_2)\partial^{\mu}(\delta\tau_2) \notag \\
+\frac{\lambda}{2\alpha\langle\tau_1\rangle^{3/2}}\partial_{\mu}(\delta\tau_3)\partial^{\mu}(\delta\tau_3)
+\frac{1}{4\langle s\rangle^2}\partial_{\mu}(\delta s)\partial^{\mu}(\delta s),
\label{Lkinleadquiver}
\end{gather}
while the subleading bit is given by:
\be
\mathcal{L}_{kin}^{sublead}=-\frac{9\gamma_2\sqrt{\langle\tau_2\rangle}}{4\langle\tau_1\rangle^{5/2}}\partial_{\mu}(\delta\tau_1)\partial^{\mu}(\delta\tau_2)
+\frac{9\xi\sqrt{\langle s\rangle}}{8\alpha\langle\tau_1\rangle^{5/2}}\partial_{\mu}(\delta\tau_1)\partial^{\mu}(\delta s)
-\frac{9\gamma_2\xi\sqrt{\langle s\rangle}\sqrt{\langle\tau_2\rangle}}{8\alpha\langle\tau_1\rangle^3}\partial_{\mu}(\delta\tau_2)\partial^{\mu}(\delta s).
\label{Lkinsubleadquiver}
\ee
The kinetic Lagrangian (\ref{Lkinleadquiver}) can be put in the canonical form defining:
\be
\delta \tau_1=\sqrt{\frac{2}{3}}\langle\tau_1\rangle \delta\phi_1,\text{ \ }\delta\tau_2=\sqrt{\frac{4}{3\gamma_2}}\langle\tau_1\rangle^{3/4}\langle\tau_2\rangle^{1/4}\delta\phi_2,
\text{ \ }\delta\tau_3=\sqrt{\frac{\alpha}{\lambda}}\langle\tau_1\rangle^{3/4}\delta\phi_3,\text{ \ }\delta s=\sqrt{2}\,\langle s\rangle\,\delta \phi_s.
\ee
However the Lagrangian (\ref{Lkinsubleadquiver}) introduces a subleading mixing between $\delta\tau_1$, $\delta\tau_2$ and $\delta s$
whose volume scaling can be derived by imposing to have a vanishing $\mathcal{L}_{kin}^{sublead}$. We can start writing:
\begin{eqnarray}
\delta \tau_1 &=&\sqrt{\frac{2}{3}}\langle\tau_1\rangle \delta\phi_1
+k_{12}\langle\tau_1\rangle^{c_{12}}\delta \phi_2+k_{1s}\langle\tau_1\rangle^{c_{1s}}\delta \phi_s, \label{Cnorm1} \\
\delta \tau_2 &=& k_{21}\langle\tau_1\rangle^{c_{21}}\delta \phi_1+\sqrt{\frac{4}{3\gamma_2}}\langle\tau_1\rangle^{3/4}
\langle\tau_2\rangle^{1/4}\delta\phi_2
+k_{2s}\langle\tau_1\rangle^{c_{2s}}\delta\phi_s,
\label{Cnorm2} \\
\delta \tau _{3} &=&\sqrt{\frac{\alpha}{\lambda}}\langle\tau_1\rangle^{3/4}\delta\phi_3, \label{Cnorm3} \\
\delta s &=&k_{s1}\langle\tau_1\rangle^{c_{s1}}\delta \phi_1+k_{s2}\langle\tau_1\rangle^{c_{s2}}\delta\phi_2+
\sqrt{2}\,\langle s\rangle\,\delta \phi_s, \label{Cnorm4}
\end{eqnarray}
where the coefficients $k_{ij}$, $i\neq j$, $i,j=1,2,s$, can involve only powers of the small moduli
$\tau_2$ and $s$ while the $c$-coefficients have to satisfy the following constraints:
\be
c_{12}=c_{1s}<1,\text{ \ \ }c_{2s}<c_{21}<3/4,\text{ \ \ }c_{s2}<c_{s1}<0,
\label{constraints}
\ee
reflecting the fact that in (\ref{Lkinsubleadquiver}) the mixing of $\delta\tau_1$ with $\delta\tau_2$ and $\delta s$ takes
places at the same order in a large volume expansion whereas the mixing between $\delta\tau_2$ and $\delta s$ occurs only at
subleading order.
Now substituting the general form for the canonical normalisation (\ref{Cnorm1})-(\ref{Cnorm4}) in (\ref{Lkinleadquiver}) and (\ref{Lkinsubleadquiver}),
and then requiring a vanishing mixing between the moduli at leading order in a large volume expansion, we obtain the following results:
\be
c_{12}=c_{1s}=1/4,\text{ \ \ }c_{21}=0,\text{ \ \ }c_{2s}=-3/4,\text{ \ \ }c_{s1}=-3/4,\text{ \ \ }c_{s2}=-3/2,
\ee
which are clearly in agreement with the constraints (\ref{constraints}). Writing now the volume as $\vo\simeq \tau_1^{3/2}$,
the expressions (\ref{Cnorm1})-(\ref{Cnorm4}) take the final form:
\begin{eqnarray}
\delta \tau_1 &\sim&\mathcal{O}(\mathcal{V}^{2/3})\delta \phi_1
+\mathcal{O}(\mathcal{V}^{1/6})\delta \phi_2+\mathcal{O}(\vo^{1/6})\delta\phi_s\sim
\mathcal{O}(\mathcal{V}^{2/3})\delta \phi_1, \\
\delta \tau_2 &\sim&\mathcal{O}(1)\delta \phi_1+\mathcal{O}(\mathcal{V}
^{1/2})\delta \phi_2+\mathcal{O}(\vo^{-1/2})\delta\phi_s\sim \mathcal{O}(\mathcal{V}
^{1/2})\delta \phi_2, \\
\delta \tau_3 &\sim&\mathcal{O}(\mathcal{V}
^{1/2})\delta \phi_3, \label{dt3}\\
\delta s &\sim&\mathcal{O}(\vo^{-1/2})\delta \phi_1+\mathcal{O}(\mathcal{V}
^{-1})\delta \phi_2+\mathcal{O}(1)\delta\phi_s\sim \mathcal{O}(1)\delta\phi_s.
\label{ds}
\end{eqnarray}
\newline
\textbf{Moduli couplings}
\newline
\newline
The couplings of $\delta\phi_1$, $\delta\phi_2$ and $\delta\phi_s$ to visible gauge bosons
can be obtained from the tree-level gauge kinetic function $\text{Re}(f)_{tree}=s$
while the coupling of $\delta\phi_3$ to visible gauge bosons can be derived
from the 1-loop gauge kinetic function $\text{Re}(f)_{1-loop}=h \, \tau_3$.
In fact the kinetic terms read:
\begin{equation}
\mathcal{L}_{gauge}=- \frac{(s+h\tau_3)}{M_P}F_{\mu\nu}F^{\mu\nu}.
\end{equation}
We then expand $s$ and $\tau_3$ around their VEVs and go to the canonically
normalised field strength $G_{\mu\nu}$ defined as:
\begin{equation}
G_{\mu\nu}=2\sqrt{\langle s \rangle}F_{\mu\nu}, \label{redefinition}
\end{equation}
and obtain:
\begin{equation}
\mathcal{L}_{gauge}=-\frac{1}{4}G_{\mu\nu}G^{\mu\nu}-\frac{\delta s}
{4 M_P\langle s\rangle}G_{\mu\nu}G^{\mu\nu}-\frac{h \delta\tau_3}
{4 M_P\langle s\rangle}G_{\mu\nu}G^{\mu\nu}. \label{EFmunuEFmunu}
\end{equation}
Now by substituting in (\ref{EFmunuEFmunu}) the expressions (\ref{ds}) for $\delta s$
and (\ref{dt3}) for $\delta\tau_3$, we obtain the
moduli couplings to the visible gauge bosons.
It turns out that $\delta\phi_3$ couples as $1/M_s$, $\delta\phi_s$ as $1/M_P$,
$\delta\phi_1$ as $1/(M_P\sqrt{\vo})$, and $\delta\phi_2$ as $1/(M_P\vo)$.
Hence the coupling of $\delta\phi_2$ to visible sector gauge bosons
is even more suppressed than in the geometric case.
This result is not surprising since it is reflecting the presence
of a singularity besides the geometrical separation between the two blow-up cycles.

We finally stress that the cycle $\tau_2$ supporting the hidden sector responsible for the generation
of the non-perturbative superpotential, is fixed at a size larger than the string scale.
Thus the moduli couplings to hidden sector particles take the same form
as in section \ref{ModCoupGR}.

\subsection{Fibre Inflation}

In the case of FI, the overall volume reads:
\be
\vo=\alpha\left(\sqrt{\tau_1}\tau_2-\sum_i \gamma_i\tau_i^{3/2}\right),
\ee
and so the only difference with the case of Swiss-cheese Calabi-Yau manifolds
used for BI is that the overall volume can be approximated as $\vo\simeq\sqrt{\tau_1}\tau_2$
instead of just $\vo\simeq\tau_1^{3/2}$. In addition we have seen that, in this case,
both $\tau_1$ and $\tau_2$ are large 4-cycles stabilised at subleading 1-loop order.

This implies that the same considerations of the previous sections for the moduli canonical normalisation and couplings apply also
in this case where we have just to work out the `fine-structure' mixing between $\tau_1$ and $\tau_2$.
In fact, the same structure of the overall volume implies that we shall obtain the same mixing between large and
small moduli.

\subsubsection{Canonical normalisation without loop corrections}

Let us derive the mixing between the two
large moduli $\tau_1$ and $\tau_2$ once all the blow-up modes have
been fixed at their VEVs: $a_i\langle\tau_i\rangle\sim\ln\vo$. We shall start by neglecting the loop
corrections. Recalling the definition of $J$ from (\ref{minhea}),
the scalar potential schematically looks like:
\begin{equation}
V\simeq\frac{-J\left(\ln\vo\right)^{3/2}+\hat{\xi}}{\vo^3}
=J\left[\frac{-\left(\ln\vo\right)^{3/2}+\tilde{\xi}}{\vo^3}\right],
\label{Partenza}
\end{equation}
where $\tilde{\xi}\equiv\hat{\xi}/J$, $\vo\simeq \sqrt{\tau_1}\tau_2$ and the inverse K\"{a}hler
metric is given by:
\begin{equation}
K_{ij}^{-1}=4\left(
\begin{array}{cc}
\langle\tau_1\rangle^2 & \frac{\sqrt{\langle\tau_1\rangle}}{2}\left(\ln\vo\right)^{3/2} \\
\frac{\sqrt{\langle\tau_1\rangle}}{2}\left(\ln\vo\right)^{3/2} & \langle\tau_2\rangle^2/2
\end{array}
\right). \label{G}
\end{equation}
The presence of a flat direction can be inferred by the fact that
the first derivative of the scalar potential with respect to
$\tau_1$, $V_1$, is proportional to the derivative with respect to
$\tau_2$, $V_2$:
\begin{equation}
V_1=\left(\frac{\tau_2}{2\tau_1}\right) V_2,
\end{equation}
therefore the vanishing of $V_2$ automatically implies also the
vanishing of $V_1$. The only direction which is fixed is the one
corresponding to the overall volume: $\vo\simeq
\exp{\left(\tilde{\xi}^{2/3}\right)}$.

In order to diagonalise simultaneously both the kinetic terms and
the mass matrix around the minimum, one has to find the
eigenvalues and eigenvectors of the following matrix:
\begin{equation}
M_{ik}^{2}=\left. K_{ij}^{-1}V_{jk}\right\vert _{\mathcal{V}=\exp
(\tilde{\xi}^{2/3})}=\frac{9}{2}\frac{\sqrt{\ln
\mathcal{V}}}{\mathcal{V}^{3}}\left(
\begin{array}{cc}
1+\epsilon & \frac{2\langle\tau_1\rangle}{\langle\tau_2\rangle}(1+\epsilon) \\
\frac{\langle\tau_2\rangle}{\langle\tau_1\rangle}(1+\epsilon) & 2+\epsilon
\end{array}
\right),\text{ \ with \ }\epsilon=\mathcal{O}\left(\frac{(\ln\vo)^{3/2}}{\vo}\right)\ll 1, \notag
\end{equation}
where this matrix has been evaluated at:
\begin{equation}
\tilde{\xi}=\frac{1}{2}\left[2\left(\ln\vo\right)^{3/2}-\sqrt{\ln\vo}\right]
\simeq \left(\ln\vo\right)^{3/2}, \label{xisol}
\end{equation}
which is the solution of $V_1\propto V_2=0$. The eigenvalues with
the corresponding eigenvectors turn out to be:
\begin{eqnarray}
m_1^2 &=&0\text{ \ \ \ \ \ \ \ \ \ \ \ \ \
}\longleftrightarrow \text{ \ \ \ }\overrightarrow{v}_{1}=\left(
\begin{array}{c}
-\frac{2\langle\tau_1\rangle}{\langle\tau_2\rangle}x \\
x
\end{array}
\right) , \\
m_2^2 &\simeq &J\frac{\sqrt{\ln \mathcal{V}}}{\mathcal{V}^{3}}
\text{ \ \ \ }\longleftrightarrow \text{ \ \ \ \ }\overrightarrow{v}
_{2}=\left(
\begin{array}{c}
\frac{\langle\tau_1\rangle}{\langle\tau_2\rangle}y(1+\epsilon) \\
y
\end{array}
\right),
\label{mass2}
\end{eqnarray}
and so the canonical normalisation around the minimum looks like:
\begin{equation}
\left(
\begin{array}{c}
\delta \tau_1 \\
\delta \tau_2
\end{array}
\right)
=\overrightarrow{v}_{1}\frac{\delta\phi_1}{\sqrt{2}}+\overrightarrow{v}_{2}\frac{\delta\phi_2}{\sqrt{2}}
=\left(
\begin{array}{c}
-\frac{2\langle\tau_1\rangle}{\langle\tau_2\rangle}x \\
x
\end{array}
\right) \frac{\delta\phi_1}{\sqrt{2}}+\left(
\begin{array}{c}
\frac{\langle\tau_1\rangle}{\langle\tau_2\rangle}y(1+\epsilon) \\
y
\end{array}
\right) \frac{\delta\phi_2}{\sqrt{2}}.
\end{equation}
Notice that from (\ref{minhea}):
\begin{equation}
J\simeq\frac{\hat{\xi}}{\left(a_i\langle\tau_i\rangle\right)^{3/2}}\simeq
\frac{\mathcal{O}(1)}{\left(g_s \ln\vo\right)^{3/2}},
\end{equation}
and so the non-vanishing mass-squared (\ref{mass2}) scales exactly as (\ref{MASSE}):
\begin{equation}
m_{2}^{2} \simeq \frac{\mathcal{O}(1)}{g_s^{3/2}\vo^3\ln\vo}.
\end{equation}
In order to work out the values of $x$ and $y$, we need to impose
$\overrightarrow{v}^T_{\alpha}\cdot\mathcal{K}\cdot\overrightarrow{v}_{\beta}=\delta_{\alpha\beta}$
which implies $x=y/\sqrt{2}=2\langle\tau_2\rangle (1+\epsilon)/\sqrt{6}$. Therefore the final
canonical normalisation looks like:
\begin{equation}
\left\{
\begin{array}{c}
\frac{\delta \tau_1}{\langle\tau_1\rangle}=-\left[\frac{2}{\sqrt{3}}(1+\epsilon)\right]\delta\phi_1+\left[\sqrt{\frac{2}{3}}
(1+\epsilon)\right]\delta\phi_2, \\
\frac{\delta \tau_2}{\langle\tau_2\rangle}=\left[\frac{1}{\sqrt{3}}(1+\epsilon)\right]\delta\phi_1+\left[\sqrt{\frac{2}{3}}
(1+\epsilon)\right]\delta\phi_2.
\end{array}
\right.
\label{fgh}
\end{equation}
The physical interpretation of these two mass-eigenstates becomes
clearer once we notice that from $\vo\simeq\sqrt{\tau_1}\tau_2$, we obtain:
\begin{equation}
\frac{1}{\sqrt{3}}\frac{\delta\vo}{\vo}=\frac{1}{2\sqrt{3}}\frac{\delta\tau_1}{\langle\tau_1\rangle}
+\frac{1}{\sqrt{3}}\frac{\delta\tau_2}{\langle\tau_2\rangle}=\delta\phi_2(1+\epsilon).
\label{CNchi2noloop}
\end{equation}
Hence $\delta\phi_2$ plays the r\^{o}le of the overall volume. This is the reason why this
mass-eigenstate has a mass-squared of the order $\vo^{-3}$ given
that the volume is stabilised at that order. On the other hand,
$\delta\phi_1$ cannot be expressed as a function of just the overall volume, since:
\begin{equation}
\delta\phi_1(1 + \epsilon)=\frac{1}{\sqrt{3}}\frac{\delta\vo}{\vo}-\frac{\sqrt{3}}{2}\frac{\delta\tau_1}{\langle\tau_1\rangle}.
\label{CNchi1noloop}
\end{equation}
Due to the dependence of $\delta\phi_1$ on $\delta\tau_1$ and the
fact that, without loop corrections, $\tau_1$ is a flat direction,
the mass-eigenstate $\delta\phi_1$ correctly turns out to be
massless.

On top of this, at this level of analysis, besides being still massless, $\delta\phi_1$
does not couple to any gauge boson living on a stack of $D7$-branes wrapping any of the small
blow-up cycles. This is because, at this stage, $\delta\phi_1$ does not mix with any blow-up mode.
In order to see this important implication, let us now expand the blow-up modes around their VEVs.
The leading order kinetic Lagrangian then looks like (setting without loss of generality $\alpha=\gamma_i=1$ $\forall i$):
\bea -{\cal L}_{kin}
&=&\frac{\partial^{\mu} (\delta\tau_1)\partial_{\mu} (\delta\tau_1)}{4 \langle\tau_1\rangle^2}
+\frac{\partial^{\mu} (\delta\tau_2)\partial_{\mu} (\delta\tau_2)}{2\langle\tau_2\rangle^2} +\frac{3
}{8}\sum_i \frac{\partial^{\mu} (\delta\tau_i)\partial_{\mu} (\delta\tau_i) }
{\vo\sqrt{\langle\tau_i\rangle}}+\frac{\sum_i\langle\tau_i\rangle^{3/2}}{2
\vo}\frac{\partial^{\mu}(\delta\tau_1)}{\langle\tau_1\rangle}
\frac{\partial_{\mu}(\delta\tau_2)}{\langle\tau_2\rangle} \nonumber  \\
&-& \frac{3}{2}\sum_i \frac{\sqrt{\langle\tau_i\rangle}}{\vo} \left(\frac12\frac{\partial^{\mu}(\delta \tau_1)}{\langle\tau_1\rangle}+
\frac{\partial^{\mu}(\delta \tau_2)}{\langle\tau_2\rangle}
 \right) \partial_{\mu}(\delta \tau_i) +
 \frac{9}{4}\sum_{j>i}\frac{\sqrt{\langle\tau_i\rangle\langle\tau_j\rangle}}{\vo^2}\,
\partial^{\mu}(\delta \tau_i)\partial_{\mu}(\delta\tau_j),
\label{op}
\eea
which in terms of the canonical normalisation (\ref{fgh}) takes the form:
\begin{gather}
-{\cal L}_{kin}=\frac{1}{2}\partial_{\mu} (\delta\phi_1)\partial^{\mu}(\delta\phi_1)(1+\epsilon)+\frac{1}{2}\partial_{\mu} (\delta\phi_2)\partial^{\mu}(\delta\phi_2)(1+\epsilon)+\epsilon \partial_{\mu} (\delta\phi_1)\partial^{\mu}(\delta\phi_2) \nonumber \\
+\frac{3}{8}\sum_i \frac{\partial^{\mu} (\delta\tau_i)\partial_{\mu} (\delta\tau_i) }
{\vo\sqrt{\langle\tau_i\rangle}}-\frac{3\sqrt{3}}{2\sqrt{2}}\sum_i \frac{\sqrt{\langle\tau_i\rangle}}{\vo} \partial^{\mu}(\delta\phi_2)
 \partial_{\mu}(\delta \tau_i)(1+\epsilon) +
 \frac{9}{4}\sum_{j>i}\frac{\sqrt{\langle\tau_i\rangle\langle\tau_j\rangle}}{\vo^2}\,
\partial^{\mu}(\delta \tau_i)\partial_{\mu}(\delta\tau_j). \nonumber
\end{gather}
Therefore we realise that there is no mixing between $\delta\phi_1$ and any of the
small blow-up modes. As we pointed out at the beginning of this section,
we notice that the structure of the mixing between the small moduli has the same form
as the case of BI studied before.

\subsubsection{Canonical normalisation with loop corrections}

In this section we shall show how the introduction of string loop
corrections to the scalar potential (\ref{Partenza}), gives a mass to
$\delta\phi_1$ and introduces a coupling of this mode to the
gauge bosons living on a stack of $D7$-branes wrapping a blow-up
cycle. More precisely string loops will produce a subleading correction in
the canonical normalisation (\ref{fgh}) such that the term:
\be
\left(\frac12\frac{\partial^{\mu}(\delta \tau_1)}{\langle\tau_1\rangle}+
\frac{\partial^{\mu}(\delta \tau_2)}{\langle\tau_2\rangle}
 \right) =\sqrt{\frac{3}{2}}\partial^{\mu}(\delta\phi_2)(1+\epsilon),
\label{re}
\ee
in the kinetic Lagrangian (\ref{op}), will not depend on just $\delta\phi_2$ anymore but
a subleading dependence on $\delta\phi_1$ will also be introduced, generating a mixing
between $\delta\phi_1$ and any blow-up mode present in the theory. Let us start from the potential:
\begin{equation}
V=J\left[\frac{-\left(\ln\vo\right)^{3/2}+\tilde{\xi}}{\vo^3}+\frac{a_1}{\vo^3\sqrt{\tau_1}}\right],
\text{ \ \ with \ \ }a_1\simeq\frac{\mathcal{O}(1)}{J}
\simeq\mathcal{O}(1)\left(g_s\ln\vo\right)^{3/2},
\label{NewPartenza}
\end{equation}
where we have schematically introduced the leading order
dependence on the $g_s$ corrections. The way we shall proceed in
order to work out the loop corrections to the canonical
normalisation is the following one:
\begin{enumerate}
\item We shall still consider the inverse of the tree-level
K\"{a}hler metric given by (\ref{G}) but now with a loop corrected
scalar potential (\ref{NewPartenza});

\item We shall evaluate the matrix $M^2_{ik}=K^{-1}_{ij}V_{jk}$
(where now $V_{jk}$ are second derivatives of (\ref{NewPartenza}))
not at (\ref{xisol}) but at the loop corrected solution for
$\tilde{\xi}$;

\item We can obtain the loop corrected solution for $\tilde{\xi}$ by
noticing that the $g_s$ corrections, besides introducing in the scalar potential a
dependence on $\tau_1$, introduce also a subleading dependence on
$\vo$. Therefore the volume minimum will acquire a tiny
$\tau_1$-dependent shift $\vo=\vo_0+\delta\vo(\tau_1)$ which has
already been derived in appendix A.1 of \cite{fiberinfl}. We can
use that result, plug the loop corrected expression for the volume
in (\ref{xisol}) and then obtain the loop corrected result for
$\tilde{\xi}$ by Taylor expansion. Following this procedure, we
obtain:
\begin{equation}
\tilde{\xi}=\tilde{\xi}_0+\frac{3 a_1}{\sqrt{\langle\tau_1\rangle}},
\label{xisolloop}
\end{equation}
where with $\tilde{\xi}_0$ we have denoted the leading order
solution given by (\ref{xisol}).
\item We will assume
that $\tau_1$ is stabilised by the remaining 1-loop terms in the potential
which we have neglected since we are just interested in the leading order volume scaling of the
corrections to canonical normalisation (\ref{fgh}).
\end{enumerate}
The new eigenvalues with the corresponding eigenvectors turn out
to be:
\begin{eqnarray}
m_1^2 &\simeq&J\frac{a_1}{\vo^{3}\sqrt{\langle\tau_1\rangle}}\sim\frac{\mathcal{O}(1)}
{\vo^{3}\sqrt{\langle\tau_1\rangle}}\text{ \ \ \
}\longleftrightarrow \text{ \ \ \ }\overrightarrow{v}_{1}=\left(
\begin{array}{c}
-\frac{2\langle\tau_1\rangle}{\langle\tau_2\rangle} x (1+\delta)\\
x
\end{array}
\right),
\label{fgt} \\
m_2^2 &\simeq &J\frac{\sqrt{\ln \mathcal{V}}}{\mathcal{V}^{3}}\sim
\frac{\mathcal{O}(1)}{g_s^{3/2}\mathcal{V}^{3}\ln\vo}
\text{ \ \ \ }\longleftrightarrow \text{ \ \ \ \ }\overrightarrow{v}%
_{2}=\left(
\begin{array}{c}
\frac{\langle\tau_1\rangle}{\langle\tau_2\rangle} y (1+\delta)\\
y
\end{array}
\right),
\end{eqnarray}
where $\delta=\mathcal{O}\left(\frac{1}{\sqrt{\langle\tau_1\rangle}\sqrt{\ln\vo}}\right)\ll 1$
since $\langle\tau_1\rangle=c \vo^{2/3}$,
and we have neglected corrections proportional to $\epsilon$ since $\epsilon\ll \delta$.
We see from (\ref{fgt}) that now $\delta\phi_1$ acquires a mass of the order $m_1^2\simeq c^{-1/2}\vo^{-10/3}$.
The values of $x$ and $y$ are now given by $x=y/\sqrt{2}=2\langle\tau_2\rangle (1+\delta)/\sqrt{6}$ and so the leading
order loop-corrected canonical normalisation around the minimum
looks like:
\begin{equation}
\left\{
\begin{array}{c}
\frac{\delta \tau_1}{\langle\tau_1\rangle}=-\left[\frac{2}{\sqrt{3}}(1+\delta/3)\right]\delta\phi_1+\left[\sqrt{\frac{2}{3}}
(1+2\delta/3)\right]\delta\phi_2, \\
\frac{\delta \tau_2}{\langle\tau_2\rangle}=\left[\frac{1}{\sqrt{3}}(1-2\delta/3)\right]\delta\phi_1+\left[\sqrt{\frac{2}{3}}
(1-\delta/3)\right]\delta\phi_2.
\end{array}
\right.
\label{fghNew}
\end{equation}
In terms of this new canonical normalisation, the expression (\ref{re}) now reads:
\be
\left(\frac12\frac{\partial^{\mu}(\delta \tau_1)}{\langle\tau_1\rangle}+
\frac{\partial^{\mu}(\delta \tau_2)}{\langle\tau_2\rangle}
 \right) =\sqrt{\frac{3}{2}}\partial^{\mu}(\delta\phi_2)+\frac{\delta}{\sqrt{3}}\partial^{\mu}(\delta\phi_1)
 \sim\mathcal{O}(1)\partial^{\mu}(\delta\phi_2)
 +\mathcal{O}\left(\frac{c^{-1/2}}{(\ln\vo)^{1/2}\vo^{1/3}}\right)\partial^{\mu}(\delta\phi_1), \notag
\ee
introducing a subleading mixing between $\delta\phi_1$ and any blow-up mode in the theory.
Therefore, for the geometric regime case, the final canonical normalisation will look like (\ref{C1})-(\ref{C3}) but now
with a subleading dependence on $\delta\phi_1$ of the form (for the case of two blow-up modes $\tau_3$ and $\tau_4$):
\begin{eqnarray}
\delta \tau_1 &\sim&\sum_{i=1}^2\mathcal{O}(\mathcal{V}^{2/3})\delta \phi_i
+\sum_{j=3}^4\mathcal{O}(\mathcal{V}^{1/6})\delta \phi_j\sim
\sum_{i=1}^2\mathcal{O}(\mathcal{V}^{2/3})\delta \phi_i, \label{D1} \\
\delta \tau_2 &\sim&\sum_{i=1}^2\mathcal{O}(\mathcal{V}^{2/3})\delta \phi_i
+\sum_{j=3}^4\mathcal{O}(\mathcal{V}^{1/6})\delta \phi_j\sim
\sum_{i=1}^2\mathcal{O}(\mathcal{V}^{2/3})\delta \phi_i, \label{D2} \\
\delta \tau_3 &\sim&\mathcal{O}(c^{-1/2}\mathcal{V}
^{-1/3})\delta \phi_1+\mathcal{O}(1)\delta \phi_2+\mathcal{O}(\mathcal{V}
^{1/2})\delta \phi_3+\mathcal{O}(\mathcal{V}^{-1/2})\delta
\phi_4\sim \mathcal{O}(\mathcal{V}
^{1/2})\delta \phi_3,
\label{D3} \\
\delta \tau_4 &\sim&\mathcal{O}(c^{-1/2}\mathcal{V}
^{-1/3})\delta \phi_1+\mathcal{O}(1)\delta \phi_2+\mathcal{O}(\mathcal{V}^{-1/2})\delta
\phi_3+\mathcal{O}(\mathcal{V}
^{1/2})\delta \phi _4\sim \mathcal{O}(\mathcal{V}
^{1/2})\delta \phi_4. \label{D4}
\end{eqnarray}
Similar considerations apply also for the case at the quiver locus.

\subsubsection{Moduli couplings}

The moduli couplings for the case of FI have the same behaviour as in the case of BI
since the canonical normalisation has the same structure. The only difference is
for the large modulus $\delta\phi_1$ since $\delta\phi_2$ behaves exactly as the large
overall volume mode of the previous case. Hence we shall just focus
on the couplings and decay rates of $\delta\phi_1$. As we have stressed before,
the largest couplings are to visible and hidden gauge bosons, and on top of this,
$\delta\phi_1$ is so light that it can decay only to those particles.
Hence we need just to derive the strength of these interactions.

Assuming that $\tau_3$ is supporting a hidden sector while
$\tau_4$ the visible sector, the fact that (\ref{D3}) and (\ref{D4}) have the same
volume dependence on $\delta\phi_1$, implies that $\delta\phi_1$ couples to hidden gauge bosons
$X_3$ and visible gauge bosons $g\equiv X_4$ with the same strength:
\be
\mathcal{L}_{\delta\phi_1 X_i X_i}\sim\left(\frac{c^{-1/2}}
{40 (\ln\vo)^{1/2}\,\vo^{1/3}}\right)\frac{\delta\phi_1}{M_P} G^{(i)}_{\mu\nu}G_{(i)}^{\mu\nu},\text{ \ \ }i=3,4.
\label{pl1}
\ee
In addition, also $\tau_1$ and $\tau_2$ are wrapped by the stacks of $D7$-branes which
source the string loop corrections that stabilise the flat direction orthogonal
to the overall volume. Thus each of them also supports a hidden sector.
The fact that (\ref{D1}) and (\ref{D2}) have the same
volume dependence on $\delta\phi_1$, implies that $\delta\phi_1$ couples to hidden gauge bosons
$X_1$ and $X_2$ with the same strength:
\be
\mathcal{L}_{\delta\phi_1 X_i X_i}\sim\left(\frac{1}
{4 M_P}\right)\delta\phi_1 (G^{hid}_{(i)})_{\mu\nu}(G^{hid}_{(i)})^{\mu\nu},\text{ \ \ }i=1,2.
\label{plo1}
\ee
The decay rates corresponding to the couplings (\ref{pl1}) are:
\be
\Gamma_{\delta\phi_1 \to g g}\sim\Gamma_{\delta\phi_1 \to X_3 X_3}
\sim\left(\frac{1}{6400\pi\ln\vo\langle\tau_1\rangle}\right)\frac{m_1^3}{M_P^2}
\sim 5\cdot 10^{-5}\frac{M_P}{c^{7/4}(\ln\vo)\vo^{17/3}},
\ee
whereas the decay rates corresponding to the couplings (\ref{plo1}) read:
\be
\Gamma_{\delta\phi_1 \to X_1 X_1}\sim\Gamma_{\delta\phi_1 \to X_2 X_2}
\sim \left(\frac{1}{64\pi}\right)\frac{m_1^3}{M_P^2}\sim 5\cdot 10^{-3}\frac{M_P}{c^{3/4}\vo^5},
\ee
and so we conclude that $\delta\phi_1$ decays to $X_1$ and $X_2$ instead
of $g$ and $X_3$ since:
\be
\frac{\Gamma_{\delta\phi_1 \to gg}}{\Gamma_{\delta\phi_1 \to X_1 X_1}}
\sim \left(\frac{10^{-2}}{\ln\vo}\right)\frac{1}{c\,\vo^{2/3}}
=\left(\frac{10^{-2}}{\ln\vo}\right)\frac{1}{\langle\tau_1\rangle}\ll 1.
\ee
%
We finally point out that when $\tau_4$ shrinks to zero size, the coupling of $\delta\phi_1$
to visible gauge bosons is even more suppressed since it scales as $c^{-1/2}/(M_P\vo^{5/6})$.


\end{document}